\documentclass{aa}  
\usepackage{amsmath, amssymb, amsfonts, amscd}
\usepackage{natbib}
\usepackage{graphicx}
\graphicspath{ {./} }
\usepackage{txfonts}
\usepackage[final]{hyperref}
\usepackage{mathrsfs}
\usepackage{color}
\usepackage{mathabx}
\usepackage{supertabular}
\usepackage{longtable}
\usepackage{ragged2e}
\usepackage{tikz}
\usepackage{multirow}
\usepackage{siunitx}
\usepackage{xspace}
\usepackage[switch]{lineno}
\usepackage{environ}
\hypersetup{
   colorlinks=true,
   citecolor=blue,
   linkcolor=blue,
   urlcolor=black
   }

\definecolor{emerald}{rgb}{0.0,0.5,0.0}
\definecolor{brown}{RGB}{160,64,0.0}
\definecolor{darkbrown}{RGB}{135,54,0.0}
\definecolor{firebrick}{RGB}{178,34,34}
\definecolor{darkgreen}{RGB}{1,102,48}
\definecolor{blue-violet}{rgb}{0.54, 0.17, 0.89}
\definecolor{royalblue}{rgb}{0.254,0.412,0.882}
\definecolor{mygreen}{rgb}{0,0.392,0}
\definecolor{orangered}{rgb}{1,0.271,0}
\definecolor{crimson}{rgb}{0.86,0.08,0.24}
\newcommand{\LEt}[1]{}

\newif\ifrevised
\revisedfalse

\ifrevised
  
  \NewEnviron{Revised}{{\color{red}\BODY}}
\else
  
  \NewEnviron{Revised}{\BODY}
\fi

\def\inumber{i}                                 % Nombre imaginaire
\def\sign{{\rm sign}}                           % Signe 
\def\scale{\, \propto \,}                       % proportionnel à

\def\cot{{\rm cot}}

\newcommand{\ddroit}{{\rm d}}                           % d droit
\newcommand{\normvar}[1]{\tilde{#1}}  
\newcommand{\dd}[2]{\partial_{#2} #1}           % dérivée partielle
\newcommand{\DD}[2]{\dfrac{\ddroit #1}{\ddroit #2}}                     % dérivée droite
\newcommand{\scal}[2]{\langle #1 , #2 \rangle}  % Produit scalaire fonctions
\newcommand{\dotprod}[1]{\cdot}
\newcommand{\abs}[1]{\left| #1 \right|}         % Valeur absolue/module
\newcommand{\ceil}[1]{\lceil #1 \rceil}
\newcommand{\conj}[1]{\overline{#1}}
\newcommand{\infvar}[1]{\ddroit #1}

\newcommand{\kron}[2]{\delta_{#1, #2}}

\def\XX{X}

\def\surface{S}

\def\sphere{\mathcal{S}}

\def\llat{l}                                            % Latitudinal wavenumber

\def\mm{m}                                      % Zonal wavenumber
\def\kk{k}                                              % Indice supplémentaire

\def\jj{j}
\def\qq{q}

\def\Legendre{P}                                % Notation polynôme/fonctions de Legendre
\def\SPH{Y}     % Notation harmoniques sphérique

\newcommand{\Ylm}[2]{\SPH_{#1}^{#2}}

\def\ratioamp{\xi}
\def\ratiopeaks{\Xi}

\def\nn{n}                                              % Indice fonctions de Hough
\def\Hough{\Theta}                              % Notation fonction de Hough
\def\spinpar{\nu}                        % Paramètre de spin
\def\spinparm{\spinpar_\mm}
\def\alphapar{\alpha}
\def\betapar{\beta}
\def\gammapar{\gamma}
\def\betasurf{\betapar_{\isurf}}

\def\Usurf{\gravpot_{\iforcing ; \isurf}}
\def\thetapar{\vartheta}
\def\mucoef{\mu}

\newcommand{\LegP}[1]{\Legendre_{#1}}                     % Fonction de Legendre associée
\newcommand{\LegFnorm}[2]{\Legendre_{#1}^{#2}}          % Fonction de Legendre associée normée
\newcommand{\expo}[1]{{\rm e}^{#1}}                             % Exponentielle 
\newcommand{\integ}[4]{\int_{#3}^{#4} #1 \infvar{#2} }   % Integrale

\newcommand{\HoughF}[3]{\Hough_{#1}^{#2,#3}}            % Fonction de Hough
\newcommand{\Houghval}[3]{\Lambda_{#1}^{#2,#3}}         % Valeur propre de Houg
\def\Ggrav{G}                                   % Constante gravitation
\def\carbondiox{{\rm CO_2}}
\def\nitrogen{{\rm N_2}}
\def\methane{{\rm CH_4}}
\def\ipla{{\empty}}                              % Indice planète
\def\istar{\star}                                       % Indice étoile

\def\xx{x}

\def\zz{z}
\def\XX{X}

\def\col{\theta}                                % colatitude
\def\lon{\varphi}                             % longitude     
\def\time{t}                                    % temps 

\def\philag{\lon_{\rm lag}}
\def\theating{\time_{\rm H}}
\def\phiheating{\lon_{\rm H}}
\def\phiatm{\lon_{\rm A}}

\def\bpgain{A_0}
\def\bpquality{Q}

\def\smaxis{a}                                  % Demi-grand axe
\def\norb{n_\istar}                             % Moyen mouvement
\def\spinrate{\Omega}                   % Vitesse angulaire de rotation spin
\def\colstar{\col_{\istar}}
\def\lonstar{\lon_{\istar}}

\def\Mbody{M}                                   % Masse d'un corps
\def\Rbody{R}                                   % Rayon d'un corps
\def\Mpla{\Mbody_\ipla}                 % Masse de la planète
\def\Mstar{\Mbody_\istar}                       % Masse de l'étoile

\def\pressure{p}

\def\Cp{C_\pressure}

\def\Rpla{\Rbody_\ipla}                 % Rayon planète
\def\freq{\sigma}                               % Notation fréquence
\def\ggravi{g}                                  % Gravité de surface
\def\fnewton{\freq_{\rm C}}

\def\temperature{T}

\def\adiabexp{\Gamma_1}

\def\psurf{\pressure_{\rm s}}
\def\Tsurf{\temperature_\isurf}

\def\Tbulk{\overline{\temperature}}
\def\Temis{\temperature_{\rm e}}

\def\height{H}

\def\Rspec{R_{\rm s}}
\def\rcp{\kappa}

\def\zenithphi{\psi}				% Zenithal angle
\def\heightsurf{\height_{\isurf}}

\def\ftide{\freq}                                       % Fréquence de marée
\def\timeint{\Delta \time}
\def\fdiurnal{\omega}
\def\fdiurnalnorm{\tilde{\fdiurnal}}
\def\gravpot{U}                                 % Notation potentiel de marée

\def\isurf{{\rm s}}                             % Indice surface océanique

\def\iforcing{{\rm T}}                          % Indice tidal force

\def\Utide{\gravpot_\iforcing}          % Potentiel excitateur de marée
\def\flux{F}
\def\Fstar{\flux_{\istar}}
\newcommand{\fluxni}[2]{\normvar{\vartide{\flux}}_{#1}^{#2}}
\newcommand{\fluxnalfi}[2]{\hat{\vartide{\flux}}_{#1}^{#2}}
\def\fluxnlm{\fluxnalfi{\llat}{\mm}}
\def\fluxnnm{\fluxni{\nn}{\mm}}

\newcommand{\vartide}[1]{\delta  #1}

\def\presstide{\vartide{\pressure}}

\def\Jtide{J}
\def\presstidesurf{\presstide_{\isurf}}

\def\presstiden{\normvar{\presstide}}

\def\Jtiden{\normvar{\Jtide}}

\newcommand{\presstidenlm}[2]{\hat{\presstide}_{\isurf; #1}^{#2, \ftide}}
\def\presstidenquad{\hat{\presstide}_{\isurf ; 2}^{2,\ftide}}
\def\presstideeq{\presstide_{\isurf ; \rm eq}}
\def\presstideneq{\presstide_{\isurf ; \rm eq}^{\ftide}}

\newcommand{\vprofilei}[3]{#1_{#2}}
\newcommand{\vprofile}[1]{\vprofilei{#1}{\nn}{\mm,\ftide}}
\def\presstidev{\vprofile{\presstiden}}
\newcommand{\presstidevsurfi}[2]{\vprofilei{\presstiden}{{\isurf ;} #1}{#2,\ftide}}
\def\presstidevsurf{\presstidevsurfi{\nn}{\mm}}

\def\Jtidev{\vprofile{\Jtiden}}

\newcommand{\deptheqi}[1]{h_{#1}}

\def\kwave{k}                           % Notation nombre d'onde
\newcommand{\kverti}[1]{\kwave_{#1}}
\def\kvert{\kverti{\nn}}
\newcommand{\fresoi}[1]{\ftide_{{\rm L;} #1}}

\def\LOD{{\rm LOD}}
\newcommand{\LODresoi}[1]{\LOD_{{\rm L;}#1}}
\def\freson{\ftide_{{\rm L;} \nn}}
\def\disspar{\varepsilon}
\def\Sn{S_{\nn}}
\def\LODimplus{\LOD_{\rm I;+}}

\def\torque{\mathcal{T}}                        % Notation couple de marée

\newcommand{\torquei}[1]{\torque_{#1}}

\def\torquez{\torquei{\zz}}

\newcommand{\tidal}[1]{\delta #1}

\def\cshape{c_f}
\def\itherm{{\rm H}}

\def\bj{b_{\itherm}}

\def\dlnbet{\zeta}
\def\lambdapar{\lambda}

\def\fluxtiden{\tidal{\flux}_{\nn}}
\newcommand{\fluxtideni}[1]{\tidal{\flux}_{#1}}

\def\imod{{\rm M}}
\def\irp{{\rm R}}
\def\iip{{\rm I}}
\def\iplus{+}
\def\iminus{-}
\def\ilowfreq{{\rm LF}}

\newcommand{\XXi}[2]{\XX_{#1 {\rm ;} #2}}
\def\Xmaxsyn{\XXi{\iip}{\ilowfreq}}
\def\imdpmaxsyn{\Im \left\{ \presstidevsurf \left(\Xmaxsyn  \right) \right\}}
\def\XIrespm{\XXi{\iip}{\pm}}
\def\imdprespm{\Im \left\{ \presstidevsurf \left( \XIrespm \right) \right\}}
\def\XIresplus{\XXi{\iip}{\iplus}}
\def\imdpresplus{\Im \left\{ \presstidevsurf \left( \XIresplus \right) \right\}}
\def\dximsep{\Delta \XX_{\rm I}}
\def\XAresplus{\XXi{\imod}{\iplus}}
\def\absdpresplus{\abs{\presstidevsurf \left(\XAresplus \right)}}
\def\XRresplus{\XXi{\irp}{\iplus}}
\def\redpresplus{\Re \left\{ \presstidevsurf \left( \XRresplus \right) \right\}}
\def\XIresminus{\XXi{\iip}{\iminus}}
\def\imdpresminus{\Im \left\{ \presstidevsurf \left( \XIresminus \right) \right\}}

\def\XRresminus{\XXi{\irp}{\iminus}}
\def\redpresminus{\Re \left\{\presstidevsurf \left( \XRresminus \right) \right\}}
\def\dxmidres{\Delta \XX_{1/2 {\rm ;} \imod}}
\newcommand{\absdpresplusi}[2]{\abs{\presstidevsurfi{#1}{#2} \left( \XAresplus \right)}}

\def\phieq{\varphi_{\rm s ;  eq}}

\def\tmax{\time_{\rm max}}
\def\cequat{c_{\rm eq}}
\def\Lambn{\Houghval{\nn}{\mm}{\spinpar}}

\def\opacity{\eta}
\def\presstidevc{\vprofilei{\presstiden}{{\rm T}}{}}

\def\mucos{\mu}

\def\gcmdata{{\color{blue} \dag}}
\def\wuscale{{\color{blue} \ddag}}

\newcommand{\eq}[1]{Eq.~(\ref{#1})}
\newcommand{\eqs}[2]{Eqs.~(\ref{#1}) and~(\ref{#2})}

\newcommand{\append}[1]{Appendix~\ref{#1}}

\newcommand{\fig}[1]{Fig.~\ref{#1}}

\newcommand{\sect}[1]{Sect.~\ref{#1}}
\newcommand{\sects}[2]{Sects.~\ref{#1} and~\ref{#2}}
\newcommand{\sectsto}[2]{Sects.~\ref{#1} to~\ref{#2}}

\newcommand{\comments}[1]{}

\newcommand{\paperone}{Paper~I\xspace}
\newcommand{\papertwo}{Paper~II\xspace}
\newcommand{\paperthree}{Paper~III\xspace}

\usepackage{etoolbox}
\usepackage[refpage]{nomencl}
\providetoggle{nomsort}
\settoggle{nomsort}{true} % true = sort by use, false = sort as usual
\makeatletter
\iftoggle{nomsort}{%
    \let\old@@@nomenclature=\@@@nomenclature        
        \newcounter{@nomcount} \setcounter{@nomcount}{0}%
        \newcommand{\threedigits}[1]{\ifnum#1<100 0\two@digits{#1} \else \number#1\fi}
        \renewcommand\the@nomcount{\threedigits{\value{@nomcount}}}% Ensure 10>01
        \def\@@@nomenclature[#1]#2#3{% Taken from package documentation
          \addtocounter{@nomcount}{1}%
        \def\@tempa{#2}\def\@tempb{#3}%
          \protected@write\@nomenclaturefile{}%
          {\string\nomenclatureentry{\the@nomcount\nom@verb\@tempa @[{\nom@verb\@tempa}]%
          \begingroup\nom@verb\@tempb\protect\nomeqref{\theequation}%
          |nompageref}{\thepage}}%
          \endgroup
          \@esphack}%
      }{}
\makeatother

\newcommand{\mynomone}[3][section]{%
  \begingroup\edef\x{\endgroup
  \unexpanded{\nomenclature{#2}}%
    {\unexpanded{#3} \hspace*{\fill}  (\csname the#1\endcsname)}}\x}

\newcommand{\mynomtwo}[4][section]{%
  \begingroup\edef\x{\endgroup
  \unexpanded{\nomenclature[#2]{#3}}%
    {\unexpanded{#4} \hspace*{\fill}  (\csname the#1\endcsname)}}\x}

\renewcommand\nomgroup[1]{%
  \item[\bfseries
  \ifstrequal{#1}{A}{Acronyms}{%
  \ifstrequal{#1}{S}{Symbols}{%
  \ifstrequal{#1}{C}{Other Symbols}{}}}%
]}
\newcounter{logglabel}

\newcommand{\mynom}[3][S]{\nomenclature[#1]{#2}{#3~}}

\makenomenclature

\makeatletter
\newcommand{\subalign}[1]{%
  \vcenter{%
    \Let@ \restore@math@cr \default@tag
    \baselineskip\fontdimen10 \scriptfont\tw@
    \advance\baselineskip\fontdimen12 \scriptfont\tw@
    \lineskip\thr@@\fontdimen8 \scriptfont\thr@@
    \lineskiplimit\lineskip
    \ialign{\hfil$\m@th\scriptstyle##$&$\m@th\scriptstyle{}##$\hfil\crcr
      #1\crcr
    }%
  }%
}
\makeatother

\newcommand{\twolabelsat}[4]{%
\begin{tikzpicture}
\path (0,0) node {\vphantom{.}};
\path (#1,0) node[inner sep=0cm, outer sep=0cm] {#2};
\path (#3,0) node[inner sep=0cm, outer sep=0cm] {#4};
\end{tikzpicture}
}

\begin{document}

% \title{Closed-form solutions for thermal atmospheric tides}
%\title{Closed-form ab initio solutions to understand atmospheric thermal tides on rocky planets}
%\title{Analytical modelling of atmospheric tides \\and rotational dynamics of rocky planets}
 \title{An analytical framework for atmospheric tides on rocky planets}
 %\title{An analytical model of atmospheric tides on rocky planets}
\subtitle{II. Semidiurnal thermotidal torque}

  %\subtitle{I. }
  \author{Pierre Auclair-Desrotour\inst{1}\fnmsep\thanks{Corresponding author: \texttt{pierre.auclair-desrotour@obspm.fr}}
  \and Jacques Laskar\inst{1} 
  \and Mohammad Farhat\inst{2,3} 
  \and Gwenaël Boué\inst{1} 
     %\and  J. Leconte\inst{1}  %\fnmsep
          }

  \institute{LTE, Observatoire de Paris, Université PSL, Sorbonne Université, Univ. Lille, Laboratoire National de Métrologie et d'Essai, CNRS, 75014 Paris, France 
  %\email{pierre.auclair-desrotour@obspm.fr} 
    \and Department of Astronomy, University of California, Berkeley, Berkeley, CA 94720-3411, USA
    \and Department of Earth and Planetary Science, University of California, Berkeley, Berkeley, CA 94720-4767, USA
  }

  \date{Received ...; accepted ...}

  \abstract
  {The thermodynamic acceleration induced by atmospheric thermal tides has a major impact on the rotation of temperate rocky planets, where gravitational and radiative tidal forcings are of comparable strength. This acceleration drives planets away from synchronous rotation, thereby affecting their climate, large-scale circulation, and surface conditions.}
  {These implications underline the need to better constrain the dependence of the thermotidal torque on the tidal frequency and on the physical parameters characterising the star-planet system, as this relationship remains insufficiently understood. In particular, the role of resonantly excited Lamb waves requires clarification due to its potential impact on Earth's length of day\mynom[A]{LOD}{length of the day} (LOD) during the Precambrian.}
  {Building on a previous work, which provided a fully analytical formulation of the atmospheric tidal response of rocky planets, we apply this framework to isentropic and isothermal atmospheres. We perform a comprehensive functional analysis of the resulting torque and compare the analytical predictions with numerical solutions from earlier general circulation model\mynom[A]{GCM}{general circulation model} (GCM) studies. }
  {
The analytical model reproduces the present-day thermotidal bulge and torque inferred from barometric observations when direct absorption is assumed. It also agrees well with GCM results, accurately capturing the frequency dependence of both surface-pressure oscillation amplitudes and tidal torque for paleo-Earth, as well as that of the tidal torque for present-day Earth. More generally, the solutions capture the torque peak near synchronous rotation and the amplification associated with Lamb-wave resonance, although incomplete GCM spectra limit some quantitative comparisons. Near resonance, the GCM results exhibit a braking torque of larger amplitude than the accelerating torque. This asymmetry is reproduced by allowing tidal heating to lag behind insolation. The agreement with GCM spectra constrains the free parameters of our model. It further suggests that thermal tides in the benchmark simulations are driven predominantly by an indirect thermal forcing characterised by a substantial time delay, in contrast to the direct absorption of sunlight that dominates semidiurnal tides on present-day Earth.
}
   {}


\maketitle
\ifrevised \else \nolinenumbers \fi
%________________________________________________________________

\section{Introduction}

\begin{figure}
\centering
\includegraphics[width=0.48\textwidth,trim=0cm 0cm 21cm 14.5cm,clip]{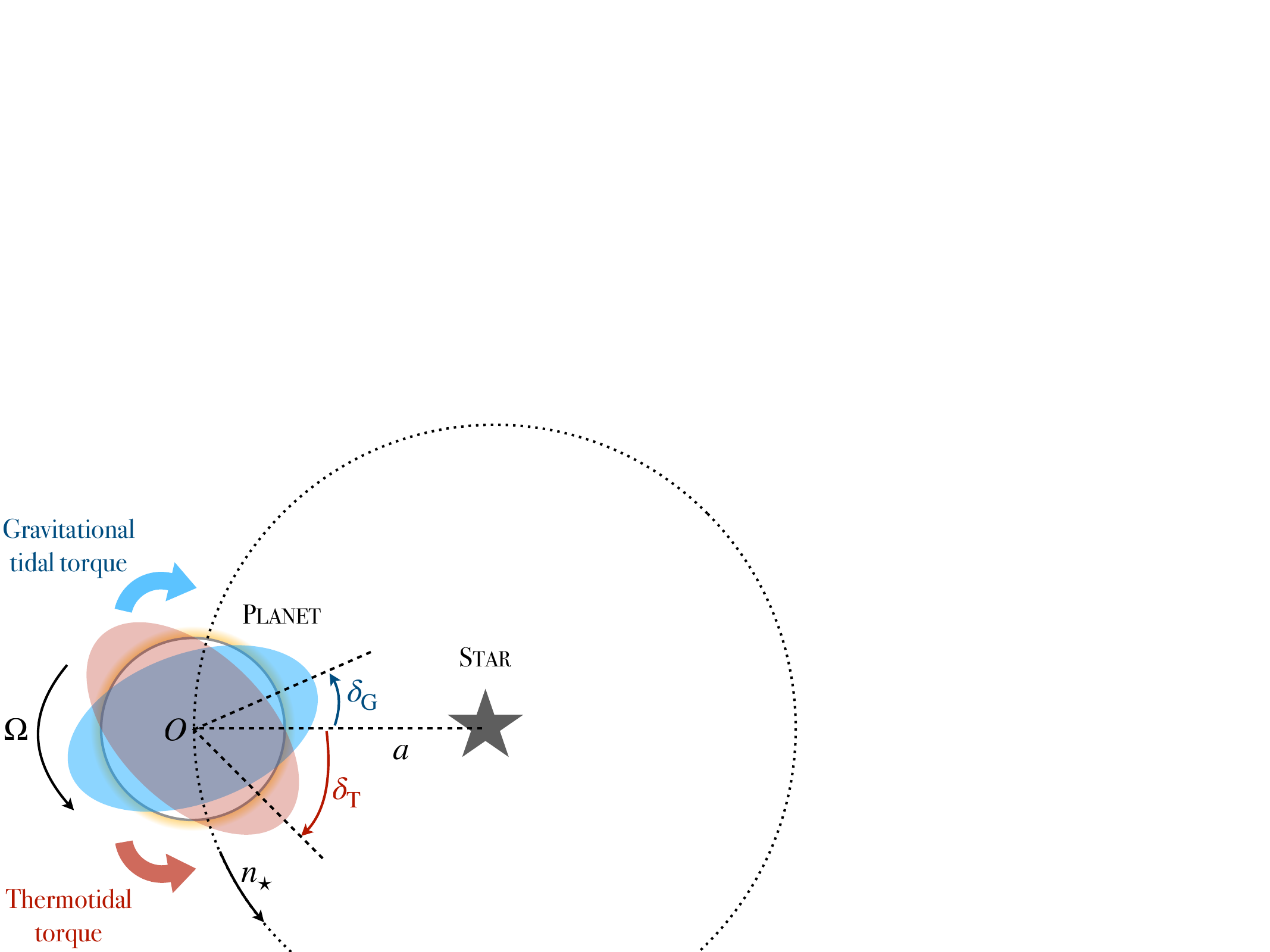}
    \caption{Schematic illustrating the competing effects of gravitational and atmospheric thermal tides on the rotation of rocky planets. As long as the pressure bulge (red ellipsoid) leads the planet-star axis, the thermal tide produces a torque that tends to accelerate the rotation, while the torque exerted on the bulge raised by gravitational tidal forces (blue ellipsoid) tends to decelerate it. Lag angles measured from the planet-star axis are denoted by $\delta_{\rm T}$ for the atmospheric thermal tide and $\delta_{\rm G}$ for the gravitational tide. }
    \label{fig:diagram_torque}
\end{figure}

The long-term rotational evolution of atmosphere-bearing rocky planets located within the habitable zone of their host stars is governed by both gravitational and thermal tides. The former arise from the differential stellar gravitational attraction, whereas the latter result from the day-night contrast in atmospheric heating by stellar radiation. In both cases, the tidal bulge is not aligned with the planet-star axis, leading to exchanges of energy and angular momentum between the orbital motion of the planet-star system and the planetary rotation. In the case of gravitational tides, this lag originates in dissipative mechanisms. By contrast, dissipation is not a prerequisite for thermal tides, which can still induce energy and angular momentum transfer even in its absence. For instance, non-dissipative isothermal models of Earth's atmosphere predict a semidiurnal surface-pressure oscillation peaking at 09:00 LST (Local Solar Time) on the dayside -- three hours before the insolation maximum at 12:00 LST \citep[see e.g.][Chapter~3, Eq.~(160)]{LC1969}\footnote{This maximum is slightly delayed in reality: barometric observations indicate that it occurs at 09:45 LST rather than 09:00 LST \citep[e.g.][]{Lindzen1978,Schindelegger2014}.}. This phase lead, illustrated by \fig{fig:diagram_torque}, gives rise to the long-recognised thermodynamic acceleration of Earth's rotation \citep[][]{Thomson1882}. 

In recent years, the role of thermal tides in the evolution of planetary rotation has attracted renewed attention in the context of Earth's rotational history. Several studies have revisited the tidal-locking hypothesis proposed by \cite{ZW1987}, who suggested that the Earth's length of day (hereafter LOD) may have remained nearly constant for several hundred million years as a result of a balance between thermal and gravitational tidal torques. This hypothesis relies on the resonant excitation of global-scale atmospheric compressibility modes known as Lamb waves \citep[e.g.][]{Farhat2024}. During the Precambrian era ($540{-}4570$~Ma), one such mode was resonantly excited, thereby amplifying thermal tides and the associated atmospheric torque. The existence of such a tidal-locking episode nevertheless remains a matter of debate, as conflicting interpretations have emerged. 

Drawing on geological constraints, several authors argue that the Earth's LOD remained stalled at approximately 19~hr for nearly one billion years \citep[][]{MK2023,Wu2023}. By contrast, other studies conclude that geological evidence does not support a tidal-locking phase and provide analytical arguments against this scenario based on linear tidal theory \citep[][]{Farhat2024,Laskar2024}. The theoretical framework underpinning these latter studies, however, remains to be further consolidated. Moreover, their analyses focus primarily on Earth, despite the fact that the underlying mechanism is expected to operate on rocky planets more generally. Robust modelling frameworks are therefore required to assess the influence of thermal tides on the rotational evolution of terrestrial planets and to properly account for tidal-orbital coupling in planetary evolution models. 
 
In a previous paper \citep[][hereafter \paperone]{AD2026a}, we developed a general analytical model of atmospheric tides on rocky planets within the framework of linear theory, starting from the primitive equations of fluid dynamics. The model is based on a series of simplifying assumptions consistent with the classical theory (see \paperone, Sects. 2 and 3). In particular, it incorporates the effective damping induced by dissipative processes through Newtonian cooling. The atmosphere is represented as a thin fluid layer at the planet's surface, with mean flows and horizontal gradients of the background equilibrium fields neglected. Tidal heating is prescribed to decay exponentially with the vertical coordinate. Under these assumptions, we derived a closed-form solution -- given by Eqs.~(85-92) in \paperone~-- that applies uniformly to two idealised atmospheric structures: (i) isentropic atmospheres, characterised by a globally uniform potential temperature and neutral stratification, and (ii) isothermal atmospheres, with globally uniform temperature and stable stratification. This solution describes the oscillations of the three-dimensional dynamical and thermodynamical fields (pressure, density, temperature, and velocities) driven by tidal gravitational forcing and radiative heating in the atmospheric shell. In particular, it captures the resonances of Lamb waves. 

In the present study (\papertwo), we extend the analytical framework developed in \paperone to investigate the influence of semidiurnal tides on planetary rotation. Our main objective is to quantify the thermotidal torque exerted on the planet's spin axis and to characterise its dependence on the tidal frequency. This requires examining the behaviour of Lamb modes resonantly excited by thermal forcing, as they play a major role in the rotational evolution of rapidly rotating planets such as Earth. More broadly, we assess the consistency between simplified analytical models and the numerical solutions obtained in previous studies from the simulations performed with general circulation models (GCM). These numerical models capture the complex coupling between atmospheric dynamics, radiative transfer, and surface heat exchanges, which lies beyond the reach of analytical theory. However, they generally do not allow the specific effects of individual physical processes to be disentangled. Owing to these limitations, it is essential to conduct a rigorous comparison between analytical and numerical approaches, and to identify the origins of their agreements and discrepancies. In a third paper (in prep.; hereafter \paperthree), it will be shown that a mass-spring-damper model can reproduce the obtained tidal solutions.

The present paper is organised as follows. In \sect{sec:thermotidal_torque}, we derive explicit formulations of the thermotidal torque as functions of the atmospheric parameters for both isentropic and isothermal atmospheres. Additionally, we apply the model to the present-day Earth, for which the semidiurnal tidal bulge is determined empirically from barometric measurements. Section~\ref{sec:frequency_behaviour} characterises the frequency dependence of the tidal response through a quantitative functional analysis. In \sect{sec:benchmark_gcm}, the analytical theory is benchmarked against recent GCM simulations addressing the cases of Earth, Venus and dry rocky planets. Finally, our main results are discussed and summarised in \sect{sec:conclusions}. Symbols and acronyms used in the study are listed in the nomenclature of \append{app:nomenclature}.

\begin{table*}[htb] 
\centering 
 \caption{\label{tab:Fl_coeff} Expansion coefficients of stellar irradiance at the substellar point as series of Legendre polynomials.  } 
\begin{tabular}{ccccccccccccccccc} 
\hline 
 \hline 
 \\[-0.3cm] 
$\llat$ & $0$ & $1$ & $2$ & $3$ & $4$ & $5$ & $6$ & $7$ & $8$ & $9$ & $10$ & $11$ & $12$ & $13$ & $14$ & $15$  \\
$\flux_\llat$ & $\dfrac{1}{4}$  & $\dfrac{1}{2}$  & $\dfrac{5}{16}$  & $0$  & $- \dfrac{3}{32}$  &  $0$ & $ \dfrac{13}{256}$  & $0$  & $- \dfrac{17}{512}$  & $0$  & $ \dfrac{49}{2048}$ & $0$  & $ -\dfrac{75}{4096}$ & $0$ & $ \dfrac{957}{65536}$ & $0$    \\[0.3cm]
 \hline 
\end{tabular} 
\tablefoot{The expansion coefficients, given in \eq{flux_Pl}, are listed for $0 \leq \llat \leq 15$.}
\end{table*}

\section{Theory}
\label{sec:thermotidal_torque}
\subsection{General formulation of the thermotidal torque}
\label{ssec:general_formulation_torque}

The small deformations of solid regions due to variations in the atmosphere's mass distribution are usually ignored in the thin layer approximation, the solid part being able to support most of the atmospheric loading without adjusting hydrostatically like fluid layers \citep[e.g.][]{DI1980,Leconte2015,ADL2019,DG2024}. However, the atmospheric loading is sometimes taken into account in refined modelling approaches as it reduces the thermotidal torque by approximately $15\%$ in Venus-like configurations \citep[e.g.][]{CL2003,Musseau2024}. Similarly, using GCM simulations, \citet{Musseau2024,Musseau2026} find that taking topography into account increases the Venusian thermal torque by a factor of $2.5$.

In the present work, we assume that solid regions are rigid and topography is ignored, consistent with the classical theory \citep[e.g.][]{CL70}. This allows the time-averaged tidal torque exerted on the atmosphere about the planet's spin axis, $\torquez$\mynom[S]{$\torquez$}{thermotidal torque about the planet's spin axis}, to be expressed as  
\begin{equation}
\label{torquez_def} 
\torquez = \lim_{\timeint \rightarrow + \infty} \frac{1}{\timeint} \integ{ \integ{  \dd{\Utide}{\lon}  \frac{\Rpla^2  \presstidesurf}{\ggravi}  }{\surface}{\sphere}{} }{\time}{0}{\timeint} ,
\end{equation}
where $\Rpla$ denotes the planet radius\mynom[S]{$\Rpla$}{planet radius}, $\ggravi$ the surface gravity\mynom[S]{$\ggravi$}{planet's surface gravity}, $\Utide$ the tide-raising gravitational potential\mynom[S]{$\Utide$}{tide-raising gravitational potential}, $\dd{}{\lon} $ the partial derivative with respect to longitude\mynom[S]{$\lon$}{longitude}\mynom[S]{$\dd{}{\lon}$}{partial derivative with respect to longitude}, $\time$ the time\mynom[S]{$\time$}{time}, $\timeint$\mynom[S]{$\timeint$}{time interval} a time interval, $\infvar{S}$ an infinitesimal surface element on the sphere of unit radius, $\sphere$\mynom[S]{$\sphere$}{sphere of unit radius}, and $\presstidesurf$\mynom[S]{$\presstidesurf$}{tidally induced surface pressure anomaly} the surface pressure anomaly generated by the gravitational and thermal tidal forcings. Owing to linearity and to the time periodicity of the tidal perturbation, the surface pressure anomaly can be expanded as a Fourier series, with the Fourier coefficients being themselves expanded in spherical harmonics. Hence, we express $\presstidesurf$ as 
\begin{equation}
\label{presstidesurf_Ylm}
\presstidesurf \left( \col, \lon, \time \right) = \psurf \Re \left\{ \sum_{\ftide} \sum_{\llat=2}^{+ \infty} \sum_{\mm=-\llat}^{\llat}  \presstidenlm{\llat}{\mm} \Ylm{\llat}{\mm} \left( \col , \lon \right) \expo{\inumber \ftide \time} \right\},
\end{equation}
where $\Re$\mynom[S]{$\Re$}{real part of a complex number} designates the real part of a complex number ($\Im$ referring to the imaginary part\mynom[S]{$\Im$}{imaginary part of a complex number}), $\inumber$\mynom[S]{$\inumber$}{imaginary unit} the imaginary unit, $\col$ the colatitude\mynom[S]{$\col$}{colatitude}, $\psurf$ the globally-averaged surface pressure\mynom[S]{$\psurf$}{globally-averaged surface pressure}, $\ftide$ the tidal frequency\mynom[S]{$\ftide$}{tidal frequency}, $\Ylm{\llat}{\mm}$ the degree-$\llat$, order-$\mm$ spherical harmonic \citep[e.g.][Sect.~5.2]{Varshalovich1988}\mynom[S]{$\Ylm{\llat}{\mm}$}{degree-$\llat$, order-$\mm$ spherical harmonic}, and $\presstidenlm{\llat}{\mm}$ the associated dimensionless expansion coefficient\mynom[S]{$\presstidenlm{\llat}{\mm}$}{nondimensional degree-$\llat$, order-$\mm$ spherical harmonic expansion coefficient of surface pressure variations}. We emphasise that the term between braces in \eq{presstidesurf_Ylm} is not a real-valued combination of pairs of complex conjugates in the adopted convention but a general complex number with non-zero imaginary part.

Thin atmospheres on rocky planets are generally presumed to rotate rigidly with the planet itself, so that any torque exerted on the atmosphere over long timescales alters the planet’s overall rotation rather than driving large‑scale zonal winds. We assume that it is the case here. For convenience, we consider the coplanar-circular configuration, where the planet moves around the host star on a circular orbit of radius~$\smaxis$\mynom[S]{$\smaxis$}{planet's orbital radius}, with its spin axis being normal to the orbital plane (no eccentricity, no obliquity). The planet's mean motion, denoted by $\norb$\mynom[S]{$\norb$}{planet's anomalistic mean motion}, is related to $\smaxis$ through Kepler's third law,
\begin{equation}
\smaxis^3 \norb^2 = \Ggrav \left( \Mstar + \Mpla \right),
\end{equation}
with $\Ggrav$ being the universal gravitational constant\mynom[S]{$\Ggrav$}{universal gravitational constant}, $\Mpla$ the mass of the planet\mynom[S]{$\Mpla$}{planet mass}, and $\Mstar$ the mass of the star\mynom[S]{$\Mstar$}{mass of the star}. In this configuration, using the hydrostatic equilibrium assumption, the time-averaged tidal torque of \eq{torquez_def} simplifies to \citep[e.g.][Eq.~(21)]{Farhat2024}\footnote{\label{footnote:factor_two}The numerical coefficient in \eq{torquez_expr}, $\sqrt{6 \pi / 5}$, differs from that of Eq.~(1) of \cite{Leconte2015}, and of Eq.~(4) of \cite{ADL2019} by a factor of 2, whereas spherical harmonics are defined similarly in all cases. This results from the used definitions of the surface pressure anomaly. In \cite{Leconte2015} and \cite{ADL2019}, the quantity expanded as series of spherical harmonics is the real surface pressure anomaly, while it is the complex surface pressure anomaly in the present study, following \cite{Farhat2024}.}
\begin{equation}
\label{torquez_expr}
\torquez = \sqrt{\frac{6 \pi}{5}} \frac{\Mstar}{\Mpla} \frac{\Rpla^6 \psurf}{\smaxis^3} \Im \left\{ \presstidenquad \right\},
\end{equation}
where $\presstidenquad$\mynom[S]{$\presstidenquad$}{degree-2, order-2 spherical expansion coefficient of surface pressure variations} denotes the degree-2, order-2 spherical expansion coefficient of \eq{presstidesurf_Ylm} and $\ftide = 2 \left( \spinrate - \norb \right)$ is the semidiurnal tidal frequency, with $\spinrate$\mynom[S]{$\spinrate$}{planet's spin angular velocity} being the planet's spin angular velocity. The coefficient $\presstidenquad$ is a complex-valued function of tidal frequency, atmospheric parameters, and the multipole expansion components of the tide-raising gravitational potential and stellar irradiance. 

The torque can also be expressed as a function of the complex Fourier coefficient of the equatorial semidiurnal surface pressure oscillation, $\presstideeq \left( \time , \lon \right)=  \Re \left\{ \presstideneq \, \expo{\inumber \left( \ftide \time + 2 \lon \right)} \right\} $\mynom[S]{$\presstideeq$}{equatorial semidiurnal surface pressure oscillation}\mynom[S]{$\presstideneq$}{Fourier coefficient of the equatorial semidiurnal surface pressure oscillation}. The relationship between the degree-2, order-2 spherical expansion coefficient of tidal surface pressure variations and $\presstideneq$ is given by 
\begin{equation}
\label{presstideneq_presstidenquad}
\presstidenquad = 4 \sqrt{\frac{2 \pi}{15}} \cshape \frac{\presstideneq }{\psurf},
\end{equation}
which yields
\begin{equation}
\label{torquez_simp}
\torquez = \frac{8\pi}{5} \cshape  \frac{\Mstar}{\Mpla} \frac{\Rpla^6 }{\smaxis^3} \Im \left\{ \presstideneq \right\}.
\end{equation}
The parameter $\cshape$\mynom[S]{$\cshape$}{shape factor characterising the latitudinal distribution of surface pressure variations} introduced in \eqs{presstideneq_presstidenquad}{torquez_simp} is a dimensionless shape factor that quantifies the degree to which the latitudinal structure of semidiurnal pressure variations resembles a quadrupolar pattern. As defined in \append{app:shape_factor}, $\cshape =1$ corresponds to a purely quadrupolar semidiurnal tidal bulge. In this limit, the classical expression for the tidal torque is recovered \citep[e.g.][]{DI1980,ZW1987}\footnote{The degree-2, order-2 spherical harmonic is expressed as $\Ylm{2}{2} \left( \col , \lon \right) = \left( 1/4 \right) \sqrt{15 / \left( 2 \pi \right)} \sin^2 \col \, \expo{2 \inumber \lon}$ \citep[][Sect.~5.2]{Varshalovich1988}. Thus, the latitudinal distributions of the tidal gravitational potential and pressure variations both scale as $\scale \sin^2 \col$, which induces a factor $\integ{\sin^5 \col}{\col}{0}{\pi} = 16/15 \approx 1.067$ close to unity in the expression of the torque. This factor appears explicitly in \cite{DI1980}, Eq. (27). However, it is omitted in \cite{ZW1987}, Eq. (3), which induces the difference in numerical factors between their equation and \eq{torquez_simp}.}. 

Under present-day Earth conditions, the global barometric observations reported by \cite{Schindelegger2014} indicate that the latitudinal structure of the semidiurnal thermal tide is well approximated by a scaled $\sin^3 \col$ function, corresponding to a shape factor of $\cshape = 75 \pi / 256 \approx 0.92$ (see \append{app:shape_factor}). To ensure consistency with terrestrial observations, this value is adopted throughout the present study. It should be emphasised, however, that the shape factor is intrinsically linked to tidal dynamics, as it implicitly incorporates effects such as rotational scattering. As a consequence, $\cshape$ is not expected to remain constant in general, but rather to depend on both the planetary spin rate and the tidal frequency. In particular, substantial variations may arise if a spherical harmonic of degree $\llat \geq 4$ were resonantly amplified to amplitudes comparable to that of the degree-2 component at the equator, as discussed in \sect{ssec:thermotidal_pressure}. The empirical estimate inferred from \cite{Schindelegger2014} further indicates that converting $\presstidenquad$ into $\presstideneq$ introduces a relative uncertainty of approximately $8\%$ in the corresponding thermotidal torque. In the following sections, we derive expressions for $\presstidenquad$ in isentropic and isothermal atmospheres using the analytical framework established in Sect.~3 of \paperone.

\subsection{Multipole expansion of stellar irradiance}
\label{ssec:multipole_expansion_irradiance}
From now on, we ignore the gravitational tidal forcing. The atmospheric tide solely originates in the absorption of the incident stellar flux impinging the top of the atmosphere. Following earlier works \citep[e.g.][]{ADL2019,Farhat2024}, this flux is defined as\mynom[S]{$\flux$}{incident stellar flux impinging the top of the atmosphere}
\begin{equation}
\label{flux_spatial_distribution}
\flux = \left\{
\begin{array}{ll}
\Fstar \cos \zenithphi & \displaystyle \mbox{if} \ 0 \leq \zenithphi \leq \frac{\pi}{2} \ \mbox{(dayside)}, \\
0 & \displaystyle \mbox{if} \ \frac{\pi}{2} < \zenithphi \leq \pi \ \mbox{(nightside)}.
\end{array}
\right.
\end{equation}
In the above equation, $\Fstar$ designates the incident stellar flux at the substellar point\mynom[S]{$\Fstar$}{incident stellar flux at the substellar point} and $\zenithphi$ the stellar zenithal angle\mynom[S]{$\zenithphi$}{stellar zenithal angle}, namely the angle between the planet-star direction and the position vector of any point on the planet's surface with respect to the planet's centre of mass. In order to compute the weighting coefficients of the flux associated with Hough functions (\paperone, Eq.~(75)), we proceed as detailed in \append{app:expansion_stellar_irradiance}. 

First, $\flux$ is expanded in series of Legendre polynomials \citep[][Chapter~22]{AS1972}. Second, by virtue of the Legendre polynomial addition theorem \citep[e.g.][Eq.~(9.189), p.~600]{Arfken2005}, we express it in terms of spherical harmonics. Finally, we simplify the obtained results by making use of the coplanar-circular approximation, namely assuming that the perturber's orbit is a circle within the planet's equatorial plane. This yields 
\begin{equation}
\label{flux_Plm}
\flux \left( \col , \lon , \time \right) = \Fstar \Re \left\{  \sum_{\llat=0}^{+ \infty} \sum_{\mm = 0}^\llat \fluxnlm \LegFnorm{\llat}{\mm} \left( \cos \col \right) \expo{\inumber \mm \left( \fdiurnal \time + \lon \right)}   \right\},
\end{equation}
where $\fdiurnal = \spinrate - \norb$ designates the diurnal tidal frequency\mynom[S]{$\fdiurnal$}{diurnal tidal frequency}, and $\LegFnorm{\llat}{\mm}$ the degree-$\llat$, order-$\mm$ normalised associated Legendre function \citep[e.g.][Appendix~A]{Auclair2019}\mynom[S]{$\LegFnorm{\llat}{\mm}$}{degree-$\llat$, order-$\mm$ normalised associated Legendre function}. The expansion coefficients of the flux\mynom[S]{$\fluxnlm$}{spherical harmonic coefficients of the flux}, $\fluxnlm$, are given by 
\begin{equation}
\fluxnlm = \left( 2 - \kron{\mm}{0} \right) \frac{2 \flux_\llat}{2 \llat+1}  \LegFnorm{\llat}{\mm} \left( 0 \right),
\end{equation}
where $\kron{\mm}{\qq}$ is the Kronecker delta function, such that $\kron{\mm}{\mm} = 1$ and $\kron{\mm}{\qq}=0$ if $\mm \neq \qq$ \citep[e.g.][Eq.~(1.20), p.~10]{Arfken2005}, and the coefficients $\flux_\llat$\mynom[S]{$\flux_\llat$}{degree-$\llat$ expansion coefficient of the flux in Legendre polynomials} are functions of the degree-$\llat$ Legendre polynomial, $\LegP{\llat} $\mynom[S]{$\LegP{\llat}$}{degree-$\llat$ Legendre polynomial}, expressed as
\begin{equation}
\label{flux_Pl}
\flux_\llat = \frac{2 \llat+1}{2} \integ{\mucos \LegP{\llat} \left( \mucos \right)  }{\mucos}{0}{1}. 
\end{equation}
The values of the latter are listed in Table~\ref{tab:Fl_coeff} for degrees ranging from 0 to 15. We note that the order-$\mm$ component of the flux is associated with the tidal frequency $\ftide_\mm = \mm \fdiurnal$. Thus, $\mm=1$ corresponds to the diurnal tide, $\mm=2$ to the semidiurnal tide, $\mm=3$ to the terdiurnal tide, $\mm = 4$ to the quaddiurnal tide, and so on. 

Finally, we introduce the order-$\mm$ spin parameter\mynom[S]{$\spinparm$}{order-$\mm$ spin parameter}, 
\begin{equation}
\spinparm \left( \spinrate , \norb \right) = \frac{2 \spinrate}{\mm \left( \spinrate - \norb \right)}, 
\end{equation}
and the coefficients\mynom[S]{$\fluxnnm$}{degree-$\nn$, order-$\mm$ Hough function expansion coefficient of the incident stellar flux}
\begin{equation}
\fluxnnm = \sum_{\llat=\mm}^{+\infty}  \scal{ \LegFnorm{\llat}{\mm}}{\HoughF{\nn}{\mm}{\spinparm} } \, \fluxnlm .
\end{equation}
In the above equations, $\HoughF{\nn}{\mm}{\spinparm}$\mynom[S]{$\HoughF{\nn}{\mm}{\spinpar}$}{degree-$\nn$ Hough function} denotes the degree-$\nn$ Hough function (see \paperone, Sect.~2.4), and $\scal{\cdot}{\cdot}$ the dot product
\begin{equation}
\label{dotproduct}
\scal{\mathcal{F}}{\mathcal{G}} = \integ{\mathcal{F} \left( \mucos \right) \mathcal{G} \left( \mucos \right) }{\mucos}{-1}{1},
\end{equation}
where $\mathcal{F}$ and $\mathcal{G}$ are two functions defined for $-1 \leq \mucos \leq 1$. This allows \eq{flux_Plm} to be reformulated as
\begin{equation}
\flux \left( \col , \lon , \time \right) = \Fstar \Re \left\{ \sum_\mm \sum_\nn \fluxnnm \HoughF{\nn}{\mm}{\spinparm} \left( \cos \col \right) \expo{\inumber \mm \left( \fdiurnal \time + \lon \right)} \right\}.
\end{equation}
Therefore, the coefficient of the flux associated with the degree-$\nn$ Hough function in the thermal forcing (\paperone, Eq.~(75)) is expressed, for the semidiurnal tide, as\mynom[S]{$\fluxtiden$}{degree-$\nn$, order-2 expansion coefficient of the flux in terms of Hough functions}
\begin{equation}
\label{fluxtiden}
\fluxtiden = \Fstar \sum_{\llat=2}^{+\infty}  \scal{\LegFnorm{\llat}{2}}{\HoughF{\nn}{2}{\spinpar_2} } \, \fluxnalfi{\llat}{2}.
\end{equation}

\subsection{Thermotidal pressure oscillations}
\label{ssec:thermotidal_pressure}

The corresponding pressure variation is obtained by setting to zero the tide-raising potential ($\Usurf=0$\mynom[S]{$\Usurf$}{tide-raising potential at the planet's surface}) in our closed-form solution (\paperone, Eq.~(90)), which yields\mynom[S]{$\presstidev$}{degree-$\nn$ expansion coefficient of the Fourier coefficient of the normalised tidal pressure oscillation}
\begin{align}
\label{presstidev_X}
\presstidev  = & \ \frac{\inumber \, \presstidevc \left( 1 + \bj \right) \expo{- \dlnbet \xx}}{\left( \XX - \inumber \disspar \right)  \left[  \kvert^2 + \left( \lambdapar + \bj \right)^2 \right]} \left\{  \left( \bj + 1 \right) \expo{-\bj \xx} \right. \\
& \left. + \frac{\left[ \Sn \left( 1 - \gammapar \right) - \left( \bj + 1 \right)  \XX^2 \right] \left( \inumber \kvert + \lambdapar - 1 \right)}{\Sn \left( 1  - \gammapar \right) + \left( \inumber \kvert + \lambdapar - 1 \right) \XX^2} \expo{ \left( \inumber \kvert + \lambdapar \right) \xx}  \right\}. \nonumber
\end{align}
In the above equation, $\xx$\mynom[S]{$\xx$}{dimensionless vertical coordinate} is the dimensionless vertical coordinate (\paperone, Eq.~(8)); $\gammapar$\mynom[S]{$\gammapar$}{reduced stratification number}, $\dlnbet$\mynom[S]{$\dlnbet$}{reduced lapse rate}, and $\lambdapar$\mynom[S]{$\lambdapar$}{dimensionless atmospheric parameter} are dimensionless parameters given by (\paperone, Eqs.~(30) and~(77))
 \begin{equation}
 \label{control_param}
 \begin{array}{lll}
 \dlnbet = \left\{
 \begin{array}{ll}
 - \rcp & \! \mbox{(isentropic)}, \\
 0  & \! \mbox{(isothermal)},
 \end{array}
 \right.
 & \displaystyle \gammapar = \dlnbet + \rcp,
 & \displaystyle \lambdapar = \frac{1 - \inumber \alphapar \left( 1 + \dlnbet \right) }{2 \left( 1 - \inumber \alphapar \right)},
 \end{array}
 \end{equation}
with $\rcp = \left( \adiabexp - 1 \right) / \adiabexp$\mynom[S]{$\adiabexp$}{adiabatic exponent}\mynom[S]{$\rcp$}{$\rcp = \left( \adiabexp-1 \right)/\adiabexp$} being a function of the adiabatic index, $\adiabexp$, and $\alphapar= \fnewton/\ftide $\mynom[S]{$\alphapar$}{ratio between the Newtonian and tidal frequencies} denoting the ratio between the Newtonian cooling frequency, $\fnewton$\mynom[S]{$\fnewton$}{Newtonian cooling frequency}, and the tidal frequency, $\ftide$. It is worth noting that, in practice, $\fnewton$ should be regarded as an effective parameter encapsulating all dissipative processes that are not explicitly represented in the model, including radiative transfer, wind-induced broadening, turbulent diffusion within the boundary layer, and spatial inhomogeneities.

The parameter $\bj$\mynom[S]{$\bj$}{vertical variation rate of the tidal heating profile} in \eq{presstidev_X} is the vertical variation rate of the considered exponentially decaying tidal heating profile (\paperone, Eq.~(74)), $\Sn$\mynom[S]{$\Sn$}{sign of the eigenvalue associated with the degree-$\nn$ Hough function} is the sign of the eigenvalue associated with the degree-$\nn$ Hough function,
\begin{equation}
\Sn = \sign \left( \Lambn \right),
\end{equation}
and $\kvert$\mynom[S]{$\kvert$}{vertical wavenumber of the degree-$\nn$ mode} is the vertical wavenumber of the degree-$\nn$ mode, such that (\paperone, Eq.~(78))
\begin{equation}
\label{kvert2}
\kvert^2 = \frac{\gammapar \betasurf \Lambn - \inumber \alphapar \dlnbet}{1 - \inumber \alphapar} - \lambdapar^2,
\end{equation}
where $\betasurf = \ggravi \heightsurf / \left( \Rpla^2 \ftide^2 \right)$\mynom[S]{$\betasurf$}{dimensionless atmospheric parameter} is a dimensionless parameter involving the pressure scale height at the planet's surface, $\heightsurf$\mynom[S]{$\heightsurf$}{pressure scale height at the planet's surface}. A limitation of the present formulation is that $\bj$ is assumed to be a real-valued parameter, such that the corresponding tidal-heating profile is everywhere in phase within the atmospheric column. The model is therefore unable to discriminate between heating induced by stratospheric ozone absorption and that arising from the surface, although these processes are known to produce tidal responses with different phase signatures \citep[e.g.][]{CL70}. Within the linear regime, this shortcoming can nevertheless be addressed to some extent by representing forcings with distinct phase lags as separate contributions.

Additionally, we have introduced in \eq{presstidev_X} the normalised tidal frequency $\XX$, damping rate $\disspar$, and thermotidal pressure variation $\presstidevc$, defined as \mynom[S]{$\XX$}{normalised tidal frequency}\mynom[S]{$\disspar$}{normalised damping rate}\mynom[S]{$\presstidevc$}{normalised scaling factor of thermotidal pressure variations}
\begin{equation}
\label{reduced_frequency}
\begin{array}{lll}
\displaystyle \XX = \frac{\ftide}{\freson} , & \displaystyle \disspar = \frac{\fnewton}{\freson} ,& \displaystyle \presstidevc = \frac{\rcp \opacity \fluxtiden}{\psurf \heightsurf \freson},
\end{array}
\end{equation}
where $\opacity$\mynom[S]{$\opacity$}{absorbed flux fraction} -- such that $0 \leq \opacity \leq 1$ -- designates the fraction of incident stellar flux absorbed by the atmosphere, and $\freson$\mynom[S]{$\freson$}{resonance frequency of the degree-$\nn$ tidally excited Lamb wave} the resonance frequency of the degree-$\nn$ tidally excited Lamb wave (\paperone, Eq.~(97)),
\begin{equation}
\label{freson_degreen}
\freson = \frac{1}{\Rpla} \sqrt{\frac{\ggravi \heightsurf \left| \Lambn \right|}{1 - \gammapar} }.
\end{equation}
We emphasise that the three nondimensional parameters given by \eq{reduced_frequency} depend on the degree of the excited tidal mode, like $\kvert$ and $\Sn$. However, the subscript $\nn$ is omitted in notations in order to lighten expressions, noting that each mode can be considered individually. It should also be noted that the parameter $\presstidevc$, as defined in \eq{reduced_frequency}, pertains specifically to the regime of direct atmospheric absorption, in which tidal heating arises from the absorption of incident stellar radiation in the visible range \citep[this corresponds to the $\mathcal{G}=1$ limit case of][]{Farhat2024}. The analysis presented below can nevertheless be readily extended to cases involving indirect thermal forcing by introducing the transfer function that relates the tidal heat flux to the incident stellar flux and applying it as a multiplicative correction to $\presstidevc$, as discussed in \sect{sec:phase_shifted_heating}.

Similar to stellar irradiance, we obtain the quadrupolar component of the surface pressure variation needed to compute the tidal torque, $\presstidenquad$ (see \eq{torquez_expr}), by expanding Hough functions in series of normalised associated Legendre functions, and noting that 
\begin{equation}
\Ylm{\llat}{\mm}\left( \col , \lon \right) = \frac{1}{\sqrt{2 \pi}} \LegFnorm{\llat}{\mm} \left( \cos \col \right) \expo{\inumber \mm \lon}. 
\end{equation}
It follows
\begin{equation}
\label{deltaps22_series}
 \presstidenquad  = \sqrt{2 \pi} \sum_\nn   \scal{\HoughF{\nn}{2}{\spinpar_2}}{\LegFnorm{2}{2}} \, \presstidevsurf.
\end{equation}
The Fourier component of the corresponding surface pressure oscillation at the equator, introduced in \eq{presstideneq_presstidenquad}, is readily deduced from \eq{deltaps22_series},
\begin{equation}
\label{presstideneq}
\presstideneq = \frac{\psurf}{\cshape} \sqrt{\frac{15}{16}} \sum_\nn   \scal{\HoughF{\nn}{2}{\spinpar_2}}{\LegFnorm{2}{2}} \, \presstidevsurf.
\end{equation}
In the above equations, $\presstidevsurf = \left. \presstidev \right|_{\xx=0}$\mynom[S]{$\presstidevsurf$}{$\presstidev$ evaluated at the planet's surface} designates the surface pressure variation induced by the degree-$\nn$ mode. 

The values of the dot products $\scal{\HoughF{\nn}{2}{\spinpar_2}}{ \LegFnorm{\llat}{2}}$ and $\scal{\LegFnorm{\llat}{2}}{\HoughF{\nn}{2}{\spinpar_2}}$ are given by Table~B.1 of \paperone for Earth's Solar semidiurnal tide, namely $\mm=2$ and $\spinpar_2 = 1$ (neglecting the difference between the Solar and sidereal days). We note that the Hough function associated with the predominating gravity mode of the Earth's semidiurnal tide, $\HoughF{2}{2}{\spinpar}$, is very similar to the corresponding normalised associated Legendre function, $\LegFnorm{2}{2} $, as their overlap coefficient is $\scal{\HoughF{2}{2}{\spinpar_2}}{ \LegFnorm{2}{2}} = 0.969613 \approx 1 $ (\paperone, Table~B.1). In the tidal torque exerted about the planet's spin axis, the contribution of the predominant gravity mode is thus weighted by the factor $\scal{\HoughF{2}{2}{\spinpar_2}}{ \LegFnorm{2}{2}}^2 =  0.940149 $, meaning that it would represents $94 \%$ of the tidally dissipated energy if the modal response were independent of frequency. The associated eigenvalue is given by $\Houghval{2}{2}{1} = 11.12895$ (\paperone, Table~B.1). As a last step, we shall establish from \eq{presstidev_X} the analytical expressions of $\presstidevsurf$ for isentropic and isothermal atmospheres.

\def\hinter{0.3cm}
\begin{table*}[h]
\centering
\caption{\label{tab:solutions} Transfer functions relating the surface pressure oscillations to the incident stellar flux in the Fourier domain. }
\begin{tabular}{ll} 
 \hline 
 \hline 
 \textsc{Isentropic} & \textsc{Isothermal} \\
 \hline \\[-0.3cm] 
 \multicolumn{2}{c}{\textit{Parameters}} \\[0.1cm]
$ \displaystyle \freson = \frac{\sqrt{ \ggravi \heightsurf \abs{\Lambn}}}{\Rpla}$ & 
$ \displaystyle \freson = \frac{\sqrt{\adiabexp  \ggravi \heightsurf \abs{\Lambn}}}{\Rpla}$  \\[\hinter] 
 & 
$ \displaystyle \thetapar =  2 \sqrt{\rcp \left( 1- \rcp \right) }$ \\[\hinter]
 \hline \\[-0.3cm] 
 \multicolumn{2}{c}{\textit{Vertical wavenumber of forced tidal waves}} \\[0.1cm]
& 
$ \displaystyle \mucoef = \frac{\Sn \XX^2}{\XX^2 + \disspar^2} \left( \frac{\vartheta}{\XX} \right)^2 - 1 + \left[ \frac{\left( \left( \frac{\vartheta}{\XX} \right)^2 - \Sn \right)^2 \XX^2 + \disspar^2  }{\XX^2 + \disspar^2}  \right]^{1/2}$ \\[\hinter]
$ \displaystyle \kvert =  \frac{\inumber}{2} \left[ \frac{\XX - \inumber \disspar \left( \rcp +1 \right)}{\XX - \inumber \disspar}  \right]$ & 
$ \displaystyle \kvert = \frac{\sqrt{\mu \left( \XX \right)}}{2 \sqrt{2}} \left[ \Sn + \inumber \frac{ \disspar \XX}{\mu \left( \XX \right) \left( \XX^2 + \disspar^2 \right)} \left( \frac{\vartheta}{\XX} \right)^2 \right]  \sign \left( \XX \right)$ \\[0.5cm]
\hline \\[-0.3cm] 
 \multicolumn{2}{c}{\textit{Transfer function of pressure and surface pressure}} \\[0.3cm]
$\begin{array}{rl}  \displaystyle \presstidev = & \displaystyle \frac{\inumber \, \presstidevc \, \expo{\rcp \xx} }{\bj \left( \XX - \inumber \disspar \right) + \inumber \rcp \disspar} \left\{ \left( \bj + 1 \right) \expo{- \bj \xx}   \right.  \\
 & \displaystyle \left.  + \frac{\left[ \Sn - \left( 1 + \bj \right) \XX^2  \right] \left[ \XX - \inumber \disspar \left( 1 + \rcp \right) \right]}{\left( \XX - \inumber \disspar \right) \left( \XX^2 - \Sn \right)  - \inumber \disspar \rcp \XX^2} \expo{\inumber \frac{\rcp \disspar}{\XX - \inumber \disspar} \xx }   \right\} \end{array}$ & 
 $ \begin{array}{rl} \displaystyle \presstidev = &  \displaystyle  \frac{\inumber \presstidevc \left( \bj +1 \right)  }{\left( \XX - \inumber \disspar \right) \left[ \kvert^2 + \left( \frac{1}{2} + \bj \right)^2 \right]} \left\{ \left( \bj +1 \right) \expo{- \bj \xx}  \right.  \\
& \displaystyle \left. + \frac{\left[ \Sn - \adiabexp \left( 1 + \bj \right) \XX^2 \right] \left( \inumber \kvert - \frac{1}{2} \right) }{\Sn + \adiabexp \left( \inumber \kvert - \frac{1}{2} \right) \XX^2} \expo{\left( \inumber \kvert + \frac{1}{2} \right) \xx}  \right\}  \end{array}$   \\[1.2cm]
$\displaystyle \presstidevsurf = \frac{\Sn \presstidevc }{ \disspar \left[ \left( 1 + \rcp \right) \XX^2 - \Sn \right] + \inumber \XX \left( \XX^2  - \Sn\right)  }$ & 
$\displaystyle \presstidevsurf = \frac{ \inumber \Sn \presstidevc  \left( \bj  + 1 \right)  \left( \inumber \kvert + \frac{1}{2} + \bj \right)  \XX  }{ \left[ \frac{1}{4} \Sn \vartheta^2  +  \bj \left( \bj + 1 \right) \left( \XX - \inumber \disspar  \right) \XX \right] \left[ \adiabexp \left(\inumber \kvert - \frac{1}{2} \right) \XX^2 + \Sn \right]  }$ \\[0.5cm] 
$\displaystyle \left. \presstidevsurf \right|_{\bj = + \infty}  =  \frac{\Sn \presstidevc  }{ \disspar \left[ \left( 1 + \rcp \right) \XX^2 - \Sn \right] + \inumber \XX \left( \XX^2  - \Sn\right)  }$ & 
$\displaystyle \left. \presstidevsurf \right|_{\bj = + \infty}  = \frac{ \inumber \Sn \presstidevc  }{  \left( \XX - \inumber \disspar  \right)  \left[ \adiabexp \left(\inumber \kvert - \frac{1}{2} \right) \XX^2 + \Sn \right]  }$ \\[0.5cm] 
\hline 
 \end{tabular}
 \tablefoot{Comparison of the analytical expressions of the quantity in the isentropic (left column) and isothermal (right column) atmospheric solutions. The resonance frequency of the degree-$\nn$ mode, $\freson$, is defined by \eq{freson_degreen}, while $\XX$, $\disspar$ and $\presstidevc$ are defined by \eq{reduced_frequency}, $\thetapar$ by \eq{thetapar}, $\mucoef$ by \eq{mucoef}, $\kvert$ by \eqs{kvert_neu}{kvert_iso}, and $\presstidev$ by \eqs{presstidev_neu}{presstidev_iso}. The surface pressure variation is given by \eq{presstidev_surf_neu} for isentropic atmospheres, and by \eqs{presstidev_surf_isotherm}{presstidev_surf_isotherm_dirac} for isothermal atmospheres.}
 \end{table*}

\subsection{Isentropic atmosphere}

The isentropic atmosphere is the simpler of the two configurations described by the closed-form solution of \paperone. Here, the control parameters associated with the atmospheric structure take the values $\dlnbet = - \rcp$ and $\gammapar = 0$ (see \eq{control_param}). Therefore, the vertical wavenumber of a tidal mode is readily deduced from \eq{kvert2}. As $\Im \left\{ \kvert \right\} > 0$ by convention, this wavenumber is expressed as
\begin{equation}
\label{kvert_neu}
\kvert = \frac{\inumber}{2} \left[  \frac{\XX - \inumber \disspar \left(1 + \rcp \right)}{\XX - \inumber \disspar} \right].
\end{equation}
The vertical profile of the degree-$\nn$ tidal pressure variations, given by \eq{presstidev_X}, thus becomes
\begin{align}
\label{presstidev_neu}
\presstidev = &  \frac{\inumber \, \presstidevc \expo{\rcp \xx}}{\bj \left( \XX- \inumber \disspar \right) + \inumber \rcp \disspar} \left\{  \left( \bj + 1 \right) \exp \left( - \bj \xx \right) \right. \\
& \left. + \frac{ \left[ \Sn -  \left( \bj + 1 \right) \XX^2 \right] \left[ \XX - \inumber \disspar \left( 1 + \rcp \right) \right]}{\left( \XX- \inumber \disspar \right) \left( \XX^2 - \Sn \right) - \inumber \rcp \disspar \XX^2} \exp \! \left( \frac{\inumber \rcp \disspar}{\XX-\inumber \disspar} \xx \right) \! \right\} \! ,\nonumber
\end{align}
and the corresponding surface pressure variation simplifies to 
\begin{equation}
\label{presstidev_surf_neu}
\presstidevsurf = \frac{\Sn \, \presstidevc }{\disspar \left[ \left(1 + \rcp \right) \XX^2 - \Sn \right] + \inumber \XX \left( \XX^2 - \Sn \right)}.
\end{equation}
The key components of this analytical solution can be retrieved in Table~\ref{tab:solutions}. Interestingly, the surface pressure oscillation induced by a tidal mode is independent of $\bj$, meaning that the vertical distribution of tidal heating does not affect the tidal torque exerted on the planet. This results from the assumed atmospheric structure. First, the Archimedean forces vanish since the atmosphere is neutrally stratified. Second, we have filtered out Coriolis and compressibility terms in the vertical momentum equation of tidal waves (\paperone, Eq.~(14)), which implies that fluid motions are solely driven by the horizontal terms of these two restoring forces. As a consequence, atmospheric tides are analogous to ocean's barotropic tides in this configuration.

\subsection{Isothermal atmosphere}

For an isothermal atmosphere, the structure control parameters take the values $\dlnbet = 0$ and $\gammapar = \rcp$, which implies $\lambdapar = 1/2$ (see \eq{control_param}). Due to stable stratification, the dependence of the vertical wavenumber on tidal frequency is significantly more complex than in the isentropic configuration. Derivations yield
\begin{equation}
\label{kvert_iso}
\kvert \left( \XX \right) = \frac{\sqrt{\mu \left( \XX \right)}}{2 \sqrt{2}} \left[ \Sn + \inumber \frac{ \disspar \XX}{\mu \left( \XX \right) \left( \XX^2 + \disspar^2 \right)} \left( \frac{\vartheta}{\XX} \right)^2 \right]  \sign \left( \XX \right) ,
\end{equation}
where we have introduced the parameter\mynom[S]{$\thetapar$}{$\rcp$-dependent dimensionless parameter of the isothermal case}
\begin{equation}
\label{thetapar}
\thetapar =  2 \sqrt{\rcp \left( 1 - \rcp \right)},
\end{equation}
and the function\mynom[S]{$\mucoef$}{calculation quantity introduced in the isothermal case}
\begin{equation}
\label{mucoef}
\mucoef \left( \XX \right) = \frac{\Sn \XX^2}{\XX^2 + \disspar^2} \left( \frac{\thetapar}{\XX} \right)^2 - 1 + \left[ \frac{\left( \left( \frac{\thetapar}{\XX} \right)^2 - \Sn \right)^2 \XX^2 + \disspar^2  }{\XX^2 + \disspar^2}  \right]^{1/2}.
\end{equation}
The tidal pressure variation associated with the degree-$\nn$ mode is thus expressed as 
\begin{align}
\label{presstidev_iso}
\presstidev = & \frac{\inumber \, \presstidevc \left( \bj +1 \right)}{\left( \XX- \inumber \disspar \right) \left[ \kvert^2 + \left( \frac{1}{2} + \bj \right)^2 \right]} \left\{ \left( \bj +1 \right) \exp \left( - \bj \xx \right)   \right.  \\
& \left.  + \frac{\left[ \Sn - \adiabexp \left( \bj +1 \right) \XX^2 \right] \left( \inumber \kvert - \frac{1}{2} \right)}{\Sn + \adiabexp \left( \inumber \kvert - \frac{1}{2} \right) \XX^2} \exp \left[ \left( \inumber \kvert + \frac{1}{2} \right) \xx \right]  \right\}, \nonumber 
\end{align}
and the corresponding surface pressure variation as
\begin{equation}
\label{presstidev_surf_isotherm}
\presstidevsurf = \frac{\inumber \, \Sn \presstidevc \left(1 + \bj \right)  \left( \inumber \kvert + \frac{1}{2} + \bj \right) \XX}{ \left[ \frac{1}{4} \Sn \thetapar^2  \!+ \! \bj \! \left( \bj \! +\! 1 \right) \! \XX \! \left( \XX \! - \! \inumber \disspar \right) \right] \left[ \Sn \! + \! \adiabexp \! \left( \inumber \kvert \! - \! \frac{1}{2} \right) \! \XX^2 \right]}.
\end{equation}
Unlike isentropic atmospheres, isothermal atmospheres permit the propagation of internal waves in the vertical direction. Because of the interplay between the vertical wavelength of these waves and the thickness of the heated region, the surface pressure oscillations explicitly depend on $\bj$ in \eq{presstidev_surf_isotherm}. To remove this dependence and thus reduce the dimensionality of the model, the heated layer must be much thinner than the vertical wavelength of the tidal modes. Since $\kvert$ tends to $+\infty$ in the zero-frequency limit, the thickness of this layer must consequently vanish, implying $\bj=+\infty$. This limiting case corresponds to the Dirac-type heating profile adopted by \citet{DI1980} in their model of Venus' thermal tides. With $\bj = + \infty$, \eq{presstidev_surf_isotherm} simplifies to 
\begin{equation}
\label{presstidev_surf_isotherm_dirac}
\presstidevsurf= \frac{\inumber \, \Sn \presstidevc }{ \left( \XX -  \inumber \disspar \right) \left[ \Sn + \adiabexp \left( \inumber \kvert - \frac{1}{2} \right) \XX^2 \right]}.
\end{equation}

\subsection{Application to the present-day Earth}
\label{ssec:application_present_earth}

We conclude this section by benchmarking the model predictions against observational constraints for semidiurnal tides on the present-day Earth. This case, which corresponds to the $\mathcal{G}=1$ limit of \cite{Farhat2024}, is representative of the regime of direct absorption that will be analytically characterised in \sect{sec:frequency_behaviour}. From an extensive analysis of global barometric records, \citet{Schindelegger2014} derived estimates for both the modulus and the phase lag of the semidiurnal surface pressure oscillations driven by Solar thermal tides. They report a value of $\abs{\presstideneq}=123.4$~Pa at the equator, with the maximum amplitude occurring at $\tmax=09{:}45$~LST. 

As these experimental estimates agree well with earlier measurements \citep[e.g.][]{haurwitz1956,HC1973,Lindzen1978,DW1999} and GCM simulations \citep[e.g.][]{Covey2011,Covey2014}, we adopt them as reference values for experimental constraints. However, these estimates shall not be considered as precise. Particularly, earlier empirical studies end up with lower amplitudes, such as $116.1$~Pa \citep{HC1973} or $107.9$~Pa \citep[][]{DW1999}, indicating a systematic uncertainty of $10{-}15\%$ in experimental calibration. Through an exhaustive comparative analysis of climate models for the Earth's diurnal and semidiurnal tides, \cite{Covey2011} and \cite{Covey2014} show that GCM results exhibit a similar variability. Using $\abs{\presstideneq}=123.4$~Pa and $\tmax=09{:}45$~LST in \eq{torquez_simp}, together with $\Rpla=6371$~km, $\smaxis=1.0$~au, and $\Mstar/\Mpla = \num{3.329e5}$ \citep{Luzum2011}, yields an associated thermotidal torque of $\torquez = \SI{3.51e15}{\newton\metre}$. This estimate is affected by the uncertainty of empirical data and GCM simulations, which makes it range between $\torquez = \SI{2.5e15}{\newton\metre}$ \citep[][]{ZW1987} and $\torquez = \SI{4.14e15}{\newton\metre}$ \citep[][Fig.~S4 in Supplementary Materials]{Wu2023}. In what follows, we assess how well the analytical model reproduces the observed modulus and phase of the surface pressure oscillations for both the isentropic and isothermal equilibria. 

To facilitate a consistent comparison, we consider atmospheres with the same total thermal energy by fixing the mass-averaged temperature of the air column -- hereafter the bulk temperature -- as defined in Eq.~(71) of \paperone. This temperature, $\Tbulk$\mynom[S]{$\Tbulk$}{atmospheric bulk temperature}, is given by 
\begin{equation}
\label{Tbulk}
\Tbulk = \frac{\Tsurf}{1 - \dlnbet} = \left\{
\begin{array}{ll}
\displaystyle \frac{\Tsurf}{1 + \rcp} & \mbox{(isentropic)}, \\
\displaystyle \Tsurf & \mbox{(isothermal)},
\end{array}
\right.
\end{equation}
where $\Tsurf$\mynom[S]{$\Tsurf$}{atmospheric temperature at the planet's surface} denotes the atmospheric temperature at the planet's surface. Adopting $\Tsurf=288$~K for present-day Earth \citep[][]{USStdAtm1976}, we obtain $\Tbulk=243$~K for the isentropic configuration via \eq{Tbulk}, which defines the corresponding equivalent isothermal atmosphere. 

We use the standard Earth values for the surface gravity and the mean Solar irradiance, namely $\ggravi = 9.81~{\rm m \ s^{-2}}$ and $\Fstar=1361~{\rm W \ m^{-2}}$ \citep[][]{Coddington2016}. For the atmospheric parameters, we adopt the prescription of the U.S. Standard Atmosphere \citep[][]{USStdAtm1976}, taking $\psurf=1.013$~bar, $\Rspec=287~{\rm J \ kg^{-1} \ K^{-1}}$, and $\adiabexp = 1.23$ (\paperone, Fig.~2). Finally, the temperature at the planetary surface sets the associated pressure scale height, 
\begin{equation}
\label{heightsurf}
\heightsurf = \frac{\Rspec \Tsurf}{\ggravi},
\end{equation}
yielding $\heightsurf=8.43$~km for the isentropic equilibrium and $\heightsurf = 7.10$~km for the isothermal case.

These parameters allow us to evaluate the Lamb-wave resonance frequency of the dominant tidal mode ($\mm=2$, $\nn=2$, $\spinpar=1$), defined in the general case by \eq{freson_degreen} and given by \mynom[S]{$\fresoi{2}$}{resonance frequency of the main mode excited by the semidiurnal tidal heating}
\begin{equation}
\label{freson_lamb_2}
\fresoi{2} = \frac{1}{\Rpla} \sqrt{\frac{\ggravi \heightsurf \abs{\Houghval{2}{2}{1}}}{1 - \gammapar}}. 
\end{equation}
In this expression, we neglect the small difference between the Solar and sidereal day, thereby removing the weak frequency dependence of the eigenvalue associated with the degree-2 mode. The latter takes the value $\Houghval{2}{2}{1} = 11.12895$ (\paperone, Table~B.1), which is positive, indicating that the mode can be resonantly excited ($\Sn=+1$ in \eqs{presstidev_surf_neu}{presstidev_surf_isotherm_dirac}). For the isentropic configuration, where $\gammapar=0$, we obtain $\fresoi{2}=\num{1.51e-4}~{\rm rad \ s^{-1}}$. In the isothermal case, $\gammapar=\rcp$ yields $\fresoi{2} = \num{1.53e-4}~{\rm rad \ s^{-1}}$. As shown in \paperone (Eq.~(100)), the ratio between the isothermal and isentropic values -- denoted $\fresoi{2}^{\rm iso}$ and $\fresoi{2}^{\rm isen}$, respectively -- is
\begin{equation}
\label{ratio_freq_iso_isen}
\frac{\fresoi{2}^{\rm iso}}{\fresoi{2}^{\rm isen}} = \sqrt{\frac{\adiabexp}{1+\rcp}} \approx 1.02.
\end{equation}
The associated length of the day,\mynom[S]{$\LODresoi{2}$}{LOD at which the resonance of the main semidiurnal tidal mode occurs}  
\begin{equation}
\label{LOD2_freso2}
\LODresoi{2} = \frac{4 \pi}{\fresoi{2}},
\end{equation}
is therefore 23.2~hr (isentropic) and 22.8~hr (isothermal). 

These values closely match the estimate reported by \cite{Wu2023} for present-day Earth, $\LODresoi{2}=22.8$~hr, which is calculated based on the change in the torque's sign in their Fig.~S4. However, they are $0.8{-}1.2$~hr larger than the 22~hr resonant LOD obtained by fitting the analytical model to the same GCM dataset (see \sect{ssec:Wu2023}) and to the dataset reported by \cite{DG2024} for a slightly warmer palaeoatmosphere with a $\methane$ mixing ratio of 30~ppm (see \sect{ssec:DG2024})\footnote{The analyses carried out in \sects{ssec:DG2024}{ssec:Wu2023} suggest that the GCM datasets of \cite{DG2024} and \cite{Wu2023} are in fact characterised by the same resonant daylength, $\LODresoi{2} = 22.0$~hr, which is 0.25 or 0.8~hr less than the values reported by the two studies, respectively.}. Additionally, our values significantly exceed the standard 21~hr estimate used as a reference for the Earth's Precambrian resonance \citep[e.g.][]{ZW1987,DG2024}. This 21~hr estimate was obtained in the framework of the classical tidal theory \citep[e.g.][]{LB1972,Salby1979}, where it can be inferred from the equivalent depth of the mode \citep[][]{CL70},
\begin{equation}
\deptheqi{2} = \frac{\Rpla^2 \ftide^2}{\ggravi \Houghval{2}{2}{1}}. 
\end{equation}
Typically, \cite{LB1972} find $\deptheqi{2} \approx 10.5$~km by assuming an isothermal profile characterised by $\adiabexp=1.4$ (ideal diatomic gas) and $\Tbulk=256$~K. Using \eqs{freson_lamb_2}{LOD2_freso2} together with these values yields $\LODresoi{2}=20.8$~hr. Similar estimates are reported by several authors solving the tidal equations numerically for realistic profiles of the Earth's atmosphere \citep[e.g.][]{LB1972,Salby1979,ZW1987}. However, the complexity of these numerical solutions do not allow for elucidating the possible causes of the 2-hour LOD difference between them and our analytical theory.  

\def\wpanel{0.48\textwidth}
\begin{figure*}
\centering
   \RaggedRight \twolabelsat{4.5cm}{Isentropic}%
                           {13.5cm}{Isothermal}\\[0.1cm]
\includegraphics[width=\wpanel,trim={0cm 0.7cm 0.5cm 0cm}, clip]{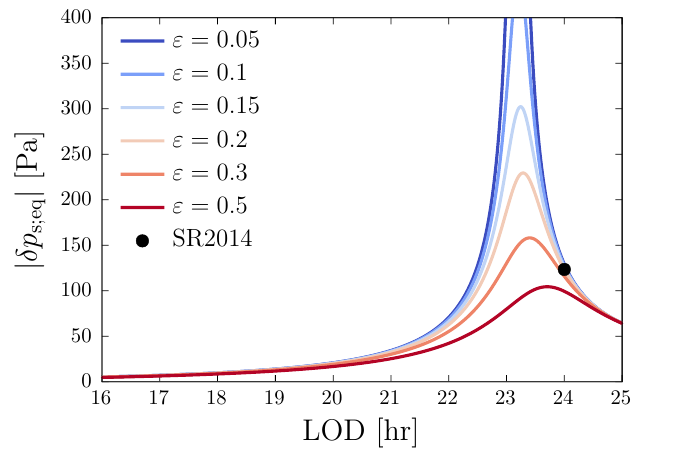} 
\includegraphics[width=\wpanel,trim={0cm 0.7cm 0.5cm 0cm}, clip]{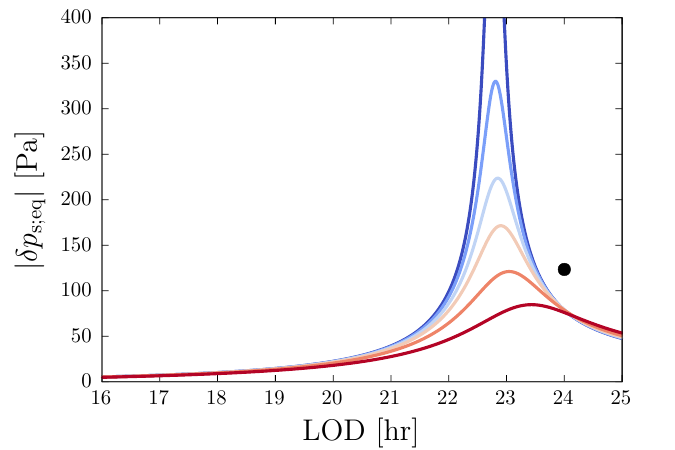} \\
\includegraphics[width=\wpanel,trim={0.0cm 0.7cm 0.5cm 0cm}, clip]{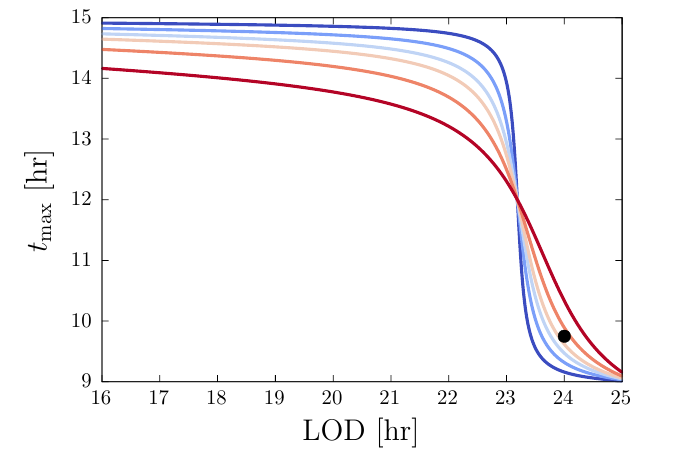}
\includegraphics[width=\wpanel,trim={0.0cm 0.7cm 0.5cm 0cm}, clip]{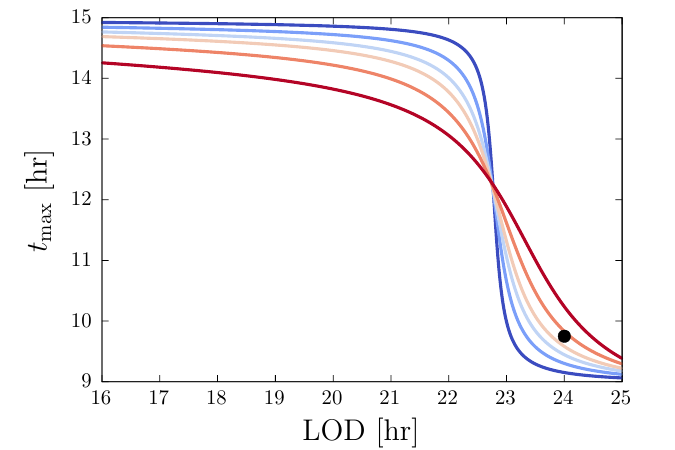} \\
\includegraphics[width=\wpanel,trim={0cm 0cm 0.5cm 0.cm}, clip]{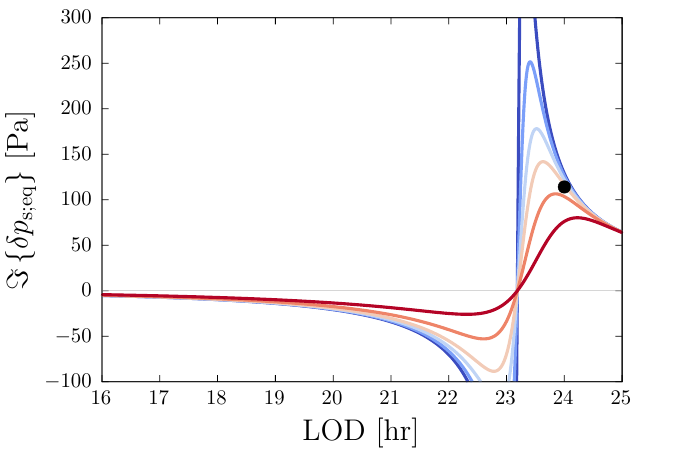}
\includegraphics[width=\wpanel,trim={0cm 0cm 0.5cm 0cm}, clip]{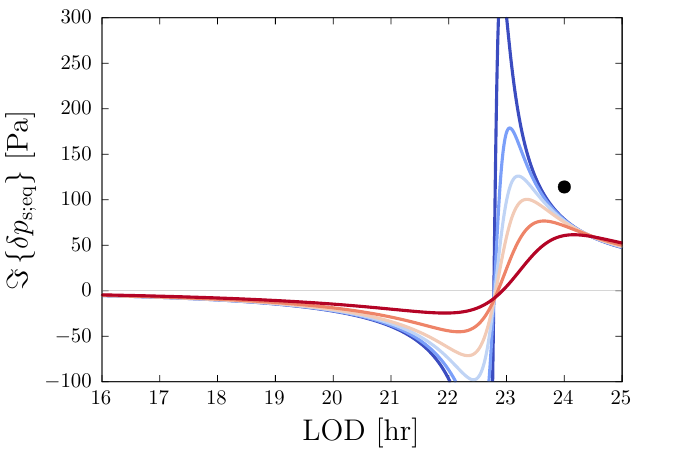}
    \caption{Predictions of the isentropic (left) and isothermal (right) models for actual Earth and various values of the damping ratio. {\it From top to bottom:} Modulus of surface pressure oscillations at the equator, Solar time at which they are maximal ($\time_{\rm max}$), and imaginary component. Solid lines denote the model predictions, which are computed using \eqs{presstidev_surf_neu}{presstidev_surf_isotherm_dirac}. The black marker labelled `SR2014' indicates the constraint inferred by \cite{Schindelegger2014} for present-day Earth on the basis of barometric measurements. Parameters are $\ggravi = 9.81~{\rm m \ s^{-2}}$, $\Rpla=6371$~km, $\adiabexp=1.23$, $\Rspec=287~{\rm J \ kg^{-1} \ K^{-1}}$, $\psurf = 1.013$~bar, $\Fstar = 1361~{\rm W \ m^{-2}}$, $\Tbulk = 243$~K.    }
    \label{fig:spectra_earth}
\end{figure*}

As a next step, we evaluate the dimensionless factor scaling the response of the degree-2 mode, $\presstidevc$. Using \eq{reduced_frequency}, it reads
\begin{equation}
\label{presstidevc2}
\presstidevc = \frac{\rcp \opacity \fluxtideni{2}}{\psurf \heightsurf \fresoi{2}},
\end{equation}
where $\opacity$ denotes the absorbed-flux fraction and $\fluxtideni{2}$ the expansion coefficient of the incident stellar flux on the degree-2 Hough function. Since tidal heating stems from the direct atmospheric absorption of the incident Solar flux by ozone and water vapour, both $\fluxtideni{2}$ and $\presstidevc$ are real-valued parameters. From the absorption profiles of \cite{Lindzen1968a} and \cite{VR2013}, \cite{Farhat2024} estimate $\opacity$ to lie between $17.6\%$ and $18.1\%$. We therefore adopt $\opacity = 18 \%$. The coefficient $\fluxtideni{2}$ is computed using \eq{fluxtiden}, yielding 
\begin{equation}
\label{deltaF2_Fstar}
\frac{\fluxtideni{2}}{\Fstar} =  \sum_{\llat =2}^{+ \infty} \fluxnalfi{\llat}{2} \scal{\HoughF{2}{2}{\spinpar_2}}{\LegFnorm{\llat}{2}} \approx 0.226502.
\end{equation}
This yields $\presstidevc = \num{8.08e-5}$ for the isentropic configuration and $\presstidevc = \num{9.42e-5}$ for the isothermal one. These values allow us to compute the complex Fourier coefficient of surface-pressure oscillations for the degree-2 mode at any tidal frequency, using \eqs{presstidev_surf_neu}{presstidev_surf_isotherm_dirac}. Finally, the tidal response at the equator, $\presstideneq$, is obtained from the degree-2 surface-pressure component, $\presstidevsurfi{2}{2}$, by retaining the dominant contribution in \eq{presstideneq},
\begin{equation}
\label{presstideneq_simp}
\presstideneq \approx \cequat \psurf  \presstidevsurfi{2}{2},
\end{equation}
where the dimensionless numerical factor $\cequat$ is \mynom[S]{$\cequat$}{dimensionless numerical factor in \eq{presstideneq_simp}}
\begin{equation}
\cequat = \sqrt{\frac{15}{16}}  \frac{\scal{\HoughF{2}{2}{\spinpar_2}}{\LegFnorm{2}{2}}}{\cshape} \approx 1.019545. 
\end{equation}

Figure~\ref{fig:spectra_earth} displays the frequency dependence of semidiurnal surface-pressure oscillations predicted at the equator by our analytical model for both the isentropic and isothermal equilibria. We plot the oscillation amplitude, the local Solar time of the dayside maximum, and the imaginary component as functions of the LOD for various values of the damping ratio, $\disspar$. For comparison, we also include the observational constraint derived by \cite{Schindelegger2014} from global barometric measurements. Interestingly, despite the markedly different functional forms of their responses (\eqs{presstidev_surf_neu}{presstidev_surf_isotherm_dirac}), the isentropic and isothermal models exhibit qualitatively similar behaviour. The isothermal curves appear as slightly shifted versions of their isentropic counterparts, consistent with the frequency ratio given by \eq{ratio_freq_iso_isen}. 

The isentropic predictions show excellent agreement with the observational data, in line with the fact that the isentropic profile provides a good description of the thermal structure of Earth's troposphere (see \paperone, Fig.~2). For $\disspar=0.25$, the model reproduces the observed modulus, phase, and imaginary component of the semidiurnal surface-pressure oscillations \citep[$\abs{\presstideneq} = 123.4$~Pa and $\tmax = 09{:}45$;][]{Schindelegger2014}. The predicted oscillation amplitude is $119.6$~Pa, and the Solar time of the dayside maximum is $09{:}46$ LST. By contrast, the isothermal predictions are less accurate owing to the shift in the resonance frequency. For all damping ratio, the oscillation amplitude is ${\sim}79$~Pa, which is $36\%$ smaller than the empirical value. We note, however, that the discrepancies between theory and measurements in the isothermal case are enhanced by the frequency dependence of $\presstideneq$.

\subsection{Consequences of phase-shifted tidal heating}
\label{sec:phase_shifted_heating}

\def\wpanel{0.48\textwidth}
\begin{figure}
\centering
\includegraphics[width=\wpanel,trim={0cm 0.7cm 0.5cm 0cm}, clip]{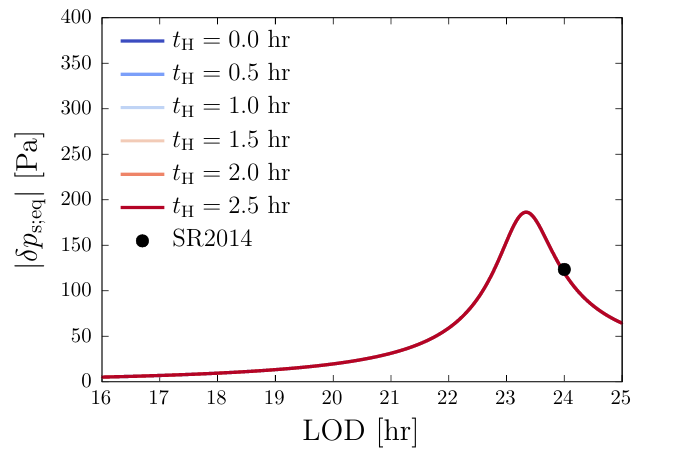} \\
\includegraphics[width=\wpanel,trim={0.0cm 0.7cm 0.5cm 0cm}, clip]{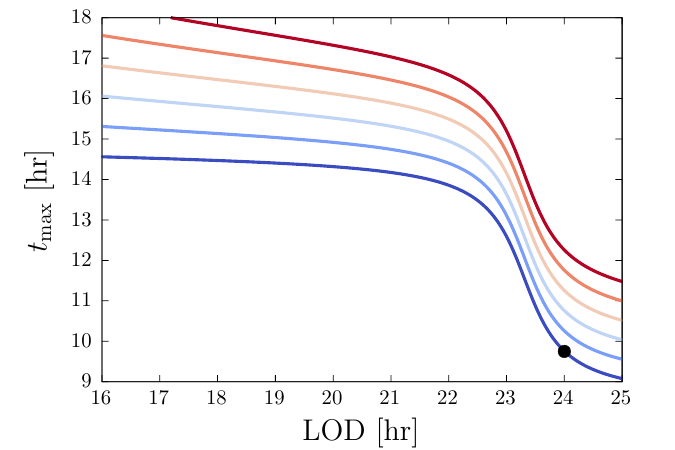} \\
\includegraphics[width=\wpanel,trim={0cm 0cm 0.5cm 0cm}, clip]{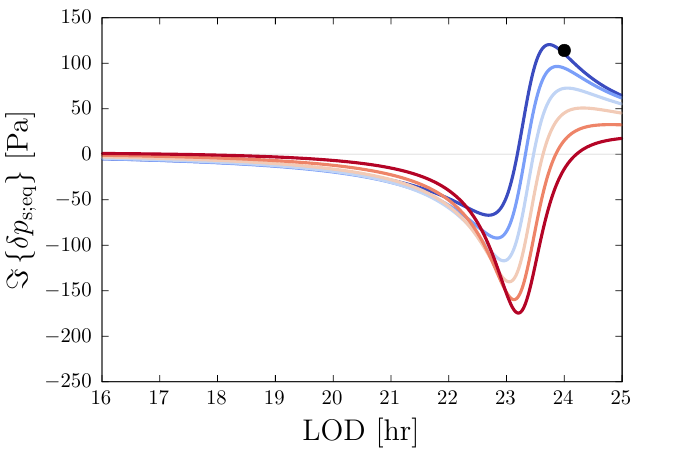}
    \caption{Influence of phase-shifted forcings on the tidal response of isentropic atmospheres for various values of the time lag. {\it From top to bottom:} Modulus of surface pressure oscillations at the equator, Solar time at which they are maximal ($\time_{\rm max}$), and imaginary component. Solid lines denote the model predictions, which are computed using \eq{presstidev_surf_neu} together with \eq{phase_shifted_forcing}. The black marker labelled `SR2014' indicates the constraint inferred by \cite{Schindelegger2014} for present-day Earth on the basis of barometric measurements. Parameters are $\ggravi = 9.81~{\rm m \ s^{-2}}$, $\Rpla=6371$~km, $\adiabexp=1.23$, $\Rspec=287~{\rm J \ kg^{-1} \ K^{-1}}$, $\psurf = 1.013$~bar, $\Fstar = 1361~{\rm W \ m^{-2}}$, $\Tsurf = 288$~K, $\disspar=0.25$. }
    \label{fig:spectra_dependence_phase}
\end{figure}

In \sect{ssec:application_present_earth}, the thermal forcing inducing the semidiurnal thermal tides is assumed to be in phase with insolation, which corresponds to the regime of direct absorption: the tidal bulge arises from the absorption of the incident stellar flux. This results in two resonant peaks of the tidal torque on either side of the Lamb-wave resonance frequency, one being negative and the other positive (see \fig{fig:spectra_earth}, bottom panels). These peaks are symmetric in amplitude in the limit of quasi-adiabatic compression ($\disspar \to 0$), and the positive peak tends to become larger than the negative one as $\disspar$ increases. However GCM simulations show the opposite.\cite{Wu2023} and \cite{DG2024} find a decelerative torque (negative peak) approximately twice larger than the accelerative one (positive peak) in absolute value (see Fig.~S4 in supplementary materials in the former, and Fig.~1c in the latter). In \cite{ADL2019}, the positive peak vanishes and the negative one only remains (see their Fig.~3, top left panel).

Multiple factors could contribute to this predominance of the negative peak. For instance, \cite{ZW1987} recover a similar behaviour by using a tidal heating with a non-zero phase that accounts for delayed latent heat releases due to cumulonimbus activity \citep[][]{Lindzen1978,Hamilton1981}. Their analytical solution exhibits a negative peak of larger amplitude than its positive counterpart (see their Fig.~4, dashed line). Similarly, this effect may result from radiative-diffusive heat exchanges at the planet's surface, as discussed by \citet{ADL2019} and \citet{Farhat2024}. \cite{Farhat2024} notably show in their Figs.~3 and~4 that a tidal model incorporating soil dissipative processes in a simplified manner can reproduce, at least qualitatively, the asymmetry observed in numerical solutions, although further refinement of the theory remains necessary \citep[see also][Fig.~7]{Laskar2024}. When soil diffusion -- or any physical process changing the phase of the forcing -- is included, the latter is delayed with respect to insolation. Consequently, the resulting atmospheric response is also delayed, which generates an offset in the imaginary part of surface pressure oscillations. 

We examine here how the phase of the tidal heating affects the dependence of the resulting thermotidal torque on tidal frequency. The physical setup is that used for the present-day Earth in \sect{ssec:application_present_earth} and the plots of \fig{fig:spectra_earth} in the isentropic configuration. The damping ratio associated with radiative cooling is set to $\disspar=0.25$. For simplicity, we adopt a forcing of the form proposed by \cite{Lindzen1978} and \cite{Hamilton1981} to account for delayed latent heat releases. The vertical profile of the heat absorbed per unit mass per unit time for the degree-$\nn$ mode is expressed as 
\begin{equation}
\label{phase_shifted_forcing}
\Jtidev \left( \xx \right) = \Jtidev^{0} \left( \xx \right) \expo{- \inumber \ftide \theating},
\end{equation}
where $\Jtidev^{0}$\mynom[S]{$\Jtidev^{0}$}{real-valued vertical profile of tidal heating in \eq{phase_shifted_forcing}} is the real-valued, exponentially decaying heating profile used in the present work, given by Eq.~(74) of \paperone, and $\theating$\mynom[S]{$\theating$}{constant time lag of tidal heating in \eq{phase_shifted_forcing}} a constant time lag. We note that the above model and that given by Eq.~(H.4) of \cite{Farhat2024} for soil diffusion have different frequency dependencies. The former results in larger phase shifts than the latter at high frequencies and smaller phase shifts at low frequencies, which may affect the functional form of analytical solutions and the shape of the associated frequency spectra.

Figure~\ref{fig:spectra_dependence_phase} displays the frequency dependence of surface-pressure oscillations at the planet's equator for values of the time lag lying between 0.~hr (no phase) and 2.5~hr (strongly delayed forcing). We plot the oscillation amplitude, the local Solar time of the dayside maximum, and the imaginary component as functions of the LOD. Similar to \fig{fig:spectra_earth}, the observational constraint derived by \cite{Schindelegger2014} from barometric measurements is included. As the phase of the tidal forcing does not affect the modulus of the response, the latter is the same for all values of $\theating$. Consequently, there is no way to infer the time lag of the forcing from the frequency dependence of the amplitude of surface pressure oscillations. However we retrieve the phase of the forcing as an offset in the phase of the tidal response. Besides, the phase influences the relative predominances of the positive and negative peaks of the resulting thermotidal torque, which is proportional to the imaginary component of $\presstideneq$. 

As the time lag increases, the positive peak responsible for the thermodynamic acceleration of the Earth's rotation diminishes, while the negative peak increases in amplitude. We remark that the shape of the numerical solutions obtained from GCM simulations by \cite{DG2024} in their Fig.~1c and \cite{Wu2023} in their Fig.~S4, where the braking peak is roughly twice larger than the accelerating one, is approximately reproduced by the analytical model for $\theating=1.0$~hr. Therefore, strong pressure oscillations do not necessarily result in a strong thermodynamic acceleration of the planet's rotation. Particularly, a tidal lock such as that hypothesised for the Precambrian Earth \citep[][]{ZW1987,BS2016,MK2023,Wu2023,DG2024} turns out to be less likely if the tidal heating is phase-shifted than in the case of direct absorption.

\def\wpanel{0.48\textwidth}
\begin{figure*}
\centering
   \RaggedRight \twolabelsat{4.5cm}{Non-resonant mode ($\Sn = -1$)}%
                           {13.5cm}{Resonant mode ($\Sn = +1$)}\\[0.1cm]
\includegraphics[width=\wpanel,trim={0cm 0.7cm 0.5cm 0cm}, clip]{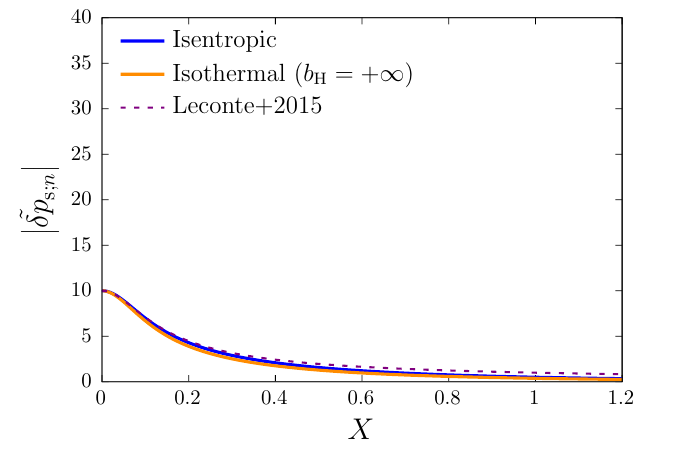} 
\includegraphics[width=\wpanel,trim={0cm 0.7cm 0.5cm 0cm}, clip]{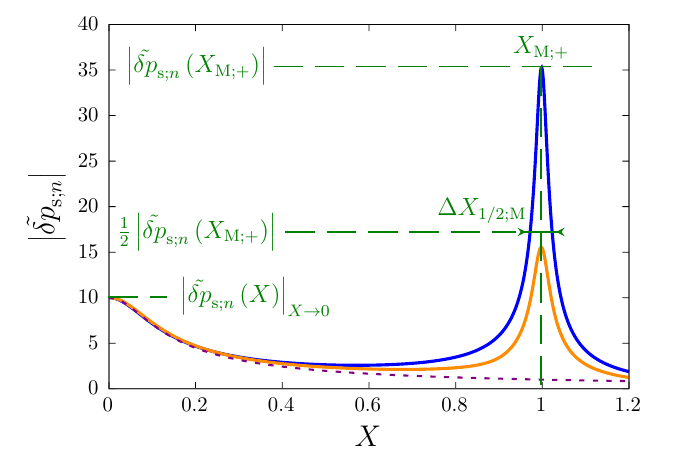} \\
\includegraphics[width=\wpanel,trim={0.0cm 0.7cm 0.5cm 0cm}, clip]{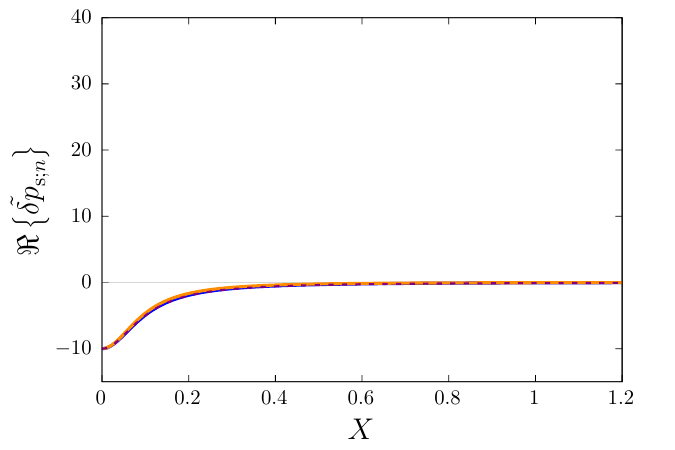}
\includegraphics[width=\wpanel,trim={0.0cm 0.7cm 0.5cm 0cm}, clip]{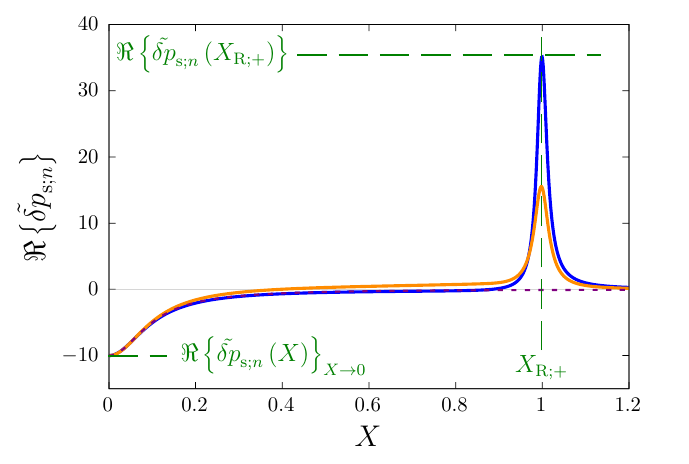} \\
\includegraphics[width=\wpanel,trim={0cm 0cm 0.5cm 0.cm}, clip]{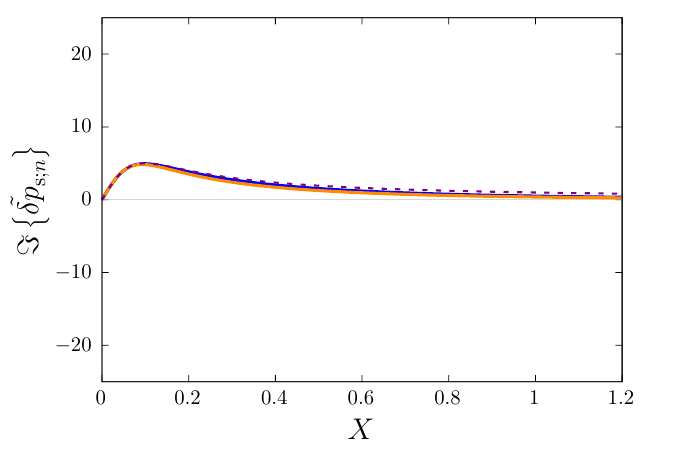}
\includegraphics[width=\wpanel,trim={0cm 0cm 0.5cm 0cm}, clip]{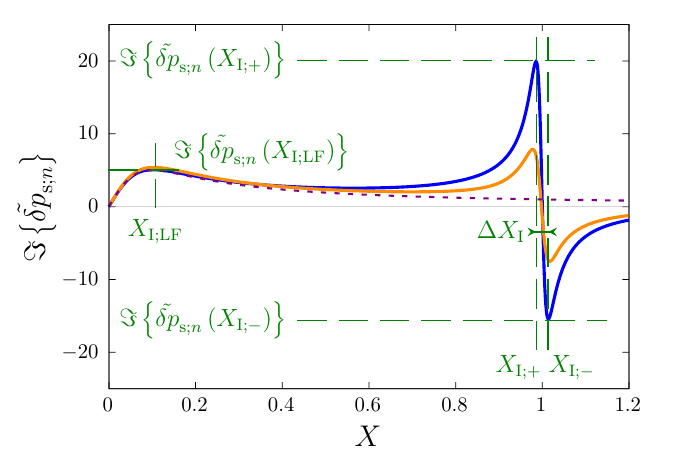}
    \caption{Frequency spectra of tidal surface pressure variations for non-resonant (left) and resonant (right) modes in the quasi-adiabatic compression regime. {\it From top to bottom:} Modulus, real part and imaginary part of $\presstidevsurf$ (vertical axis) as functions of the reduced tidal frequency, $\XX$ (horizontal axis). The blue line represents the solution for isentropic atmospheres (\eq{presstidev_surf_neu}); the orange line corresponds to the isothermal case with $\bj=+\infty$ (\eq{presstidev_surf_isotherm_dirac}); and the thin dashed purple line shows the empirical model proposed by \cite{Leconte2015} based on GCM simulations (\eq{model_leconte2015}). Spectra are computed using $\presstidevc=1$, $\adiabexp = 7/5$ (i.e. $\rcp = 2/7$, diatomic gas), and $\disspar=0.1$. Green labels denote the main spectral features quantified in Table~\ref{tab:solution_properties}.}
    \label{fig:spectra_deltap}
\end{figure*}

\begin{figure}
\centering
\includegraphics[width=0.48\textwidth]{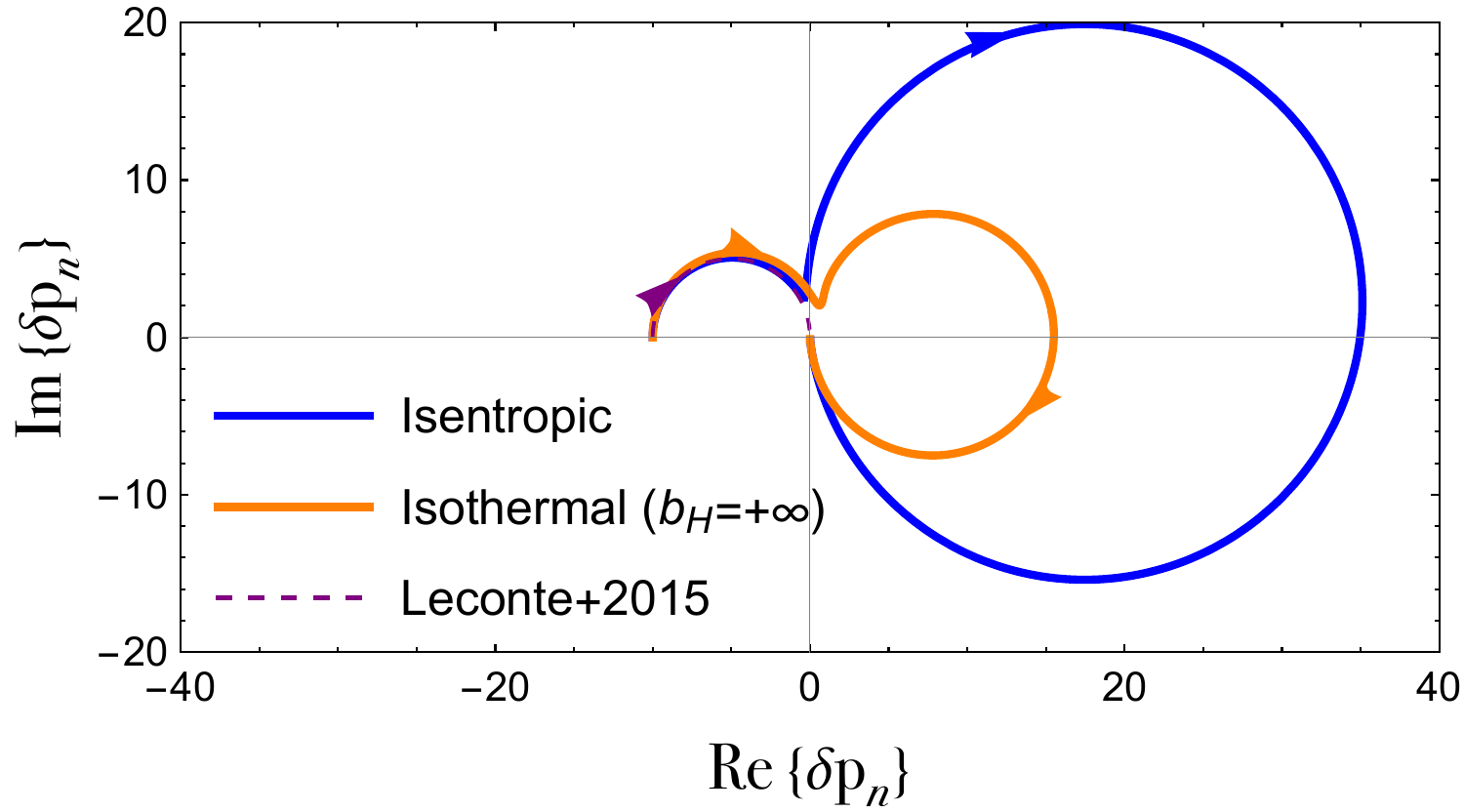}
    \caption{Nyquist plot of surface pressure variations induced by resonant modes ($\Sn = +1$). The trajectory of the atmospheric frequency response is shown in the complex plane, with the real (horizontal axis) and imaginary (vertical axis) parts of $\presstidevsurf$ plotted for $\XX \geq 0$. The blue line represents the solution for isentropic atmospheres (\eq{presstidev_surf_neu}); the orange line corresponds to the isothermal case with $\bj=+\infty$ (\eq{presstidev_surf_isotherm_dirac}); and the thin dashed purple line shows the empirical model proposed by \cite{Leconte2015} based on GCM simulations (\eq{model_leconte2015}). Parameters are $\presstidevc=1$, $\adiabexp = 7/5$ (i.e. $\rcp = 2/7$, diatomic gas), and $\disspar=0.1$. Arrows indicate the direction of increasing frequency.}
    \label{fig:nyquist_solutions}
\end{figure}

\section{Frequency behaviour}
\label{sec:frequency_behaviour}

\subsection{Pressure variations spectra}

The analytical expressions of tidal surface pressure variations established in \sect{sec:thermotidal_torque} allow the frequency behaviour of the thermotidal torque to be explored exhaustively. Here, we focus on fast rotating Earth-like planets, as the present study builds on a series of works dedicated to the past evolution of the Earth-Moon system \citep[][]{Farhat2022a,Farhat2022b,Farhat2024,Farhat2025,Auclair2023,Auclair2025,Laskar2024}. 

It is worth noting that we use the expression `fast rotation'  to designate configurations in which the central body’s spin angular velocity far exceeds any other characteristic frequency of the orbital system -- for example, the perturber’s mean motion ($\norb \ll \spinrate$). In the case of Earth, the fast-rotation approximation is generally well justified for semidiurnal tides, provided that the rotation period remains much shorter than one year, a condition that has been satisfied throughout the planet's history. By contrast, this criterion is not necessarily met in close-in star-planet systems, where a planet approaching tidal lock may rotate rapidly while its spin rate remains only moderately larger than its orbital mean motion. Nevertheless, in the context of semidiurnal tidal torques, this restriction appears to be primarily a theoretical limitation. Indeed, the torque predicted by the analytical model remains consistent with GCM results in the near-synchronous regime, where the fast-rotation approximation formally breaks down. In particular, the torque decreases linearly to zero as $\XX \to 0$, in agreement with the GCM simulations reported by \cite{Leconte2015} in the Venus' case, as discussed in \sect{ssec:lowpass_filter}. 

These findings suggest that, in practice, the frequency analysis developed in this section may remain applicable beyond the formal limits of the fast-rotation regime, despite having been derived under the assumption that $\norb \ll \spinrate$. Whether this apparent robustness extends to near-synchronously rotating planets in close-in systems nevertheless remains an open question. To our knowledge, the frequency dependence of the thermotidal torque has not yet been investigated using GCM simulations for such planets. Consequently, the applicability of the analytical model in this regime remains to be established, and its predictions should therefore be interpreted cautiously outside the formal domain of validity of the fast-rotation approximation.

When fast rotation is assumed, the parameters $\spinpar$, $\Lambn$, $\freson$, $\Sn$, $\disspar$ and $\presstidevc$ become independent of the tidal frequency. Consequently, $\presstidevsurf$ can be expressed explicitly as a function of $\XX$ in \eqs{presstidev_surf_neu}{presstidev_surf_isotherm_dirac}. Also, the asymmetric frequency dependence of the tidal response with respect to synchronisation ($\XX=0$), which is caused by Coriolis terms, vanishes. This implies
\begin{align}
\abs{\presstidevsurf \left( - \XX \right)}  & = \abs{\presstidevsurf \left(  \XX \right)} , \\
\Re \left\{ \presstidevsurf \left( - \XX \right) \right\} & = \Re \left\{ \presstidevsurf \left( \XX \right) \right\} , \\
\label{imdeltap_sym}
\Im \left\{ \presstidevsurf \left( - \XX \right) \right\} & = - \Im \left\{ \presstidevsurf \left(  \XX \right) \right\}, 
\end{align}
and the study can therefore be reduced to positive frequencies ($\XX \geq 0$). Moreover, in the viscous regime ($\disspar \gg1$), radiative cooling damps the tidal response across the entire frequency spectrum, effectively suppressing the thermodynamic acceleration of the planet's rotation. Thus, we restrict our analysis to the quasi-adiabatic compression regime ($\disspar \ll1$). Finally, we note that $\presstidevc$ may be assigned any value in the calculations because it serves merely as a scaling factor. We therefore set it to unity in this section. 

In Figure~\ref{fig:spectra_deltap}, the modulus, real part, and imaginary part of $\presstidevsurf$ are plotted as functions of the reduced tidal frequency for isentropic (solid blue line) and isothermal (solid orange line) atmospheres. For each atmospheric structure, the figure shows the behaviours of non-resonant ($\Sn = -1$, left panels) and resonant ($\Sn = +1$, right panels) modes. The plotted quantities were calculated from \eqs{presstidev_surf_neu}{presstidev_surf_isotherm_dirac} for diatomic gases and weak dissipative processes, assuming $\presstidevc = 1$, $\adiabexp = 7/5$ (i.e. $\rcp = 2/7$), and $\disspar = 0.1$. Remarkably, both solutions exhibit qualitatively similar dependences of the tidal surface-pressure oscillations on the tidal frequency. In the high-frequency regime, resonant modes display a single peak in $\abs{\presstidevsurf}$ and in $\Re \left\{ \presstidevsurf \right\} $, and a double peak in $\Im \left\{ \presstidevsurf \right\}$. The two models, however, predict different peak amplitudes, underscoring the influence of atmospheric structure on the tidal response. For instance, at resonance, the maximum amplitude of $\abs{\presstidevsurf}$ in the isothermal case is roughly half that obtained with the isentropic model. In the Nyquist plot \citep[e.g.][]{Faulkner1969book} shown in \fig{fig:nyquist_solutions}, the resonance produces a circular trajectory in the complex plane (see the positive real part). Similarly, the atmospheric response in the low-frequency range is represented by a semicircle. These patterns correspond to second-order bandpass and first-order lowpass filters, respectively.

\subsection{Second-order bandpass filter}
\label{ssec:bandpass_filter}

The circular trajectory observed in the Nyquist plot is characteristic of second-order bandpass filters, whose transfer function is expressed as 
\begin{equation}
\label{bandpass_tf}
H = \frac{\bpgain}{1 + \inumber \bpquality \left( \XX - \frac{1}{\XX} \right)},
\end{equation}
with $\bpgain$\mynom[S]{$\bpgain$}{gain coefficient of the equivalent second-order bandpass filter} being the gain coefficient and $\bpquality$\mynom[S]{$\bpquality$}{quality factor of the equivalent second-order bandpass filter}  the quality factor. For the sake of generality, we include to this expression the phase-shifted tidal heating given by \eq{phase_shifted_forcing}. Introducing the lag angle at resonance, $\phiheating = \freson \theating $\mynom[S]{$\phiheating$}{lag angle of tidal heating at resonance}, the latter becomes
\begin{equation}
\label{phase_forcing_normalised}
\Jtidev \left(\xx \right) = \Jtidev^{0} \left( \xx \right) \expo{- \inumber  \phiheating \XX}.
\end{equation}
Besides, it is noteworthy that the circular pattern associated with the bandpass filter in the Nyquist plot is not symmetric with respect to the horizontal axis, as described by \eq{bandpass_tf}, but slightly rotated anticlockwise with respect to the origin. This indicates a positive phase angle inherent to the atmospheric tidal response and depending on $\adiabexp$ and $\disspar$ solely. To account for this phase angle, denoted by $\phiatm$\mynom[S]{$\phiatm$}{phase angle of surface pressure variations at resonance in the regime of direct absorption}, we just multiply \eq{bandpass_tf} by a factor $\exp \left( \inumber \phiatm \XX \right)$ similar to that describing phase-shifted tidal heating in \eq{phase_forcing_normalised}. The total lag angle, $\philag$\mynom[S]{$\philag$}{total lag angle of the equivalent bandpass filter at resonance}, is thus given by
\begin{equation}
\philag = \phiheating-\phiatm.
\end{equation} 
Since the dependence of the phase on $\XX$ can be neglected in the vicinity of the Lamb-wave resonance ($\abs{\XX- 1}\ll 1$), the isentropic and isothermal solutions given by \eqs{presstidev_surf_neu}{presstidev_surf_isotherm_dirac} in this regime are both approximated by 
\begin{equation}
\label{presstidevsurf_bandpass}
\presstidevsurf =  \frac{\bpgain \expo{- \inumber \philag}}{1 + \inumber \bpquality \left( \XX - \frac{1}{\XX} \right)}.
\end{equation}

The parameters $\bpgain$ and $\bpquality$ in \eq{presstidevsurf_bandpass} have different definitions in the two models, which we may establish by expressing the extrema of the modulus, real part and imaginary part of $\presstidevsurf$ and their associated frequencies as functions of the two parameters. These derivations were carried out by assuming that $\bpquality \gg 1$, which corresponds to the quasi-adiabatic compression regime ($\disspar \ll1$), and retaining the dominant terms only. In order to lighten expressions, we adopt the following notations: subscripts $\imod$, $\irp$ and $\iip$ refer to quantities associated with the modulus, real part, and imaginary part of $\presstidevsurf$, respectively; the plus ($\iplus$) and minus ($\iminus$) symbols denote the maxima and minima linked to the resonance. For example, $\XX_{{\rm I};+}$ indicates the position of the maximum attained by the resonant peak of $\Im \left\{ \presstidevsurf \right\}$ in the frequency spectrum. 

The frequency of the maximum of the modulus and its amplitude are simply given by 
\begin{align}
\label{absmax_lag}
& \XAresplus = 1, & \absdpresplus = \bpgain. 
\end{align}
As discussed in \sect{sec:phase_shifted_heating}, they are both independent of the phase lag of tidal heating. Similarly, the frequencies at which the real part of $\presstidevsurf$ is maximal and minimal are expressed as
\begin{align}
& \XRresplus = 1 - \frac{\tan \left( \frac{\philag}{2} \right)}{2 \bpquality},
& \XRresminus = 1 + \frac{\cot \left( \frac{\philag}{2} \right)}{2 \bpquality},
\end{align}
with the corresponding maximum and minimum being given by 
\begin{align}
& \redpresplus = \bpgain \cos^2 \left( \frac{\philag}{2} \right), \\
& \redpresminus = - \bpgain \sin^2 \left( \frac{\philag}{2} \right).
\end{align}
The imaginary part becomes zero for 
\begin{equation}
\XX = 1 - \frac{\tan \philag}{2 \bpquality},
\end{equation}	
and exhibits two extrema in the vicinity of the resonance frequency. The frequencies at which these extrema occur are given by
\begin{align}
& \XIresplus = 1 - \frac{\cot \left(\frac{\pi}{4}- \frac{\philag}{2} \right)}{2 \bpquality} , 
& \XIresminus = 1 + \frac{\tan \left( \frac{\pi}{4} - \frac{\philag}{2} \right)}{2 \bpquality} . 
\end{align}
The maximum of the positive peak is expressed as
\begin{equation}
\imdpresplus = \bpgain \sin^2 \left( \frac{\pi}{4}  - \frac{\philag}{2} \right),
\end{equation}
and the minimum of the negative peak as
\begin{equation}
\imdpresminus = - \bpgain \cos^2 \left( \frac{\pi}{4}  - \frac{\philag}{2} \right).
\end{equation}

We remark that the phase of the forcing uniformly shifts both the zero and the two extrema of the imaginary part of $\presstidevsurf$ by $\tan \philag / \left( 2 \bpquality \right)$ towards the lower frequencies. The frequency difference between the two peaks, $\dximsep$, is expressed as 
\begin{equation}
\label{dximsep_lag}
\dximsep = \frac{1}{\bpquality \cos \philag},
\end{equation}
and the ratio between the maximum and the minimum in absolute value as\mynom[S]{$\ratiopeaks$}{ratio between the torque's maximum and minimum near resonance}
\begin{equation}
\label{ratio_extrema_imag}
\ratiopeaks = \abs{\frac{\imdpresplus}{\imdpresminus}} = \tan^2 \left( \frac{\pi}{4} - \frac{\philag}{2} \right), 
\end{equation}
which explicitly relates the asymmetry observed in \fig{fig:spectra_dependence_phase} to the forcing phase lag. 

We express the gain coefficient and the quality factor of the bandpass filter in terms of the model parameters by considering the regime of direct absorption treated in \sect{ssec:quantitative_analysis}. In this regime, $\philag= - \phiatm$, and \eq{dximsep_lag} thus simplifies to $\dximsep = \bpquality^{-1} + O \left( \phiatm^2 \right)$ assuming that $\phiatm \scale \disspar \ll 1$ and neglecting the order-2 term. Using \eqs{absmax_lag}{dximsep_lag} together with the expressions of Table~\ref{tab:solution_properties}, we obtain
\begin{equation}
\label{bpfilter_gain_quality}
\begin{array}{lll}
\displaystyle \bpgain = \frac{\presstidevc}{\rcp \disspar}  & \displaystyle \bpquality = \frac{1}{\rcp \disspar} & \mbox{(isentropic)}, \\[0.2cm]
\displaystyle \bpgain =  \frac{\presstidevc \left( 2 - \adiabexp \right)}{\left( \adiabexp - 1 \right) \disspar} & \displaystyle \bpquality = \frac{1}{\rcp \adiabexp \disspar} &  \mbox{(isothermal)}. 
\end{array}
\end{equation}
Similarly, the phase angle inherent to atmospheric tidal dynamics, $\phiatm$, is readily deduced from the summation of the two extrema of the imaginary part,
\begin{equation}
\imdpresplus + \imdpresminus = - \bpgain \sin \left( \philag \right),
\end{equation}
which, in the regime of direct absorption ($\phiheating=0$) and quasi-adiabatic compression ($\disspar \ll1$), yields
\begin{align}
\phiatm & = \arcsin \left( \frac{\imdpresplus + \imdpresminus}{\bpgain} \right) \\ 
 & \approx  \frac{\imdpresplus + \imdpresminus}{\bpgain}. \nonumber
\end{align} 
Using the above formula together with the expressions provided by Table~\ref{tab:solution_properties}, we end up with
\begin{equation}
\label{phiatm_expr}
\phiatm = \left\{
\begin{array}{ll}
\displaystyle \left( 1 + \rcp \right) \disspar & \mbox{(isentropic)}, \\[0.1cm]
\displaystyle \frac{2-\adiabexp^2}{\left(2-\adiabexp \right)^2} \disspar & \mbox{(isothermal)}. 
\end{array}
\right.
\end{equation}

\subsection{First-order lowpass filter}
\label{ssec:lowpass_filter}

In the low-frequency range, the imaginary part of $\presstidevsurf$ behaves similarly to its numerical counterpart obtained from GCM simulations, with $\Im \left\{ \presstidevsurf \right\}  $ tending to zero at $\XX = 0$ and reaching a maximum near synchronous rotation \citep[see Figs.~1 and~3 in][respectively]{Leconte2015,ADL2019}. It is worth noting that $\Im \left\{ \presstidevsurf \right\}$ is an odd function of $\ftide$ in these numerical results, which mitigates the impact of Coriolis effects on the tidal torque for $\XX \ll1$. In particular, our ab initio analytical solutions both converge towards the empirical model proposed by \cite{Leconte2015} from GCM simulations (see their Eq.~(2)),
\begin{equation}
\label{model_leconte2015}
\presstidevsurf = - \frac{\presstidevc}{\disspar + \inumber \XX},
\end{equation}
represented by the dashed purple line in \fig{fig:spectra_deltap}, which corresponds to the characteristic transfer function of first-order lowpass filters. In \eq{model_leconte2015}, the imaginary part of $\presstidevsurf$ linearly tends to zero in the zero-frequency limit and decays as $\Im \left(\presstidevsurf \right) \scale \XX^{-1} $ as $\XX \to \infty$. It reaches a maximum for $\XX = \Xmaxsyn$. The frequency at which the maximum occurs and the corresponding value are given by 
\begin{align}
& \Xmaxsyn = \disspar, & \imdpmaxsyn = \frac{\presstidevc}{2 \disspar}.
\end{align}

This asymptotic behaviour indicates that the assumed fast-rotation approximation remains valid in practice at low frequencies. Consequently, the solutions computed here with frequency-independent spin parameters can be applied over the entire frequency range, even though $\spinpar$, $\Lambn$, and $\HoughF{\nn}{\mm}{\spinpar}$ are, in principle, expected to vary with $\ftide$ in the zero-frequency limit. Furthermore, in the fast-rotation limit, the stellar semidiurnal forcing primarily excites the degree-2 mode, implying that $\presstiden_2$ largely determines the frequency spectrum of the thermotidal torque for $\XX \gtrsim 1$. One may reasonably assume that this predominance extends to lower frequencies. Under this assumption we neglect the summation in \eq{deltaps22_series} and retain only the degree-2 term. The simplified analytical solutions therefore yield the pressure coefficient $\presstidenquad$ appearing in \eq{torquez_expr},\mynom[S]{$\presstidenquad$}{semidiurnal Fourier coefficient of the tidal surface-pressure oscillation}
\begin{equation}
\label{deltaps22_deltaps2h}
\presstidenquad  \approx \sqrt{2 \pi} \scal{\LegFnorm{2}{2}}{\HoughF{2}{2}{\spinpar_2}} \,  \presstidevsurfi{2}{2}.
\end{equation}

\def\hsep{\hspace{0.3cm}}
\def\hinter{0.3cm}
\begin{table*}[h]
\centering
\caption{\label{tab:solution_properties} Analytical scaling laws characterising the main features of the two solutions for $\spinpar = {\rm cst}$, $\bj = + \infty$, $\Sn = +1$, and $\disspar \ll 1$. }
\begin{tabular}{ll} 
 \hline 
 \hline 
 \textsc{Isentropic} & \textsc{Isothermal} \\
 \hline \\[-0.3cm] 
 \multicolumn{2}{c}{\textit{Asymptotic behaviour near synchronisation ($\XX \ll \disspar$)}} \\[0.1cm]
  % =============================
  $\displaystyle \abs{\presstidevsurf \left( \XX \right)}_{\XX \rightarrow 0} \sim \frac{\presstidevc}{\disspar}$ & 
  $\displaystyle \abs{\presstidevsurf \left( \XX \right)}_{\XX \rightarrow 0} \sim \frac{\presstidevc}{\disspar}$ \\[\hinter]
   % =============================
 $\displaystyle \Re \left\{ \presstidevsurf\left( \XX \right) \right\}_{\XX \rightarrow 0} \sim -\frac{\presstidevc}{\disspar}$ & 
 $\displaystyle \Re \left\{ \presstidevsurf\left( \XX \right) \right\}_{\XX \rightarrow 0} \sim -\frac{\presstidevc}{\disspar}$ \\[\hinter]
 % =============================
$\displaystyle \left. \DD{}{\XX} \left(  \Im \left\{ \presstidevsurf \left( \XX \right) \right\} \right) \right|_{\XX=0} \sim \frac{\presstidevc}{\disspar^2}$ & $\displaystyle \left. \DD{}{\XX} \left(  \Im \left\{ \presstidevsurf \left( \XX \right) \right\} \right) \right|_{\XX=0} \sim \frac{\presstidevc}{\disspar^2}$ \\[\hinter]
 % =============================
 \hline \\[-0.3cm] 
\multicolumn{2}{c}{\textit{Behaviour in the low-frequency range ($\XX \ll 1$)} } \\[0.1cm]
$\displaystyle \Xmaxsyn = \disspar \left[ 1 + \left(2 + \rcp \right) \disspar^2 \right] + O \left( \disspar^4 \right)$\mynom[S]{$\Xmaxsyn$}{frequency at which the maximum of $\Im \left\{ \presstidevsurf \right\}$ is reached in the near-synchronous regime}  & $\displaystyle \Xmaxsyn = \disspar \left[ 1 + \frac{385 \sqrt{\adiabexp-1}}{512 \sqrt{2}} \disspar \right] + O \left( \disspar^3 \right) $  \\[\hinter]
 % =============================
 $\displaystyle \imdpmaxsyn = \frac{\presstidevc}{2 \disspar} \left[ 1+ \left( 1 + \rcp \right) \disspar^2  \right]  + O \left( \disspar^2 \right)$ & $\displaystyle \imdpmaxsyn = \frac{\presstidevc}{2 \disspar} \left[ 1 + \frac{19 \sqrt{\adiabexp-1}}{4\sqrt{15}} \disspar \right] + O \left( \disspar \right)$  \\[\hinter] 
   % =============================
   \hline \\[-0.3cm] 
\multicolumn{2}{c}{\textit{Behaviour at the Lamb-wave resonance ($\XX \approx 1$)}} \\[0.1cm]
 % =============================
$\displaystyle \XAresplus =  1 - \frac{1}{2} \rcp \left( 1 + \rcp \right) \disspar^2 + O \left( \disspar^3 \right) $\mynom[S]{$\XAresplus$}{frequency at which the maximum of $\left| \presstidevsurf \right|$ associated with the Lamb-wave resonance is reached}  & $\displaystyle \XAresplus = 1 + \frac{1 - 2 \adiabexp^2 + \adiabexp^3}{2 \left(2 - \adiabexp \right)^2} \disspar^2 + O \left( \disspar^4 \right) $ \\[\hinter]
 % =============================
 $\displaystyle \absdpresplus =  \frac{\presstidevc}{\rcp \disspar} \left[ 1 + \frac{1}{2} \left( 1 + \rcp \right)^2 \disspar^2  \right] + O \left( \disspar^2 \right)$ & $\displaystyle \absdpresplus =  \frac{\presstidevc \left( 2 - \adiabexp \right)}{\left( \adiabexp - 1 \right) \disspar} + O \left( \disspar \right) $  \\[\hinter]
 % =============================
$\displaystyle \dxmidres =   \sqrt{3} \rcp \disspar + O \left( \disspar^3 \right)  $\mynom[S]{$\dxmidres$}{FWHM of the resonant peak of $\left| \presstidevsurf \right|$} & $\displaystyle \dxmidres = \sqrt{3} \rcp \adiabexp \disspar + O \left( \disspar^3 \right) $  \\[0.2cm]
 % =============================
$\displaystyle \XRresplus = 1 - \frac{1}{4} \rcp \left( 1 + \rcp \right) \disspar^2 + O \left( \disspar^4 \right) $\mynom[S]{$\XRresplus$}{frequency at which the maximum of $\Re \left\{ \presstidevsurf \right\} $ associated with the Lamb-wave resonance is reached}  & $ \displaystyle \XRresplus = 1 -  \frac{ \adiabexp \left(\adiabexp -1\right) }{4 \left( 2 - \adiabexp \right)} \disspar^2 + O \left( \disspar^4 \right)$  \\[\hinter]
 % =============================
$\displaystyle \redpresplus =   \frac{\presstidevc}{\rcp \disspar} \left[ 1 + \frac{1}{4} \left( 1 + \rcp \right)^2 \disspar^2  \right] + O \left( \disspar^2 \right)$  & $\displaystyle \redpresplus = \frac{\presstidevc \left( 2 - \adiabexp \right) }{\left( \adiabexp - 1 \right) \disspar}  + O \left( \disspar \right) $  \\[\hinter]
 % =============================
$\displaystyle \XIrespm = 1 + \frac{\rcp \disspar}{2} \left( \mp 1 + \frac{3}{4} \rcp \disspar \right) + O \left( \disspar^3 \right) $\mynom[S]{$\XIrespm$}{frequencies at which the minimum ($-$) and maximum ($+$) of $\Im \left\{ \presstidevsurf \right\}$ associated with the Lamb-wave resonance are attained} & $\displaystyle \XIrespm = 1 + \frac{\left( \adiabexp - 1 \right) \disspar}{2}  \left[ \mp 1 - \frac{\left( \adiabexp-1 \right) \left( 6 + \adiabexp \right)}{4 \left( 2 - \adiabexp \right)} \disspar \right] + O \left( \disspar^3 \right) $ \\[\hinter]
 % =============================
 $\displaystyle \imdprespm =  \frac{\presstidevc}{2 \rcp \disspar} \left[ \pm 1 + \left( 1 + \rcp \right) \disspar \right] + O \left( \disspar \right) $ & $\displaystyle \imdprespm = \frac{\presstidevc \left( 2 - \adiabexp \right)}{2 \left( \adiabexp-1 \right) \disspar} \left[ \pm 1 + \frac{\left( 2 - \adiabexp^2 \right)}{\left( 2 - \adiabexp \right)^2 } \disspar  \right] + O \left( \disspar \right) $ \\[\hinter]
  % =============================
$\displaystyle \dximsep =  \rcp \disspar + O \left( \disspar^3 \right) $\mynom[S]{$\dximsep$}{frequency interval between the minimum and maximum of $\Im \left\{ \presstidevsurf \right\}$ associated with the Lamb-wave resonance} & $\displaystyle \dximsep =  \rcp \adiabexp \disspar + O \left( \disspar^3 \right)$ \\[\hinter]
  % =============================
$ \displaystyle \left.\DD{}{\XX} \left( \Im \left\{ \presstidevsurf \left( \XX \right) \right\} \right) \right|_{\XX = 1} \sim - \frac{2 \presstidevc}{\rcp^2 \disspar^2} $ & $\displaystyle \left.\DD{}{\XX} \left( \Im \left\{ \presstidevsurf \left( \XX \right) \right\} \right) \right|_{\XX = 1} \sim  - \frac{2 \left( 2 - \adiabexp \right) \presstidevc}{\left( \adiabexp -1 \right)^2 \disspar^2}  $ \\[\hinter]
% =============================
 \hline \\[-0.3cm] 
\multicolumn{2}{c}{\textit{Asymptotic behaviour in the high-frequency range ($\XX \rightarrow +\infty$)}} \\[0.1cm]
 $ \abs{\presstidevsurf  \left( \XX \right)} \sim  \presstidevc \XX^{-3}$ & $\displaystyle \abs{\presstidevsurf  \left( \XX \right)} \sim \frac{ \presstidevc}{\adiabexp} \XX^{-3}$ \\[\hinter]
 % =============================
$ \displaystyle \Re \left\{ \presstidevsurf \left( \XX \right) \right\} \sim \presstidevc \left(1 + \rcp \right) \disspar \XX^{-4}$ & $\displaystyle \Re \left\{ \presstidevsurf \left( \XX \right) \right\} \sim \frac{\presstidevc \disspar}{\adiabexp} \XX^{-4} $ \\[\hinter]
 % =============================
$\displaystyle \Im \left\{ \presstidevsurf  \left( \XX \right) \right\} \sim - \presstidevc \XX^{-3}$ & $\displaystyle \Im \left\{ \presstidevsurf  \left( \XX \right) \right\} \sim - \frac{ \presstidevc}{\adiabexp}  \XX^{-3}$ \\[0.3cm]
\hline
 \end{tabular}
  \tablefoot{Comparison of the analytical expressions of the quantity in the isentropic (\eq{presstidev_surf_neu}, left column), and isothermal (\eq{presstidev_surf_isotherm_dirac}, right column) atmospheric solutions. The subscripts $\imod$, $\irp$, and $\iip$ refer to the modulus, real part, and imaginary part of the solutions for the surface pressure oscillation, respectively. The plus ($\iplus$) and minus ($\iminus$) symbols in subscripts refer to maxima and minima, respectively. Formulae are obtained from \eq{presstidev_surf_neu} (isentropic case) and \eq{presstidev_surf_isotherm_dirac} (isothermal case) using the method described in \sect{ssec:quantitative_analysis}.}
 \end{table*}

\subsection{Quantitative functional analysis}
\label{ssec:quantitative_analysis}

We end this theoretical section by carrying out a quantitative functional analysis of the two solutions given by \eqs{presstidev_surf_neu}{presstidev_surf_isotherm_dirac} for resonant tidal modes ($\Sn = +1$) in the regime of quasi-adiabatic compression ($\disspar \ll1$). For convenience, we assume that tidal heating arises from direct absorption ($\phiheating=0$), which amounts to setting $\mathcal{G}=1$ in Appendix~H of \cite{Farhat2024}. In this framework, we progressively explore the transition from low to high tidal frequencies. The obtained results are summarised in Table~\ref{tab:solution_properties}. 

To find the extrema of the functions, we expand the frequencies at which they occur, $\XX_\pm$, as power series in $\disspar$,
\begin{equation}
\label{Xpm_series}
\XX_{\pm} = \XX_0 +  \sum_{\jj=1}^N a_\jj \disspar^\jj + o \left( \disspar^{N} \right),
\end{equation}
where $N$ is the truncation order of the expansion\mynom[S]{$N$}{truncation order of the power series in \eq{Xpm_series}}, $\XX_0$ is the position of the extremum in the adiabatic limit ($\disspar \to 0$), and $a_\jj$\mynom[S]{$a_\jj$}{$\jj$-th coefficient of the power series in \eq{Xpm_series}} is a real coefficient depending solely on $\adiabexp$. The coefficients $a_\jj$ are obtained analytically by substituting \eq{Xpm_series} into the derivatives of the functions and identifying the zeros of the resulting equations. Computations were performed both with \texttt{Mathematica} \citep[][]{Mathematica} and with the \texttt{TRIP} dedicated computer algebra system \citep[][]{GL2011}.

We first investigate the asymptotic behaviour of $\presstidevsurf$ near synchronisation ($\XX \ll\disspar$, see Table~\ref{tab:solution_properties}, top section). Consistent with \fig{fig:spectra_deltap}, there is no discrepancy between the two models in the zero-frequency limit. Besides, the tendencies at $\XX = 0$ only depends on $\presstidevc$ and $\disspar$. In the low-frequency range, the local maximum of $\Im \left\{  \presstidevsurf \right\}$ is asymptotic to $ \presstidevc/ \left( 2 \disspar \right)$ as $\disspar \rightarrow 0$ for both isentropic and isothermal atmospheres. Similarly, the frequency at which this maximum is reached, $\Xmaxsyn$, is asymptotic to $\disspar$ in the two models. Then we move to the high-frequency range and examine the local maxima and minima associated with the Lamb wave resonance near $\XX=1$. Peak positions in frequency spectra are not affected much by dissipation, while peak heights are inversely proportional to $\disspar$, which is a classical outcome of linear tidal theory \citep[e.g.][]{OL2004,ADMLP2015}. 

Interestingly, the difference in peak heights that can be observed in \fig{fig:spectra_deltap} between isentropic and isothermal atmospheres is quantified by an explicit function of $\adiabexp$ in Table~\ref{tab:solution_properties}. The ratio of amplitudes between the resonant peak and the maximum reached at low frequencies, $\ratioamp = \imdpresplus / \imdpmaxsyn$\mynom[S]{$\ratioamp$}{ratio of amplitudes between the maximum of $\Im \left\{  \presstidevsurf \right\}$ associated with the Lamb-wave resonance and that reached in the low-frequency range}, is asymptotic to $\rcp^{-1}$ as $\disspar \rightarrow 0$ for isentropic atmospheres and to $\left( 2 - \adiabexp \right) / \left( \adiabexp-1 \right)$ for isothermal atmospheres. With $\adiabexp = 7/5$, this yields $\ratioamp \approx 3.5$ in the former case, and $\ratioamp \approx 1.5$ in the latter, consistent with the spectra of \fig{fig:spectra_deltap}. Also, the full width at half maximum (FWHM)\mynom[A]{FWHM}{full width at half maximum} of the modulus' peak, $\dxmidres$, and the frequency separation between two peaks of the imaginary part, $\dximsep$, are simple functions of $\adiabexp$ and $\disspar$ in both cases. Finally, calculations highlight the rapid decay of the atmospheric response beyond the resonance frequency. Both the modulus and imaginary part of $\presstidevsurf$ are asymptotic to $\XX^{-3}$ as $\XX \rightarrow + \infty$.

\section{Benchmark against GCM simulations}
\label{sec:benchmark_gcm}

The quantitative analysis performed in \sect{ssec:quantitative_analysis} yields explicit relationships between spectral features of thermotidal surface pressure oscillations in the Fourier domain and physical parameters (see Table~\ref{tab:solution_properties}). These relationships can be used to benchmark the analytical solutions of the present work against numerical solutions obtained from GCM simulations in former studies. 

To our knowledge, such studies remain relatively scarce owing to their substantial computational cost. Evaluating the semidiurnal tidal torque at a given tidal frequency requires running a GCM simulation until a steady state is reached. The integration must then be extended over a large number of tidal cycles, after which the tidal fields are phase-folded over a stellar day in order to filter out meteorological variability. The resulting surface-pressure distribution is subsequently decomposed into spherical harmonics, and the torque is computed from the degree-2, order-2 component using \eq{torquez_expr}\footnote{If spherical harmonics of both positive and negative orders are employed, the expression given by \eq{torquez_expr} must be multiplied by a factor of 2, following \cite{Leconte2015} and \cite{ADL2019}.}. This procedure must be repeated for each sampled tidal frequency, thereby requiring a large ensemble of simulations.

\cite{Leconte2015} brought this approach to prominence by applying it to the case of Venus' asynchronous rotation. Since then, it has been employed to investigate the equilibrium rotation states of Venus-like rocky exoplanets \citep[e.g.][]{ADL2019,SW2024}, as well as the past evolution of Earth's rotation \citep[e.g.][]{Wu2023,DG2024}. In this section, we place our analytical model in the context of these numerical studies, showing that the free parameters of the former can be constrained by comparison with the latter. The resulting estimates are summarised in Table~\ref{tab:parameter_constraints}.

\begin{table*}[h]
\centering
\caption{\label{tab:parameter_constraints} Estimates of analytical solutions' parameters deduced from GCM simulations in various cases.}
\begin{small}
\begin{tabular}{l l l c c c c} 
 \hline 
 \hline 
 & Notation & Units & \multicolumn{4}{c}{Values} \\
  \hline \\[-0.3cm] 
% Reference
Case & & 
 & (a) 
 & (b) 
 & (c)
 & (d)
 \\
% Case
Planet & & 
 & Venus  
 & Past Earth 
 & Actual Earth
 & Dry planet
 \\
Frequencies & & 
 & low
 & high 
 & high 
 & all
 \\
 \hline \\[-0.3cm] 
% Gas
 \multirow{11}{*}{Parameters} & \textsc{Gas}  & --
 & ${\sim} \carbondiox$ 
 & ${\sim} \nitrogen$ 
 & ${\sim} \nitrogen$
 & $ \nitrogen$
 \\
% Adiabatic exponent
 & $\adiabexp$ & --
 & 1.20
 & 1.23
 & 1.23 
 & 1.4
 \\
% Specific gas constant
 & $\Rspec$ & ${\rm J \ kg^{-1} \ K^{-1}}$ 
 & 191
 & 287
 & 287 
 & 297
 \\
 % Surface pressure 
 & $\psurf$ & ${\rm bar}$ 
 & 92
 & 1
 & 1
 & 10
 \\
% Planet radius
 & $\Rpla$ & km 
 & 6051.8 
 & 6371.0
 & 6371.0
 & 6051.8
 \\
% Gravity
 & $\ggravi $ & ${\rm m \ s^{-2}}$ 
 & 8.87 
 & 9.81
 & 9.81
 & 8.87
 \\
% Incident stellar flux
 & $\Fstar$ & ${\rm W \ m^{-2}}$ 
 & 2610 
 & 1225 
 & 1361
 & 2610
 \\
% Sidereal orbital period 
& $\time_{\rm orb}$ & days
& 224.701
& 365.256
& 365.256
& 224.701
\\
% Planet mass
& $\Mpla$ & ${10^{24}}$ kg
& $\num{4.8673}$
& $\num{5.9722}$
& $\num{5.9722}$
& $\num{4.8673}$
\\
% Mass of the star
& $\Mstar$ & ${10^{30}}$ kg
& $\num{1.9884}$
& $\num{1.9884}$
& $\num{1.9884}$
& $\num{1.9884}$
\\
% Mass of the star
& $\smaxis$ & au
& 0.72333
& 1.00000
& 1.00000
& 0.72333
\\
\hline \\[-0.3cm] 
 % ------------------------------------------------
 % NEUTRALLY STRATIFIED
 \multirow{17}{*}{Isentropic} 
% Pressure q0 at maximum near synchronisation
& $q_0$ & Pa
& 201 \tablefootmark{$\gcmdata$}
& 29.3  %30.9
& 28.1 \tablefootmark{$\wuscale$}
& 1790
\\ 
 % Frequency Omega-n at maximum near synchronisation
& $\omega_0$ & ${\rm rad \ s^{-1}}$
& $\num{3.77e-7}$ \tablefootmark{$\gcmdata$}
& $\num{2.04e-5}$ %$\num{1.92e-5}$
& $\num{4.03e-5}$
& $\num{1.58e-6}$
\\ 
 % Rotation period at which maximum occurs near synchronisation
& $\LOD_0$ & days 
& 193 \tablefootmark{$\gcmdata$}
& 3.57 %3.78
& 1.80
& 46.0
\\
 & $\fresoi{2}$ & ${\rm rad \ s^{-1}}$ 
 & $\num{2.07e-4}$ 
 & $\num{1.59e-4}$ %$\num{1.58e-4}$ \tablefootmark{$\gcmdata$}
 & $\num{1.59e-4}$
 & $\num{1.69e-4}$
 \\ 
 % Daylength at which resonance occurs
 & $\LODresoi{2}$ & hr  
 & 16.9 
 &  22.0 %22.1 \tablefootmark{$\gcmdata$}
 &  22.0
 &  20.7
 \\
  % Daylength of the maximum of positive torque
 & $\LODimplus$ & hr  
 & 16.9
 & 22.9 %22.6
 & 23.9
 & --
 \\
 % Normalised dissipation parameter
 & $\disspar$&  --
 & $\num{3.64e-3}$ 
 &  0.257 %0.243
 &  0.508
 & $\num{1.87e-2}$
 \\
 % Newtonian cooling frequency
 & $\fnewton$&  ${\rm rad \ s^{-1}}$
 & $\num{7.54e-7}$
 & $\num{4.07e-5}$ %$\num{3.84e-5}$ 
 & $\num{8.06e-5}$
 & $\num{3.16e-6}$
 \\
% Normalised pressure constant
 & $\presstidevc$ &  --
 & $\num{6.55e-8}$ %$6.99 \times 10^{-8}$ 
 & $\num{1.50e-4}$
 & $\num{2.85e-4}$ \tablefootmark{$\wuscale$}
 & $\num{6.70e-5}$
 \\
% Temperature at surface deduced from the Lamb resonance frequency
 & $\Tsurf$ &  K
 & 737
 & 319 %317
 & 320
 & 316
 \\
% Bulk temperature deduced from the Lamb resonance frequency
 & $\Tbulk$ &  K
 & 632 
 & 269 %267 
 & 270
 & 246
 \\
% Pressure height at surface deduced from the Lamb resonance frequency
 & $\heightsurf$ &  km
 & 15.9
 & 9.34 %9.26 
 & 9.36
 & 10.6
 \\
% Opacity deduced from the Lamb resonance frequency
 & $\opacity$ & $\%$
 & 2.01
 & 42.9 
 & 73.5
 & 70.9
 \\
% Gain coefficient of the equivalent 2nd-order bandpass filter
 & $\bpgain$ & --
 & $\num{1.08e-4}$
 & $\num{3.13e-3}$ %$\num{3.31e-3}$
 & $\num{3.00e-3}$ \tablefootmark{$\wuscale$}
 & $\num{1.25e-2}$
 \\
 % Quality of the equivalent 2nd-order bandpass filter
 & $\bpquality$ & --
 & $\num{1650}$
 & 20.8 %22.0
 & 10.5
 & 187
 \\
% Lag angle of the equivalent 2nd-order bandpass filter
 & $\philag$ & rad
 & $\num{-4.25e-3}$
 & 0.228 %0.224
 & 0.345
 & 1.49
 \\
% Phase angle of tidal heating
 & $\phiheating$ & rad
 & 0
 & 0.533 %0.513
 & 0.948
 & 1.51
 \\
 \hline \\[-0.3cm] 
% ------------------------------------------------
% ISOTHERMAL
\multirow{17}{*}{Isothermal} 
% Pressure q0 at maximum near synchronisation
& $q_0$ & Pa
& 201 \tablefootmark{$\gcmdata$}
& 46.7 %49.4
& 46.4 \tablefootmark{$\wuscale$}
& 1755
\\ 
% Frequency Omega-n at maximum near synchronisation
& $\omega_0$ & ${\rm rad \ s^{-1}}$
& $\num{3.77e-7}$ \tablefootmark{$\gcmdata$}
& $\num{1.62e-5}$ %$\num{1.56e-5}$
& $\num{2.83e-5}$
& $\num{1.67e-6}$
\\ 
 % Rotation period at which maximum occurs near synchronisation
& $\LOD_0$ & days 
& 193 \tablefootmark{$\gcmdata$}
& 4.50 %4.66
& 2.57
& 43.5
\\
% Lamb resonance frequency from given surface temperature
& $\fresoi{2}$ & ${\rm rad \ s^{-1}}$ 
& $\num{2.10e-4}$ 
& $\num{1.58e-4}$ %\tablefootmark{$\gcmdata$}
& $\num{1.58e-4}$
& $\num{1.68e-4}$
\\
% Daylength at which resonance occurs
& $\LODresoi{2}$  & hr  
& 16.6
& 22.0 %22.1  \tablefootmark{$\gcmdata$}
& 22.1
& 20.7
\\
  % Daylength of the maximum of positive torque
 & $\LODimplus$ & hr  
 & 16.6
 & 22.8 %22.6
 & 23.9
 & 21.3
 \\
% Normalised dissipation parameter
& $\disspar$& --
& $\num{3.59e-3}$ 
& $\num{0.204}$ % $\num{0.198}$ 
& $\num{0.359}$
& $\num{1.99e-2}$
\\
 % Newtonian cooling frequency
 & $\fnewton$&  ${\rm rad \ s^{-1}}$
 & $\num{7.54e-7}$
 & $\num{3.23e-5}$ %$\num{3.12e-5}$ 
 & $\num{5.66e-5}$
 & $\num{3.35e-6}$
 \\
% Normalised pressure constant
& $\presstidevc$ &  --
& $\num{6.46e-8}$ 
& $\num{1.91e-4}$ %$\num{1.95e-4}$
& $\num{3.33e-4}$ \tablefootmark{$\wuscale$}
& $\num{6.98e-5}$
\\
% Temperature at surface deduced from the Lamb resonance frequency
 & $\Tsurf$ &  K
 & 632
 & 259 %257
 & 257
 & 224
 \\
% Bulk temperature deduced from the Lamb resonance frequency
 & $\Tbulk$ &  K
 & 632 
 & 259 %257
 & 257
 & 224
 \\
% Pressure height at surface deduced from the Lamb resonance frequency
 & $\heightsurf$ &  km
 & 13.6
 & 7.58 %7.53
 & 7.52
 & 7.51
 \\
 % Opacity deduced from the Lamb resonance frequency
 & $\opacity$ & $\%$
 & 1.73
 & 44.1 
 & 68.5 \tablefootmark{$\wuscale$}
 & 52.2
 \\
% Gain coefficient of the equivalent 2nd-order bandpass filter
 & $\bpgain$ & --
 & $\num{7.19e-5}$
 & $\num{3.13e-3}$ %$\num{3.31e-3}$
 & $\num{3.11e-3}$ \tablefootmark{$\wuscale$}
 & $\num{5.26e-3}$
 \\
 % Quality of the equivalent 2nd-order bandpass filter
 & $\bpquality$ & --
 & $\num{1390}$
 & 21.3 %22.0
 & 12.1
 & 126
 \\
% Lag angle of the equivalent 2nd-order bandpass filter
 & $\philag$ & rad
 & $\num{-3.14e-3}$
 & 0.231 %0.224
 & 0.396
 & 1.29
 \\
% Phase angle of tidal heating
 & $\phiheating$ & rad
 & 0
 & 0.399 %0.387
 & 0.691
 & 1.29
 \\
\hline
 \end{tabular}
 \end{small}
 \tablefoot{{\it From top to bottom:} the fixed parameters and the corresponding estimates of free parameters in analytical solutions for the two studied atmospheric structures (isentropic and isothermal atmospheres). Calculations are detailed in \sect{sec:benchmark_gcm}. 
\tablebib{ (a) \citet{Leconte2015}, Table~1, Venus; (b) \citet{DG2024}, Fig.~1c, ExoCAM, 30 ppm ${\rm CH_4}$; (c) \citet{Wu2023}, Supplementary Materials, Fig.~S4; (d) \citet{ADL2019}, Table~1 and Fig.~3.}
\tablefoottext{$\gcmdata$}{Quantities constrained by GCM data in case (a), from which the remaining quantities are inferred.} \\
\tablefoottext{$\wuscale$}{Quantities affected by the torque renormalisation applied by \cite{Wu2023} in their Fig.~S4, where the torque scale is assumed rather than derived from the GCM simulations (see Footnote~\ref{footnote:wu_rescaling} in \sect{ssec:Wu2023}).}
  } 
 \end{table*}

\subsection{\cite{Leconte2015}}
\label{ssec:leconte2015}
To start with, we consider the Venus case in \cite{Leconte2015}. In this study, GCM simulations were performed in the low-frequency range. As a consequence, they do not explore the resonant excitation of the thermally forced Lamb wave that occurs at higher tidal frequencies. Nevertheless, the resonance frequency can be predicted to a certain extent from the planet's thermal state, which is known. As the deep layers of Venus' atmosphere are neutrally stratified due to strong convection \citep[e.g.][]{Seiff1985}, the isentropic profile represents well Venus' atmospheric structure (see \paperone, Fig.~2). In this configuration, we set the surface temperature to the planet's average surface temperature, namely $\Tsurf = 737$~K \citep[e.g.][]{Seiff1985}. We deduce the bulk temperature from $\Tsurf$ using \eq{Tbulk}, which yields $\Tbulk= 632~K$. This value of $\Tbulk$ is also adopted for the isothermal configuration as we consider model atmospheres with the same heat content. Pressure scale heights at the planet's surface ($\heightsurf$) are then evaluated from $\Tsurf$ using \eq{heightsurf}.

Recalling that $\Houghval{2}{2}{1} = 11.12895 $ (\paperone, Table~B.1), \eq{freson_lamb_2} yields the resonance frequency ($\fresoi{2}$) for the isentropic and isothermal atmospheres. We also obtain the associated length of the day, $\LODresoi{2}$, given by \eq{LOD2_freso2}, which takes the values 16.6~hr (isothermal) and 16.9~hr (isentropic). In the low-frequency range, \cite{Leconte2015} fit to GCM simulations a model similar to that given by \eq{model_leconte2015}. This model is parametrised by the diurnal tidal frequency at which the maximum near synchronisation occurs, $\omega_0$\mynom[S]{$\omega_0$}{diurnal tidal frequency at which the maximum of the torque near synchronisation occurs}, and a factor dimensionally consistent with a pressure, $q_0$\mynom[S]{$q_0$}{constant pressure factor in \cite{Leconte2015}'s model}, which take the values $\omega_0 = \num{3.77e-7} \ {\rm rad \  s^{-1}}$ and $q_0=201 \ {\rm Pa}$ for Venus \citep[see][Table~1]{Leconte2015}. By comparing the formulation of the torque given by Eqs~(1-2) of \cite{Leconte2015} with ours (\eqs{torquez_expr}{model_leconte2015}), we observe that the maximum is reached for $\ftide= \fnewton$ in our model and we obtain the correspondence relations linking $\omega_0$ and $q_0$ to the parameters of this work,
\begin{align}
\label{Imdeltaps_q0}
& \fnewton = 2 \omega_0, & \Im \left\{ \presstidenquad \right\}_{\ftide =\fnewton} = \frac{q_0}{\psurf}.
\end{align}
The LOD at the maximum is thus given by\mynom[S]{$\LOD_0$}{LOD maximising the torque near synchronisation}
\begin{equation}
\label{LODzero}
\LOD_0 = \frac{4 \pi}{\fnewton},
\end{equation}
yielding $\LOD_0 = 193$~days, and the value of the dissipation parameter is deduced from those calculated for $\fnewton$ and $\fresoi{2}$ in both configurations. We get $\disspar=\num{3.64e-3} $ for the isentropic profile and $\disspar = \num{3.59e-3} $ for the isothermal one.

Following \cite{Leconte2015}, we note that $\omega_0$ is inversely proportional to the radiative cooling time scale of the atmosphere. Therefore, this parameter should scale as 
\begin{equation}
\label{scaling_omega0}
\omega_0 \scale \frac{\ggravi \Temis^3}{\Cp \psurf},
\end{equation} 
where $\Cp$ designates the atmospheric heat capacity per unit mass and $\Temis$ the emission temperature of the atmosphere, namely the temperature of the deepest layers radiating energy towards space. This scaling law highlights the strong dependence of the radiative timescale on temperature and pressure, implying a corresponding sensitivity of the damping ratio, $\disspar$, and, by extension, of the amplitude of the Lamb-wave resonance. However, it should be borne in mind that the dependence of $\omega_0$ on atmospheric parameters is in reality more complex than implied by \eq{scaling_omega0}. In particular, $\omega_0$ is governed by the atmospheric physics and composition, neither of which is explicitly accounted for in \eq{scaling_omega0}. Besides, this effective parameter encapsulates the coupled contributions of all dissipative mechanisms, as discussed in \sect{ssec:thermotidal_pressure}, and not only radiative cooling.

To calculate $\presstidevc$, we need to consider the relationship between spherical harmonics and Hough functions. In the zero-frequency limit, the phase of the forcing in \eq{phase_forcing_normalised} becomes zero. Consequently, we assume that the direct-absorption approximation holds in the considered frequency range and set $\phiheating=0$. For this configuration, the coefficient $\presstidevsurfi{2}{2}$ given by \eq{deltaps22_deltaps2h} is expanded in Table~\ref{tab:solution_properties} as a power series of $\disspar$ near the maximum of $\Im \left\{  \presstidevsurfi{2}{2} \right\}$. Retaining only the dominant term of this series, we get
\begin{equation}
\Im \left\{  \presstidevsurfi{2}{2} \left( \Xmaxsyn \right) \right\} \approx \frac{\presstidevc}{2 \disspar} ,
\end{equation}
which, combined with \eq{Imdeltaps_q0}, yields
\begin{equation}
\presstidevc = \sqrt{\frac{2}{\pi}} \frac{\disspar q_0}{\psurf \scal{\LegFnorm{2}{2}}{\HoughF{2}{2}{\spinpar_2}}}.
\end{equation}
It should be noted that retaining only the dominant term of the series expansion introduce a relative uncertainty of order ${\sim} \disspar^2$ in the inferred value of $\presstidevc$. While this approximation error is negligible in the present case ($\disspar^2 \approx \num{e-5}$), it may no longer be insignificant when $\disspar$ approaches unity, as is the case in Sections~\ref{ssec:DG2024} and~\ref{ssec:Wu2023}.

The gain and quality factor characterising the equivalent bandpass filter introduced in \sect{ssec:bandpass_filter} are computed from $\adiabexp$, $\presstidevc$, and $\disspar$ using \eq{bpfilter_gain_quality}. The resulting quality factors are found to be large, indicating a pronounced Lamb-wave resonance. Although this outcome is consistent with the long radiative cooling timescale of Venus' atmosphere, the corresponding values are derived within the framework of linear theory. Consequently, the resonance strength may be substantially reduced in reality by nonlinear dissipative processes. 

We deduce the absorbed flux fraction from $\presstidevc$ using \eq{presstidevc2}, 
\begin{equation}
\label{opacity_deltap}
\opacity = \frac{\psurf \heightsurf \fresoi{2} \presstidevc}{\rcp  \fluxtideni{2} },
\end{equation}
where the resonance frequency, $\fresoi{2}$, and the expansion coefficient $\fluxtideni{2}$ are given by \eqs{freson_lamb_2}{deltaF2_Fstar}, respectively. Interestingly, the values obtained for the absorbed flux fraction, $\opacity \approx 2 \%$, are close to estimates of the average flux impinging Venus' surface, which represents nearly $2.5\%$ of the total incident Solar flux \citep[e.g.][]{Tomasko1980a}. This is consistent with findings of former theoretical works showing that the effect of atmospheric tides on Venus' spin evolution mainly originates in semidiurnal oscillations due to ground heating \citep[][]{DI1980,SZ1990}.

As a final step, we compute the LOD maximising the tidal torque in the high-frequency range\mynom[S]{$\LODimplus$}{LOD maximising the tidal torque in the high-frequency range}, 
\begin{equation}
\LODimplus = \frac{\LODresoi{2}}{\XIresplus},
\end{equation}
which is slightly greater than the LOD at which the resonance of the tidally forced Lamb wave occurs, given by \eq{LOD2_freso2}. The normalised frequency of the positive peak of the imaginary part, $\XIresplus$, is provided by Table~\ref{tab:solution_properties} as a function of $\adiabexp$ and $\disspar$. For simplicity, we truncate the power series of $\disspar$ to first order, which yields
\begin{align}
\label{XIresplus_leconte}
& \XIresplus \approx 1 - \frac{\rcp \disspar}{2}  && \mbox{(isentropic),} \\
& \XIresplus \approx 1 - \frac{\rcp \adiabexp \disspar}{2} && \mbox{(isothermal).}
\end{align}
It is noteworthy that $\LODimplus$ does not differ much from $\LODresoi{2}$ in this case since $\disspar$ is very small.

In addition to Venus, \cite{Leconte2015} report values of $q_0$ and $\omega_0$ for a range of idealised configurations intended to represent the inner and outer boundaries of the habitable zone (see Table~1 of their study). These scenarios encompass Venus-like planets with either $\nitrogen$- or $\carbondiox$-dominated atmospheres, surface pressures of 1~bar or 10~bar, and incident stellar fluxes of $450~{\rm W \ m^{-2}}$ or $1366~{\rm W \ m^{-2}}$. While these configurations constitute compelling cases for further investigation, the corresponding surface temperatures are not provided. As a result, the resonance frequency of the tidally forced Lamb wave cannot be constrained, preventing the inference of the model parameters except in the case of Venus, where the temperature profile is well established.

\begin{figure*}
\centering
\includegraphics[width=0.48\textwidth,trim=0cm 0cm 0.5cm 0cm,clip]{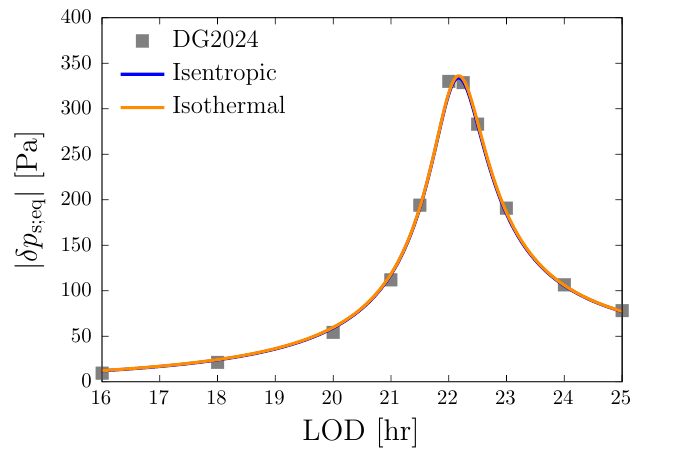}
\includegraphics[width=0.48\textwidth,trim=0cm 0cm 0.5cm 0cm,clip]{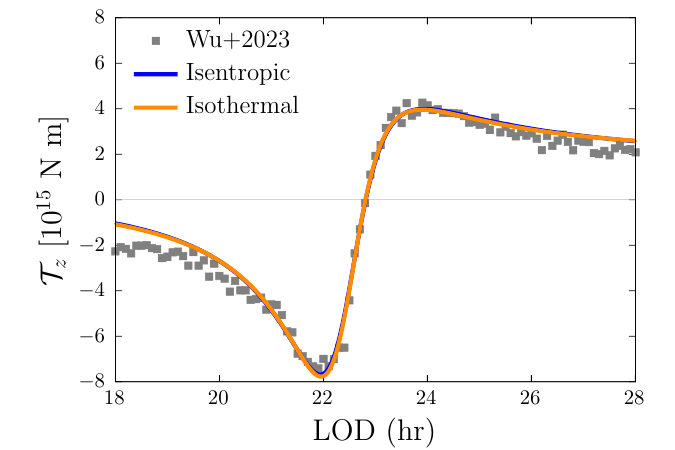}\\
\includegraphics[width=0.48\textwidth,trim=0cm 0cm 0.5cm 0cm,clip]{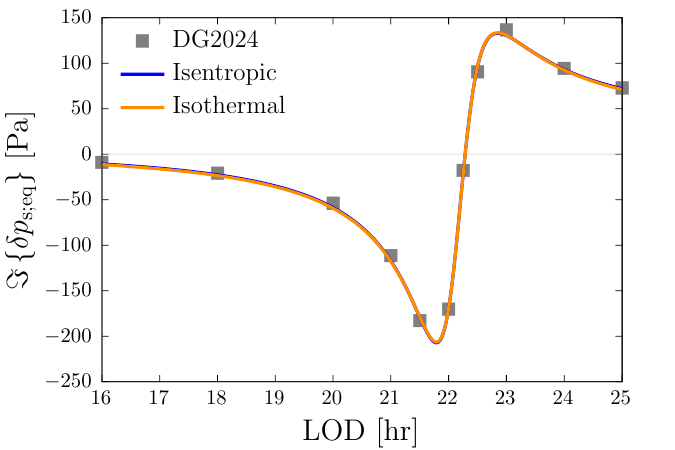}
\includegraphics[width=0.48\textwidth,trim=0cm 0cm 0.4cm 0cm,clip]{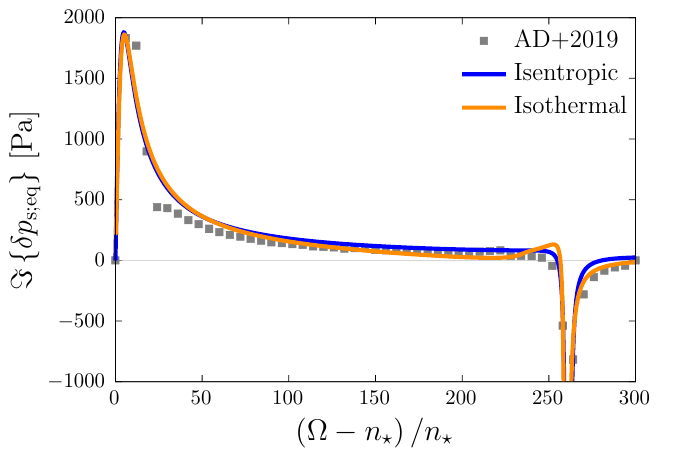} 
\caption{Frequency dependence of the semidiurnal pressure bulge at the equator. {\it Left:} Amplitude (top) and imaginary part (bottom) of surface pressure oscillations as a function of the LOD ($2 \pi / \abs{\spinrate - \norb}$) in the case labelled `ExoCAM, 30 ppm ${\rm CH_4}$' of \citet[][Figs.~1c and~3]{DG2024}. {\it Right:} Thermotidal torque as a function of the LOD in the Earth's case treated by \cite{Wu2023} in their Fig.~S4 (top), and imaginary part of surface pressure oscillations as a function of the normalised frequency $\fdiurnalnorm = \left( \spinrate - \norb \right) / \norb$ in the reference case of \citet[][bottom]{ADL2019}. In the three cases, plots show the numerical solutions obtained using GCM simulations (gray markers), and the closed-form solutions given by our analytical model for isentropic (solid blue line) and isothermal (solid orange line) atmospheres. The two analytical solutions, given by \eqs{presstidev_surf_neu}{presstidev_surf_isotherm_dirac}, are fitted to GCM data points using the parameter values of Table~\ref{tab:parameter_constraints}, as detailed in Sects~\ref{ssec:DG2024}, \ref{ssec:Wu2023} and~\ref{ssec:ADL2019}.}
    \label{fig:comparison_DG2024}
\end{figure*}

\subsection{\cite{DG2024}}
\label{ssec:DG2024}

The influence of atmospheric tides on the evolution of planetary spin has also been investigated in the context of Earth's geological history. Recent studies by \citet{Wu2023} and \citet{DG2024} address this problem using GCM simulations. Although the latter was published more recently, we begin by discussing it first, as it provides a more detailed analysis of the numerical results. \cite{DG2024} explore the role of tidally induced Lamb-wave resonances in shaping Earth's climate and surface conditions during the Proterozoic Eon ($539{-}2500$~Ma). Their calculations were performed with a range of GCMs and atmospheric compositions, with particular emphasis on the effects of varying the methane volume mixing ratio in the simulated palaeo-atmospheres. We focus here on the modulus and imaginary part of the semidiurnal surface-pressure Fourier coefficient at the equator reported by \cite{DG2024} in their 30~ppm $\methane$ configuration (Fig.~3 of the article, bottom-left panel, and Fig.1c, respectively), for which R.~Deitrick kindly provided the data tables (Deitrick 2025, priv. comm.)\footnote{The maximal amplitude of the semidiurnal component ($\mm=2$) should not be confused with the maximum of the total pressure anomaly, for which \cite{DG2024} also report estimates inferred from GCM simulations. The former peaks at approximately 330~Pa, while the latter reaches approximately 440~Pa.}. These GCM results are reproduced in \fig{fig:comparison_DG2024} (left panels; grey markers).

As in the Venus case treated by \citet{Leconte2015}, the simulations of \citet{DG2024} do not cover the entire frequency range, but only a narrow interval around the resonance. Consequently, they do not constrain the near-synchronous maximum. Nevertheless, we can still infer all the model parameters, this time from the features of the resonant peaks characterising the imaginary part of the surface-pressure Fourier coefficient. This analysis may be carried out using the relations established in \sect{sec:frequency_behaviour}, which connect peak extrema and widths to model parameters. In \append{app:scalings_DG2024}, we show that good fits of \cite{DG2024}'s GCM datasets are obtained by retaining only the leading-order terms of the power-series expansions in $\disspar$ reported in Table~\ref{tab:solution_properties}. However, for a quantitative purpose, we shall adopt an alternative approach here, as well as in \sects{ssec:Wu2023}{ssec:ADL2019}.

 In this approach, the model parameters are inferred by fitting the imaginary part of the isentropic and isothermal solutions given by \eqs{presstidev_surf_neu}{presstidev_surf_isotherm_dirac} to data points numerically with a minimisation method. A delayed tidal heating of the form given by \eq{phase_forcing_normalised} is included in both fitted models. The four adjusted free parameters are the resonant length of the day, $\LODresoi{2}$, the normalised pressure coefficient, $\presstidevc$, the damping ratio, $\disspar$, and the phase angle of tidal heating at resonance, $\phiheating$. We emphasise that \cite{DG2024} report values for the Fourier component of equatorial surface-pressure oscillations ($\presstideneq$) while our closed-form solutions characterise the Hough modes forming the global surface-pressure anomaly, especially the degree-2 component ($\presstidevsurfi{2}{2}$), which is predominant in semidiurnal tides. To relate the two, one uses \eq{presstideneq_simp}, yielding
 \begin{equation}
 \label{deltaps22_deltapseq}
\presstidevsurfi{2}{2} \approx \sqrt{\frac{16}{15}}  \frac{\cshape  \, \presstideneq}{\scal{\HoughF{2}{2}{1}}{ \LegFnorm{2}{2}} \psurf}.
\end{equation}
 Calculations were conducted using the Levenberg-Marquardt algorithm, as implemented in the \texttt{LsqFit.jl} library for the \texttt{Julia} programming language \citep[][]{WM2017LsqFit}\footnote{\url{https://github.com/JuliaNLSolvers/LsqFit.jl}}. 

For the isentropic model, the best-fitting parameter values with $2 \sigma$-errors are $\LODresoi{2} = 22.018 \pm 0.027 \ {\rm hr} $, $\presstidevc = \left( 1.503 \pm 0.074 \right) \times 10^{-4} $, $\disspar = 0.257 \pm 0.015$, and $\phiheating = 0.533 \pm 0.040 \ {\rm rad}$. The associated coefficient of determination is $R^2 = 99.9 \%$. For the isothermal model, the best-fitting parameters are $\LODresoi{2} = 22.049 \pm 0.028 \ {\rm hr}$, $\presstidevc = \left( 1.907 \pm 0.099 \right) \times 10^{-4}  $, $\disspar = 0.204 \pm 0.012 $, and $\phiheating= 0.399 \pm 0.042 \ {\rm rad}$ with $R^2 = 99.9 \%$. It is worth noting that the error margins provided above are just formal fit errors. Consequently, these errors do not account for modelling uncertainties or uncertainties associated with the frequency sampling of GCM data, which may be substantially larger, as discussed in \sect{ssec:ADL2019}. 

The two fitted models are shown by \fig{fig:comparison_DG2024} both for the modulus (top left panel) and the imaginary part (bottom left panel) of the equatorial surface-pressure component. Both match the GCM data with an excellent accuracy, demonstrating that our analytical solutions successfully reproduce the frequency dependence of semidiurnal tidal surface-pressure oscillations and the associated tidal torque for Earth. In particular, the curves agree well with the GCM dataset for the modulus, whilst only the imaginary part was used to constrain the parameters. The plots also illustrate the fact that the frequency dependence of the tidal bulge near the Lamb-wave resonance does not allow the isentropic and isothermal structures to be distinguished from each other, as both lead to the very same behaviour in the considered frequency range.

The resonance occurs for a daylength of 22.0~hr, which corresponds to $\fresoi{2} = \num{1.59e-4} \ {\rm rad \ s^{-1}}$. This frequency is lower than that evaluated for Venus mainly because of the temperature difference between the two planets and the scaling law $\fresoi{2} \scale \sqrt{\Tsurf} $ (see \eq{freson_lamb_2}). The values found for the damping ratio -- $\disspar=0.257$ (isentropic) and $\disspar=0.204$ (isothermal) -- are close to that inferred from barometric measurements for the present-day Earth ($\disspar=0.25$, see \fig{fig:spectra_earth}), suggesting dissipative mechanisms of similar efficiencies in the present-day atmosphere and the palaeoatmosphere modelled by \cite{DG2024} in their GCM simulations.

From the damping ratio, we infer the Newtonian cooling frequency, $\fnewton=\disspar \fresoi{2}$ (see \eq{reduced_frequency}), the diurnal tidal frequency at which the torque reaches its maximum near synchronisation, $\omega_0 = \fnewton/2$, and the associated daylength, given by \eq{LODzero}. The estimated values of $\fnewton$ are approximately two orders of magnitude greater than those found for the Venus' case treated in \sect{ssec:leconte2015}, indicating a dependence on surface pressure qualitatively consistent with the rough scaling relation given by \eq{scaling_omega0} for the radiative cooling timescale. Although this consistency might suggest that $\fnewton$ is predominantly controlled by radiative cooling, one shall bear in mind that it is an effective parameter depending on atmospheric pressure, temperature and composition in a complex way. Characterising the dependence of $\fnewton$ to the atmospheric physics thus requires further investigation, which lies beyond the scope of this study.

The pressure scale height at the planet's surface is then determined from \eq{freson_lamb_2},
\begin{equation}
\label{heightsurf_sigmares}
\heightsurf = \frac{\left( 1 - \gammapar \right) \left(\Rpla \fresoi{2} \right)^2}{\ggravi \abs{\Houghval{2}{2}{1}}},
\end{equation}
which in turn allows us to compute the atmospheric surface and bulk temperatures (see \eq{Tbulk}). The surface temperature obtained in GCM simulations \citep[$\Tsurf \approx 295$~K; see][Fig.~1a]{DG2024} lies between the two values predicted by the analytical models (259~K and 319~K), even though one would expect it to lie closer to the isentropic estimate given the excellent agreement between the isentropic model and present-day barometric measurements shown in \fig{fig:spectra_earth}. This discrepancy may arise either from differences in the vertical atmospheric structure or from qualitative differences in the thermotidal response between the present-day Earth and the palaeo-atmosphere modelled by \cite{DG2024}.

The absorbed flux fraction is finally inferred from $\presstidevc$ using \eqs{deltaF2_Fstar}{opacity_deltap}, following the same procedure as in \sect{ssec:leconte2015}. The resulting values, $\opacity = 42.9{-}44.1\%$, are roughly 2.4 times greater than the estimates derived from absorption profiles in the visible for the present-day Earth \citep[$\opacity \approx 17.6{-}18.1\%$; see][Appendix~J]{Farhat2024}. Although multiple factors could contribute to this discrepancy, the significant phase angle found for tidal heating suggests that the energy excess primarily stems from another heating source delayed with respect to the incident Solar flux, as discussed in \sect{sec:phase_shifted_heating}. Typically, the opacity parameter should not be interpreted as a physical absorbed flux fraction but rather as an effective parameter that compensates for the omission of boundary layer physics in the model. The estimated values for $\opacity$ may therefore account for the infrared radiative tidal heating of the planet's surface, which emits part of the incident flux reaching it upwards, as hypothesised by \citet{Farhat2024}. They may also reflect cumulonimbus activity and the associated latent heat transport \citep[e.g.][]{Lindzen1978}. The latter mechanism seems more likely in the studied case, as \cite{DG2024} show that clouds and precipitation are organised by the semidiurnal tide in their GCM simulations, thus suggesting a strong coupling between tidal dynamics and latent heat release.

\subsection{\cite{Wu2023}}
\label{ssec:Wu2023}

Like \citet{DG2024}, \citet{Wu2023} investigate the thermotidal response of Earth's palaeoatmosphere and its influence on the planet's rotation. In their study, GCM simulations serve mainly as a secondary tool for exploring the parameter space, as the authors primarily adopt a simplified ad hoc model that reproduces the frequency dependence of the tidal torque predicted by linear theory \citep[][Eqs.~(3–4)]{Wu2023}. The free parameters of this model are calibrated by fitting it to geological data. Nonetheless, the tidal torque is also quantified directly from GCM simulations, as presented in the Supplementary Materials of the article (Sect.~S2). In particular, we focus on the frequency spectrum of the tidal torque shown in their Fig.~S4, reconstructed from GCM data points near the Lamb-wave resonance. These data are plotted in the top right panel of \fig{fig:comparison_DG2024}.

It is worth noting that the numerical solution obtained by \cite{Wu2023} from GCM simulations behaves similarly to that shown in Fig.~1c of \cite{DG2024} in the vicinity of the resonance. In both cases, the torque exhibits a negative peak whose amplitude significantly exceeds that of the corresponding positive peak. As discussed in \sect{sec:phase_shifted_heating}, such behaviour is indicative of delayed tidal heating, which may arise either from latent heat transport \citep[][]{Lindzen1978,Hamilton1981} or from soil diffusion \citep[][]{Farhat2024}. In particular, \cite{Laskar2024} approximately reproduce, in their Fig.~7, the numerical solution displayed in Fig.~S4 of \cite{Wu2023} by including an indirect thermotidal forcing of the atmosphere through the planet's surface that incorporates radiative-conductive heat transfer.  

To fit our models to the GCM data point of \cite{Wu2023}, we must first establish the relationship between the torque exerted on the planet with respect to its spin axis, $\torquez$, and the degree-2 Hough coefficient of the normalised semidiurnal surface pressure component, $\presstidevsurfi{2}{2} $. This relationship is obtained by combining \eqs{torquez_expr}{deltaps22_deltaps2h} together, yielding
\begin{equation}
\torquez \approx  \pi \sqrt{\frac{12 }{5}} \scal{\LegFnorm{2}{2}}{\HoughF{2}{2}{\spinpar_2}} \frac{\Mstar}{\Mpla} \frac{\Rpla^6 \psurf}{\smaxis^3} \Im \left\{  \presstidevsurfi{2}{2}  \right\} .
\end{equation}

For the isentropic model, the best-fitting parameter values with $2 \sigma$-errors are $\LODresoi{2} = 21.995 \pm 0.089 \ {\rm hr} $, $\presstidevc = \left( 2.852 \pm 0.163 \right) \times 10^{-4} $, $\disspar = 0.508 \pm 0.054$, and $\phiheating = 0.948 \pm 0.066 \ {\rm rad}$. The associated coefficient of determination is $R^2 = 98.1 \%$. For the isothermal model, the best-fitting parameters are $\LODresoi{2} = 22.137 \pm 0.060 \ {\rm hr}$, $\presstidevc = \left( 3.333 \pm 0.135 \right) \times 10^{-4}  $, $\disspar = 0.359 \pm 0.026 $, and $\phiheating= 0.691 \pm 0.045 \ {\rm rad}$ with $R^2 = 98.2 \%$. The fit reveals a noticeable difference between the resonant LOD and the LOD of the change in the torque's sign, which are sometimes confused with each other. The resonance actually occurs near the minimum of the negative peak, at a 0.8~hr shorter LOD than the change in the torque's sign. This difference is not negligible, as it approximately corresponds to a 100~Myr gap in geological time for Earth \citep[e.g.][]{Daher2021,Farhat2022b}.

Using \eq{LOD2_freso2}, we deduce the resonance frequency from the resonant length of the day, $\LODresoi{2}$, which yields $\fresoi{2} = \num{1.59e-4}~{\rm rad \ s^{-1}} $ for the isentropic atmosphere and $\fresoi{2} = \num{1.58e-4}~{\rm rad \ s^{-1}}$ for the isothermal one. The associated values of $\fnewton$, $\omega_0$, $\heightsurf$, $\Tsurf$, and $\Tbulk$ are then obtained following the same steps as in \sect{ssec:DG2024}. They are close to those calculated for the GCM simulations of \cite{DG2024} in order of magnitude (see Table~\ref{tab:parameter_constraints}). However, we remark that the values of $\fnewton$ derived from \cite{Wu2023} for present-day Earth and from \cite{DG2024} for ancient Earth differ by a factor of ${\sim}2$, despite modelling similar conditions. This discrepancy indicates that radiative cooling is not the only dissipative process controlling $\fnewton$, as discussed in \sect{ssec:thermotidal_pressure}.

The absorbed-flux fraction, $\opacity$, is inferred from $\presstidevc$ using \eq{opacity_deltap}. The estimated values are of order of ${\sim}70 \%$, exceeding those obtained from the GCM simulations of \cite{DG2024} for the ancient Earth. This difference, however, cannot lead to physical interpretation since the absorbed flux fraction inherits the scale of the torque assumed by \cite{Wu2023} in their Fig.~S4\footnote{\label{footnote:wu_rescaling}The caption of Fig.~S4 in \cite{Wu2023} states that the torque results are normalised to the estimated present-day value, $\torquez=\SI{4.14e15}{\newton \meter}$, indicating that this value was assumed rather than derived from the GCM simulations. This rescaling of the GCM data may account for the discrepancy between the amplitude of the surface-pressure oscillations obtained from the GCM simulations in Fig.~S2 of the same study and the estimate inferred here from the torque shown in Fig.~S4.}. The substantial phase lag of tidal heating derived from the fit, for its part, is consistent with that found for case (b) in \sect{ssec:DG2024}, supporting the interpretation that thermal tides in the simulations of \cite{Wu2023} are driven predominantly by indirect thermal forcing rather than by direct atmospheric absorption alone, as discussed in \sect{sec:phase_shifted_heating}. When interpreted through the lens of analytical models, the GCM results therefore appear to challenge the classical assumption underlying linear tidal theory, according to which thermal tides on Earth are generated by the absorption of incident Solar radiation in the visible range \citep[e.g.][]{CL70}.

In particular, we note that the amplitude of the equatorial surface-pressure oscillations inferred from the fit for present-day Earth, $\abs{\presstideneq}$, is substantially larger than the standard empirical estimates derived from barometric observations \citep[e.g.][]{HC1973,DW1999,Schindelegger2014}. This amplitude can be readily inferred from \eq{torquez_simp}, by writing $\Im \left\{\presstideneq \right\} = \abs{\presstideneq} \sin \left( \phieq \right) $. In the fitted isentropic model, the phase angle of the surface-pressure oscillations, $\phieq$, is 0.7125~rad, corresponding to a pressure maximum at $10{:}38$~LST, approximately one hour later than the observed value \citep[$09{:}45$~LST ; e.g.][]{CL70,Schindelegger2014}. This gives $\abs{\presstideneq} = 205.8 $~Pa, almost twice the empirical estimate of $123.4$~Pa reported by \citet{Schindelegger2014}. Although part of this discrepancy comes from \cite{Wu2023}'s normalisation exceeding the empirical constraint derived in \sect{ssec:application_present_earth}, it also arises from the strong sensitivity of the thermotidal torque to the phase lag of tidal heating. The latter reduces the torque by shifting the tidal bulge towards the zero-torque configuration, in which it is aligned with the planet-star direction. Consequently, a larger amplitude of the surface-pressure oscillations is required to compensate for this reduction in torque.

Finally, we determine the LOD corresponding to the maximum resonant thermotidal torque, $\LODimplus$, by maximising the two fitted models. In both cases, the GCM results of \cite{Wu2023} yield $\LODimplus = 23.9$~hr (see \fig{fig:comparison_DG2024}), implying that the Solar thermotidal torque acting on Earth should currently be close to its maximum value. This theoretical prediction emphasises the difference between the resonant LOD and the LOD maximising the tidal torque, which should be clearly distinguished from each other. Here, this difference in LOD almost reaches two hours given that $\LODresoi{2} \approx 22 \ {\rm hr}$ in both fitted models. This two-hour gap is particularly significant in the context of the Earth's rotational evolution, as it corresponds to a $500{-}1000$~Myr time separation between the resonant maximum and the induced tidal locking \citep[see e.g.][Fig.~5]{Farhat2022b}. This effect adds to the increase of the resonant LOD induced by the tidal heating lag, both of them tending to bring the accelerating peak closer to the present day.

It is noteworthy that a stalled day length of approximately 19~hr, as inferred by \cite{MK2023} and \cite{Wu2023} from their interpretation of geological data, would require a resonant LOD ranging between 17~hr and 18~hr, depending whether tidal heating is delayed or in phase (see \sect{ssec:application_present_earth}) with insolation, respectively. This value is substantially shorter than the 21~hr reference value of \cite{ZW1987}, and than the smallest values obtained by \cite{Wu2023} and \cite{DG2024} from GCM simulations. Typically, in the GCM results shown by \cite{Wu2023} in their Fig.~S2, the shortest resonant LOD, $21.525$~hr, is obtained in the limiting case of a 2~bar-atmosphere with a 100~mbar-$\carbondiox$ partial pressure. Such extreme conditions are consistent with the 60$^\degree {\rm C}$ Earth found by these authors for a 18~hr resonant LOD in their reconstruction of the resonant LOD as a function of mean temperature \citep[see Fig. 5 in][]{Wu2023}. The gap observed between model predictions and the scenario proposed by \cite{MK2023} and \cite{Wu2023} tends to corroborate the conclusions by \cite{Laskar2024}, who note that there is no geological evidence for a stalled day length of 19~hr. Together with the discrepancy identified in the phase of tidal heating, the above features of the Lamb-wave resonance highlight the need for further GCM investigations of the frequency dependence of the thermotidal torque for both present-day and early Earth conditions.

\subsection{\cite{ADL2019}}
\label{ssec:ADL2019}

As a last example, we consider the reference case of \cite{ADL2019}. In this scenario, a dry, bare rock Venus-sized planet orbits a Sun-like star circularly with an orbital radius equal to the present-day Venus-Sun distance. The physical setup of this last case is the same as that adopted by \cite{Leconte2015} to study the inner and outer edges of the habitable zone. Particularly, the 10-bar $\nitrogen$ atmosphere of our idealised planet contains a $\carbondiox$ mixing ratio equivalent to that of the Earth at the turn of the millenium \citep[${\sim} 370$~ppmv; e.g.][]{Etheridge1996}, which represents the only greenhouse gas taken into account in GCM simulations. 

The GCM data of \cite{ADL2019} are reproduced in the bottom right panel of \fig{fig:comparison_DG2024}. The figure displays the imaginary part of the semidiurnal surface pressure component at the equator, $\Im \left\{ \presstideneq \right\}$, plotted as a function of the normalised tidal frequency, $\fdiurnalnorm = \left( \spinrate - \norb \right) / \norb $, with $\norb = \SI{3.237e-7}{\radian \ \second^{-1}} $. We note that \cite{ADL2019} provide values for the degree-2, order-2 coefficient of the surface pressure anomaly, $\presstidenquad$. For convenience, this coefficient is converted into $\presstideneq$ using \eq{deltaps22_deltaps2h} and taking into account the factor of two difference between the conventions adopted in the two papers for spherical harmonic expansions (see Footnote~\ref{footnote:factor_two} in \sect{ssec:general_formulation_torque}).   

In \citet{ADL2019}, the thermotidal torque was computed over a wide range of frequencies, from spin-orbit synchronisation to rapid rotation \citep[see][Fig.~3]{ADL2019}. Consequently, the GCM data capture both the near-synchronous maximum and the resonance associated with the tidally forced Lamb wave. The intermediate frequency dependence agrees well with the scaling law derived from linear theory, $\torquez \scale \ftide^{-1}$. However, in the low-frequency regime, the torque obtained from GCM simulations departs substantially from the radiating-slab model adopted by \cite{Leconte2015} to fit their GCM results (\eq{model_leconte2015}). In particular, it exhibits sharp transitions in the vicinity of the near-synchronous maximum \citep[see][Fig.~3, top right panel]{ADL2019}.

Moreover, the frequency dependence of the torque near resonance deviates markedly from analytical predictions derived under the assumption of direct atmospheric absorption akin to that observed on present-day Earth. Instead of displaying two peaks of comparable amplitude and opposite sign, the spectrum is characterised solely by the negative peak, while the positive counterpart is absent. This behaviour appears to correspond to an asymptotic limit of the symmetry breaking mechanism discussed in \sect{sec:phase_shifted_heating} and identified in the numerical solutions of \cite{DG2024} and \cite{Wu2023}. 

In this section, we conduct a quantitative analysis of the GCM results reported by \cite{ADL2019}, following the same methodology as that applied in \sects{ssec:DG2024}{ssec:Wu2023} to the GCM data of \cite{DG2024} and \cite{Wu2023}. As a first step, the imaginary components of the analytical isentropic and isothermal solutions were fitted to the corresponding data points. They are plotted in \fig{fig:comparison_DG2024} (bottom right panel). For the isentropic model, the best-fitting parameters are $\LODresoi{2} = 20.675 \pm 0.00066~{\rm hr}$, $\presstidevc = \left( 6.7027 \pm 0.0051 \right) \times 10^{-5} $, $\disspar = \left( 1.8719 \pm 0.0025 \right) \times 10^{-2} $, and $\phiheating = 1.5100 \pm 0.0011~{\rm rad}$, yielding a coefficient of determination of $R^2 = 94.0 \%$. For the isothermal model, the corresponding values are $\LODresoi{2} = 20.742 \pm 0.00061~{\rm hr}$, $\presstidevc = \left( 6.9775 \pm 0.0061 \right) \times 10^{-5} $, $\disspar = \left( 1.9879 \pm 0.0031 \right) \times 10^{-2} $, $\phiheating = 1.2891 \pm 0.0018~{\rm rad}$, with a coefficient of determination of $R^2 = 94.3 \%$. The parameter estimates reported in Table~\ref{tab:parameter_constraints} are derived directly from these fitted values. 

Both models successfully reproduce the near-synchronous maximum and the resonant peak, although the quality of the fit remains inferior to that obtained for the GCM results reported by \cite{DG2024} (see \sect{ssec:DG2024}). The inferred absorption fraction and phase lag of tidal heating at resonance are consistent with the estimates derived in \sect{ssec:DG2024} and \sect{ssec:Wu2023}. These results support the interpretation that thermal tides are driven predominantly by an indirect heating mechanism rather than by direct absorption of stellar radiation in the visible range. In particular, the phase lag of tidal heating is sufficiently large to entirely offset the resonant advance of the thermotidal bulge, thereby suppressing the positive torque peak and leaving only its negative counterpart in the spectrum.

Owing to the coarse frequency resolution of the GCM dataset, the maximum amplitude of the resonant peak is considerably less well constrained than in the preceding cases. As a result, the corresponding estimates of the quality factor ($\bpquality \approx 126 {-}187$) and damping ratio ($\disspar \approx 1.87 {-} \num{1.99e-2} $) are subject to substantial uncertainties and should therefore be interpreted with caution. Similar uncertainties affect the model prediction for the near-synchronous peak, which is poorly resolved by GCM data. The normalised diurnal frequency obtained from the fit in the isentropic model, $\fdiurnalnorm = 4.88 $, is less than that of the maximum GCM value, $\fdiurnalnorm = 6.0$, which also corresponds to the frequency interval separating two GCM data points (the difference is less in the isothermal model, for which the inferred value is $\fdiurnalnorm = 5.16 $). As a consequence, the uncertainty associated with $\omega_0$ and $\LOD_0$ is greater than 23\% in these calculations, and propagate to other fitted parameters. Typically, we observe a $52\%$ relative difference between the prediction of the fitted model ($\fdiurnalnorm = 4.88 $) and that obtained by \cite{ADL2019} using a cubic-spline interpolation of a larger set of GCM data ($\fdiurnalnorm = 7.4$). Furthermore, it is worth noting that the near-synchronous peak described by GCM simulations is flatter than that predicted by the analytical theory in the vicinity of the maximum, resulting in a ill-defined peak frequency \citep[see][Fig.~3, top right panel]{ADL2019}.

Despite these limitations, the inferred parameters remain consistent with the scaling relation given by \eq{scaling_omega0}. In particular, \eq{scaling_omega0} predicts the radiative-cooling frequency of the 10-bar atmosphere to be approximately one order of magnitude lower than that of present-day Earth, consistent with the inferred range, $\omega_0 = 1.58{-}\num{1.67e-6}~{\rm rad \ s^{-1}}$. However, as discussed in \sect{ssec:DG2024}, the dependence of $\omega_0$ on surface pressure warrants further investigation, as this parameter is governed by atmospheric physics in a complex manner. For instance, \cite{Auclair2019} inferred from GCM simulations of idealised Venus-like planets a power-law dependence $\omega_0 \scale \psurf^{-0.30}$, which departs substantially from a linear scaling.

Finally, it is worth noting that \cite{Farhat2024} achieved a comparably good fit to the same dataset by assuming that atmospheric tides are forced through radiative-diffusive heat exchange with the planetary surface \citep[see][Fig.~K.1]{Farhat2024}. Within the framework developed here, the corresponding ab initio model can be incorporated by replacing the factor $\exp \left(- \inumber  \phiheating \XX \right)$ in \eq{phase_forcing_normalised} with the transfer function given by Eq.~(H.4) of \cite{Farhat2024}. Comparable results may likewise be obtained using an equivalent mass-spring-damper formulation, as will be discussed in \paperthree. Regardless of the specific modelling approach adopted, the inclusion of delayed tidal heating appears essential for accurately reproducing GCM results, which departs significantly from the conventional assumption of direct atmospheric absorption. Further work is therefore required to revisit and strengthen the numerical modelling framework applied to dry rocky planets. 

\section{Conclusions}
\label{sec:conclusions}

This study continues the theoretical work initiated in \paperone, which aimed to analytically characterise the influence of atmospheric thermal tides on the rotation of rocky planets. In \paperone, starting from the primitive equations of fluid dynamics, we derived a closed-form solution describing the three-dimensional atmospheric response to gravitational and thermal tidal forcings (\paperone, Eqs.~85-92). This solution, obtained within linear theory, incorporates dissipative effects through Newtonian cooling and assumes a vertically decaying distribution of tidal heating. It applies uniformly to two atmospheric structures of practical relevance: isentropic atmospheres (uniform potential temperature) and isothermal atmospheres (uniform temperature). In particular, we explicitly identified the forcing frequencies at which the main tidal modes -- namely Lamb waves -- are resonantly excited. 

In the present work, we have extended this framework by focusing on semidiurnal thermal tides and their specific role in the long-term evolution of planetary rotation. Using the solution from \paperone, we have expressed the thermotidal torque as a function of tidal frequency and a reduced set of physical parameters defining the star-planet system. This has been done for both the convective and isothermal equilibria, assuming a Dirac distribution of tidal heating for the latter \citep[in line with][]{DI1980}. The resulting expressions (\eqs{presstidev_surf_neu}{presstidev_surf_isotherm_dirac} and Table~\ref{tab:solutions}) enable a detailed characterisation of the frequency dependence of the torque in the regime of quasi-adiabatic compression, where dissipative processes remain weak. We have derived scaling laws describing key properties of this dependence -- such as the frequencies, amplitudes, and widths of resonant peaks -- as functions of dimensionless control parameters (Table~\ref{tab:solution_properties}). Finally, we have benchmarked the analytical solutions against numerical results from GCM simulations for the Earth and Venus (Table~\ref{tab:parameter_constraints}; \fig{fig:comparison_DG2024}).

Although the analytical expressions of the tidal torque differ markedly between isentropic and isothermal atmospheres, they exhibit several robust similarities. Both predict the coexistence of resonant modes, which grow in amplitude near specific forcing frequencies, and non-resonant modes, which gradually diminish as frequency increases. Despite their simplicity, both formulations reproduces the main features of numerical solutions with notable accuracy. In particular, they asymptotically converge towards the functional behaviour empirically inferred in the zero-frequency limit from GCM studies of Venus-like, near-synchronous planets. This agreement holds for both resonant and non-resonant modes. The most significant results, however, concern the high-frequency regime.

The analytical model reproduces with high accuracy the Precambrian resonant peak obtained in Earth GCM simulations for surface pressure oscillations, thereby demonstrating its ability to capture the influence of Lamb-wave resonance on the planet's past rotational evolution. The agreement between analytical and numerical results provides explicit relationships between resonance characteristics and key atmospheric parameters, including the adiabatic index, and enables the estimation of free parameters -- such as the radiative cooling time or the absorbed flux fraction -- from numerical solutions. Moreover, this consistency indicates that the thermal tides produced in simulations of Earth's palaeoatmospheres operate within the quasi-adiabatic regime, where the scaling laws derived here remain valid, albeit near the critical limit. Typically, our analysis suggests that the past Earth's effective radiative cooling frequency lies between $\num{3.23e-5}~{\rm rad \ s^{-1}}$ and $\num{8.06e-5}~{\rm rad \ s^{-1}}$. Besides, the amplitude and lag of semidiurnal surface pressure oscillations predicted by the isentropic direct-absorption model for present-day Earth conditions are in excellent agreement with estimates derived from barometric observations.

Similarly, by fitting the analytical model to numerical simulations, we have recovered previous estimates of Venusian semidiurnal tidal heating and derived quantitative constraints on both the frequency and amplitude of the associated thermotidal resonance. The same analysis also yields physically consistent results for the idealised 10-bar rocky-planet configuration considered in this study. Across most configurations examined, the inferred tidal heating is characterised by substantial phase lags and large effective absorption fractions, departing markedly from the commonly adopted assumption of direct atmospheric absorption. This systematic discrepancy suggests that semidiurnal thermal tides in the analysed GCM simulations are driven predominantly by indirect thermal forcing mechanisms, such as heat diffusion within the soil or latent heat transport, as discussed in previous studies.

The phase lag of tidal heating has important implications for the rotational evolution of rocky planets, as it can strongly attenuate -- or even suppress entirely -- the accelerative tidal torque that may counteract tidal deceleration in the vicinity of the Lamb-wave resonance. Delayed tidal heating offers less favourable configurations for resonant torque amplification than the direct-absorption regime, which characterises semidiurnal tides on present-day Earth. This effect notably weakens the Precambrian tidal-locking hypothesis in scenarios where the Precambrian atmosphere differed from the actual one. More generally, the results demonstrate that the amplitude of pressure oscillations alone is insufficient to quantify the thermotidal torque: pronounced resonant peaks in the pressure response do not necessarily translate into strong thermodynamical accelerations of planetary rotation. These conclusions should nevertheless be interpreted with appropriate caution, as the comparative analysis presented here relies on a limited set of benchmark simulations. Additional GCM studies are therefore needed to provide a more comprehensive characterisation of the thermotidal response of planetary atmospheres. 

Because the analytical solution developed here can be readily incorporated into planetary evolution models, it represents a valuable tool for investigating equilibrium states and the long-term rotational evolution of rocky planets. Its excellent agreement with Earth GCM results makes it particularly suitable for studying the role of thermal tides in Earth's rotational length-of-day history. In particular, it offers a versatile framework for quantifying the influence of Lamb-wave resonance during the Precambrian Era and for exploring the associated parameter space. More broadly, the cases of Earth and Venus highlight the intimate coupling between planetary rotation and thermal tides and underscore the potential importance of the latter in shaping the rotational evolution of rocky exoplanets in the habitable zone of their host stars. In the forthcoming \paperthree, we will show that the complex frequency behaviour of thermotidal surface pressure oscillations described by our analytical solutions and GCM simulations is accurately reproduced by a simple mass-spring-damper model.

\begin{acknowledgements}
The authors are grateful to the anonymous referee for their insightful comments, which substantially improved the quality and clarity of the manuscript. M.~Farhat is supported by the Miller Institute for Basic Research in Science at UC Berkeley through a Miller Research Fellowship. This work was supported by the European Research Council (ERC) under the European Union’s Horizon 2020 research and innovation programme (Advanced Grant AstroGeo-885250). This research has made use of NASA's Astrophysics Data System. 
\end{acknowledgements}

%=========%=========%=========%=========%=========%=========%=========%=
%BIBLIOGRAPHIE
%=========%=========%=========%=========%=========%=========%=========%=
\bibliographystyle{aa}  % A&A bibliography style file (aa.bst)
\bibliography{aa58544-25} % references .bib file

\begin{appendix}
\ifrevised \else \nolinenumbers \fi
\section{Nomenclature}
\label{app:nomenclature}

The notations introduced in the main text are listed below in order of appearance.

\vspace{-1.2cm}

\printnomenclature

\section{Shape factor characterising the latitudinal structure of the tidal bulge}
\label{app:shape_factor}

In \eq{presstideneq_presstidenquad}, the amplitude of semidiurnal surface pressure oscillations at the equator is connected to the corresponding degree-2, order-2 spherical harmonic by introducing the shape factor $\cshape$. This parameter quantifies the deviation of the latitudinal structure of semidiurnal surface pressure variations from the quadrupole, the latter being shaped by the degree-2, order-2 normalised associated Legendre function. The latitudinal distribution of the semidiurnal component of surface pressure is assumed to be described by a function $f \left( \mucos \right)$, with $\mucos = \cos \col$. The shape factor associated with $f$ is formulated using the dot product defined in \eq{dotproduct} as 
\begin{equation}
\label{cshape_definition}
\cshape = \scal{\LegFnorm{2}{2}}{f} \frac{\LegFnorm{2}{2} \left( 0 \right)}{f \left( 0 \right)},
\end{equation}
where $\LegFnorm{2}{2}$ designates the degree-2, order-2 normalised associated Legendre function introduced in \eq{flux_Plm}, given by \citep[e.g.][Appendix~A]{Auclair2019}
\begin{equation}
\LegFnorm{2}{2} = 3 \sqrt{\frac{5}{48}} \left( 1 - \mucos^2 \right).
\end{equation}
The scaling function $\sin^3 \col$ reported by \cite{Schindelegger2014} for the Earth's semidiurnal thermal tides corresponds to 
\begin{equation}
f \left( \mucos \right) = \left( 1 - \mucos^2 \right)^{3/2}.
\end{equation}
Using this expression in \eq{cshape_definition}, we obtain
\begin{equation}
\cshape = \frac{75 \pi}{256} \approx 0.92,
\end{equation}
which is the value adopted in calculations throughout the study.

\section{Multipole expansion of stellar irradiance}
\label{app:expansion_stellar_irradiance}

We detail here the derivations leading to \eq{flux_Plm}. As a first step, the spatial distribution of the incident stellar flux, given by \eq{flux_spatial_distribution}, is expanded as a series of Legendre polynomials,
\begin{equation}
\label{flux_legendre_polynomial}
\flux = \Fstar \sum_{\llat=0}^{+\infty} \flux_\llat \LegP{\llat} \left( \cos \zenithphi \right),
\end{equation}
where the expansion coefficients are expressed as 
\begin{equation}
\flux_\llat = \frac{2 \llat+1}{2} \integ{\mucos \LegP{\llat} \left( \mucos \right)  }{\mucos}{0}{1}.
\end{equation}
The Legendre polynomial addition theorem \citep[e.g.][Eq.~(9.189), p.~600]{Arfken2005} allows for separating the local coordinates $\left( \col , \lon \right)$ from the coordinates of the star $\left( \colstar, \lonstar \right)$ in $\LegP{\llat} \left( \cos \zenithphi \right)$, which is expressed as 
\begin{equation}
\LegP{\llat} \left( \cos \zenithphi \right) = \frac{4 \pi}{2 \llat + 1} \sum_{\mm=-\llat}^\llat \Ylm{\llat}{\mm} \left( \col, \lon \right) \conj{\Ylm{\llat}{\mm}} \left( \colstar, \lonstar \right).
\end{equation}
In the above equation, $\conj{\Ylm{\llat}{\mm}}$ designates the complex conjugate of the degree-$\llat$, order-$\mm$ spherical harmonic, $\Ylm{\llat}{\mm}$ \citep[e.g.][Sect.~5.2]{Varshalovich1988}. Writing the spherical harmonics in terms of the normalised associated Legendre functions, $\LegFnorm{\llat}{\mm}$, we obtain
\begin{equation}
\Ylm{\llat}{\mm} \left( \col, \lon \right) = \frac{1}{\sqrt{2 \pi}} \LegFnorm{\llat}{\mm} \left( \cos \col \right) \expo{\inumber \mm \lon},
\end{equation}
and \eq{flux_legendre_polynomial} thus becomes
\begin{equation}
\label{flux_alf_series}
\flux = \Fstar \sum_{\llat=0}^{+ \infty} \sum_{\mm=-\llat}^\llat \frac{2 \flux_\llat}{2 \llat+1} \LegFnorm{\llat}{\mm} \left( \cos \colstar \right) \LegFnorm{\llat}{\mm} \left( \cos \col \right) \expo{\inumber \mm \left( \lon - \lonstar \right)}.
\end{equation}
In the coplanar-circular configuration, where a zero-obliquity planet orbits its host star circularly, the colatitude of the star is $\colstar = \pi/2$ and its longitude is $\lonstar = - \fdiurnal \time$, with $\fdiurnal = \spinrate - \norb$ denoting the diurnal tidal frequency. Besides, the normalised associated Legendre functions of positive and negative orders are linked together by the symmetry relation
\begin{equation}
\LegFnorm{\llat}{-\mm} \left( \cos \col \right) = \left( -1 \right)^{\mm} \LegFnorm{\llat}{\mm} \left( \cos \col \right),
\end{equation}
which allows for gathering pairs of terms of orders $\mm$ and $-\mm$ in \eq{flux_alf_series}. It follows
\begin{equation}
\flux \left( \col , \lon , \time \right) = \Fstar \Re \left\{  \sum_{\llat=0}^{+ \infty} \sum_{\mm = 0}^\llat \fluxnlm \LegFnorm{\llat}{\mm} \left( \cos \col \right) \expo{\inumber \mm \left( \fdiurnal \time + \lon \right)}   \right\},
\end{equation}
where the expansion coefficients are defined as 
\begin{equation}
\fluxnlm = \left( 2 - \kron{\mm}{0} \right) \frac{2 \flux_\llat}{2 \llat+1}  \LegFnorm{\llat}{\mm} \left( 0 \right).
\end{equation}
It is noteworthy that $\LegFnorm{\llat}{\mm} \left( 0 \right)$ is a special value of the normalised associated Legendre function given by \citep[e.g.][Sect.~5.14, Eq.~(2)]{Varshalovich1988}
\begin{equation}
\LegFnorm{\llat}{\mm} \left( 0 \right)  =  \left\{
\begin{array}{ll}
\! \! \left( -1 \right)^{\frac{\llat+\mm}{2}} \! \! \sqrt{\frac{\left(2 \llat+1\right) \left( \llat+\mm-1 \right)!! \left( \llat-\mm-1 \right)!!}{2\left(\llat+\mm \right)!! \left( \llat-\mm \right)!!} } & \! \! \mbox{if $\llat+\mm$ is even}, \\
\! \! 0 & \! \! \mbox{if $\llat+\mm$ is odd}.
\end{array}
\right.
\end{equation}
In the above equation, the double factorial of a positive integer $\nn$ is defined as
\begin{equation}
\nn !! = \prod_{\kk=0}^{\ceil{\frac{\nn}{2}} -1} \left( \nn - 2 \kk \right) = \nn \left( \nn-2 \right) \left( \nn - 4 \right) \ldots,
\end{equation}
with $\ceil{\nn/2}$ denoting the ceiling of $\nn/2$.

\section{Benchmark of analytical scalings against \cite{DG2024}'s GCM data}
\label{app:scalings_DG2024}

\begin{table}[h]
\centering
\caption{\label{tab:parameter_constraints_ana} Estimates of analytical solutions' parameters deduced from GCM simulations using analytical (Ana.) or numerical (Num.) methods for the case treated in \sect{ssec:DG2024}.}
\begin{small}
\begin{tabular}{l l l c c } 
 \hline 
 \hline 
 & Notation & Units & \multicolumn{2}{c}{Values} \\
  \hline \\[-0.3cm] 
% Reference
Method & & 
 & Ana.
 & Num.
 \\
 \hline \\[-0.3cm] 

 % ------------------------------------------------
 % NEUTRALLY STRATIFIED
 \multirow{17}{*}{Isentropic} 
% Pressure q0 at maximum near synchronisation
& $q_0$ & Pa
& 30.9
& 29.3
\\ 
 % Frequency Omega-n at maximum near synchronisation
& $\omega_0$ & ${\rm rad \ s^{-1}}$
& $\num{1.92e-5}$
& $\num{2.04e-5}$
\\ 
 % Rotation period at which maximum occurs near synchronisation
& $\LOD_0$ & days 
& 3.78
& 3.57
\\
 & $\fresoi{2}$ & ${\rm rad \ s^{-1}}$ 
 & $\num{1.58e-4}$ \tablefootmark{$\gcmdata$}
 & $\num{1.59e-4}$
 \\ 
 % Daylength at which resonance occurs
 & $\LODresoi{2}$ & hr  
 &  22.1 \tablefootmark{$\gcmdata$}
  & 22.0
 \\
  % Daylength of the maximum of positive torque
 & $\LODimplus$ & hr  
 & 22.6
 & 22.9
 \\
 % Normalised dissipation parameter
 & $\disspar$&  --
 &  0.243
 & 0.257
 \\
 % Newtonian cooling frequency
 & $\fnewton$&  ${\rm rad \ s^{-1}}$
 & $\num{3.84e-5}$
 & $\num{4.07e-5}$
 \\
% Normalised pressure constant
 & $\presstidevc$ &  --
 & $\num{1.50e-4}$
 & $\num{1.50e-4}$
 \\
% Temperature at surface deduced from the Lamb resonance frequency
 & $\Tsurf$ &  K
 & 317
 & 319
 \\
% Bulk temperature deduced from the Lamb resonance frequency
 & $\Tbulk$ &  K
 & 267
 & 269
 \\
% Pressure height at surface deduced from the Lamb resonance frequency
 & $\heightsurf$ &  km
 & 9.26 
 & 9.34
 \\
% Opacity deduced from the Lamb resonance frequency
 & $\opacity$ & $\%$
 & 42.4
 & 42.9
 \\
% Gain coefficient of the equivalent 2nd-order bandpass filter
 & $\bpgain$ & --
 & $\num{3.31e-3}$
 & $\num{3.13e-3}$
 \\
 % Quality of the equivalent 2nd-order bandpass filter
 & $\bpquality$ & --
 & 22.0
 & 20.8
 \\
% Lag angle of the equivalent 2nd-order bandpass filter
 & $\philag$ & rad
 & 0.224
 & 0.228
 \\
% Phase angle of tidal heating
 & $\phiheating$ & rad
 & 0.513
 & 0.533
 \\
 \hline \\[-0.3cm] 
% ------------------------------------------------
% ISOTHERMAL
\multirow{17}{*}{Isothermal} 
% Pressure q0 at maximum near synchronisation
& $q_0$ & Pa
& 49.4
& 46.7
\\ 
% Frequency Omega-n at maximum near synchronisation
& $\omega_0$ & ${\rm rad \ s^{-1}}$
& $\num{1.56e-5}$
& $\num{1.62e-5}$
\\ 
 % Rotation period at which maximum occurs near synchronisation
& $\LOD_0$ & days 
& 4.66
& 4.50
\\
% Lamb resonance frequency from given surface temperature
& $\fresoi{2}$ & ${\rm rad \ s^{-1}}$ 
& $\num{1.58e-4}$ \tablefootmark{$\gcmdata$}
& $\num{1.58e-4}$
\\
% Daylength at which resonance occurs
& $\LODresoi{2}$  & hr  
& 22.1  \tablefootmark{$\gcmdata$}
& 22.0
\\
  % Daylength of the maximum of positive torque
 & $\LODimplus$ & hr  
 & 22.6
 & 22.8
 \\
% Normalised dissipation parameter
& $\disspar$& --
& $\num{0.198}$
& $\num{0.204}$
\\
 % Newtonian cooling frequency
 & $\fnewton$&  ${\rm rad \ s^{-1}}$
 & $\num{3.12e-5}$
 & $\num{3.23e-5}$
 \\
% Normalised pressure constant
& $\presstidevc$ &  --
& $\num{1.95e-4}$
& $\num{1.91e-4}$
\\
% Temperature at surface deduced from the Lamb resonance frequency
 & $\Tsurf$ &  K
 & 257
 & 259
 \\
% Bulk temperature deduced from the Lamb resonance frequency
 & $\Tbulk$ &  K
 & 257
 & 259
 \\
% Pressure height at surface deduced from the Lamb resonance frequency
 & $\heightsurf$ &  km
 & 7.53
 & 7.58
 \\
 % Opacity deduced from the Lamb resonance frequency
 & $\opacity$ & $\%$
 & 44.8
 & 44.1
 \\
% Gain coefficient of the equivalent 2nd-order bandpass filter
 & $\bpgain$ & --
 & $\num{3.31e-3}$
 & $\num{3.13e-3}$
 \\
 % Quality of the equivalent 2nd-order bandpass filter
 & $\bpquality$ & --
 & 22.0
 & 21.3
 \\
% Lag angle of the equivalent 2nd-order bandpass filter
 & $\philag$ & rad
 & 0.224
 & 0.231
 \\
% Phase angle of tidal heating
 & $\phiheating$ & rad
 & 0.387
 & 0.399
 \\
\hline
 \end{tabular}
 \end{small}
 \tablefoot{{\it From top to bottom:} the fixed parameters and the corresponding estimates of free parameters in analytical solutions for the two studied atmospheric structures (isentropic and isothermal atmospheres). Calculations are detailed in \append{app:scalings_DG2024} (Ana.) and \sect{sec:benchmark_gcm} (Num.). \\
\tablefoottext{$\gcmdata$}{Quantities constrained by cubic-spline interpolation of the GCM data, from which the remaining quantities are inferred.}
  } 
 \end{table}

\begin{figure}
\centering
\includegraphics[width=0.48\textwidth,trim=0cm 0cm 0.5cm 0cm,clip]{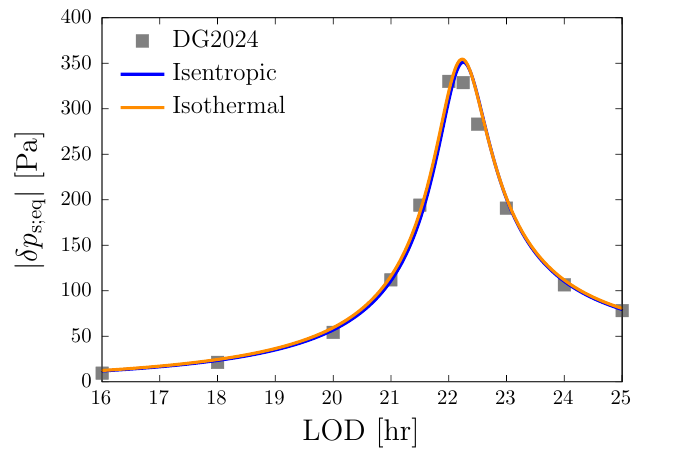} \\
\includegraphics[width=0.48\textwidth,trim=0cm 0cm 0.5cm 0cm,clip]{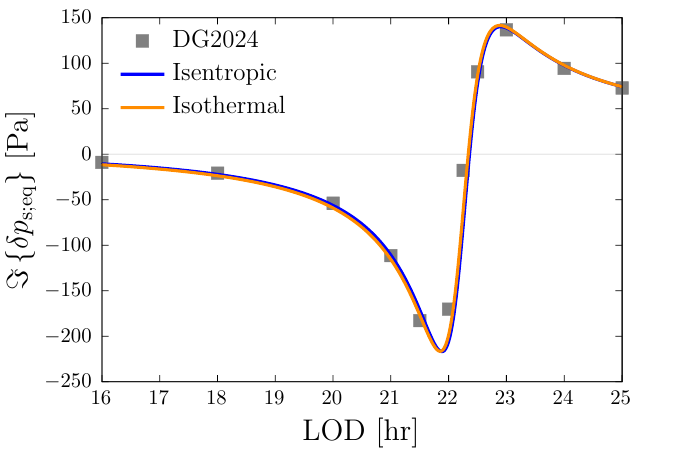}
\caption{Modulus (top) and imaginary part (bottom) of surface pressure oscillations as a function of the LOD ($2 \pi / \abs{\spinrate - \norb}$) in the case labelled `ExoCAM, 30 ppm ${\rm CH_4}$' of \citet[][Figs.~1c and~3]{DG2024}. Plots show the numerical solutions obtained using GCM simulations (gray markers), and the closed-form solutions given by our analytical model for isentropic (solid blue line) and isothermal (solid orange line) atmospheres. The two analytical solutions, given by \eqs{presstidev_surf_neu}{presstidev_surf_isotherm_dirac}, are fitted to GCM data points using the formulae provided by Table~\ref{tab:solution_properties} and the parameter values of Table~\ref{tab:parameter_constraints}, as detailed in \append{app:scalings_DG2024}.}
    \label{fig:comparison_DG2024_ana}
\end{figure}

In \sectsto{ssec:DG2024}{ssec:ADL2019}, the isentropic and isothermal models are fitted to GCM data numerically using a minimisation algorithm. However, this is not the only way to proceed. Alternatively, one may fit the models analytically using the formulae established in \sect{sec:frequency_behaviour} and reported in Table~\ref{tab:solution_properties}. In this appendix, we detail this analytical method for the case of \cite{DG2024}, which is treated in \sect{ssec:DG2024}. Analytical results are benchmarked against numerical ones in Table~\ref{tab:parameter_constraints_ana}. The corresponding solutions are plotted in \fig{fig:comparison_DG2024_ana} together with GCM data, analogous to \fig{fig:comparison_DG2024}.

To begin with, we derive the parameters $\fresoi{2}$, $\presstidevc$, and $\disspar$ from the resonant peak LOD ($\LODresoi{2}$), amplitude ($\abs{\presstideneq}_{\rm max}$), and FWHM ($\dxmidres$). These three quantities are obtained by performing a cubic-spline interpolation of GCM data for the modulus of $\presstideneq$, which are shown by \fig{fig:comparison_DG2024_ana} (top panel). In particular, the resonant LOD is approximated to the LOD at which the amplitude of semidiurnal tidal surface-pressure oscillations is maximal, and the normalised FWHM is expressed as
\begin{equation}
\dxmidres = \frac{\LODresoi{2} \left( \LOD_2 - \LOD_1 \right)}{\LOD_1 \LOD_2},
\end{equation}
with $\LOD_1$ and $\LOD_2$ being the two LODs at which the modulus is half its maximum value.

For the resonant daylength, we find $22.1$~hr, which is just $0.5\%$ larger than best-fitting values in \sect{ssec:DG2024}. For the maximal pressure amplitude at the equator and the normalised FWHM of the peak, we get $\abs{\presstideneq}_{\rm max} = 337$~Pa and $\dxmidres = \num{7.88e-2}$, respectively. We deduce $\disspar$ from $\dxmidres$ in the two models by retaining only the leading-order terms of the power-series expansions in $\disspar$ reported in Table~\ref{tab:solution_properties},
\begin{align}
& \disspar \approx \frac{\dxmidres}{\sqrt{3} \rcp}  && \mbox{(isentropic),} \\
& \disspar \approx \frac{\dxmidres}{\sqrt{3} \rcp \adiabexp} && \mbox{(isothermal)}.
\end{align}
The estimated values -- $\disspar=0.243$ (isentropic) and $\disspar=0.198$ (isothermal) -- are $5\%$ smaller than those obtained numerically using the minimisation method, consistent with the error of order ${\sim} \disspar^2$ introduced by reducing series to their leading-order terms. The Newtonian cooling frequency and the pressure scale height at the planet's surface are then deduced from $\disspar$ and $\LODresoi{2}$ using \eqs{reduced_frequency}{heightsurf_sigmares}, respectively.

We note that the accuracy of the inferred parameters is inherently limited by the frequency resolution of the GCM datasets. Their relative uncertainty may be approximated by the ratio of the LOD spacing between adjacent data points to the resonant LOD. For the dataset considered here, this ratio is 0.023, corresponding to an uncertainty of approximately 2.3\% in the determination of the resonant LOD. The resulting error propagates to all inferred quantities and adds to the aforementioned $\disspar^2$-uncertainty.

As discussed in \sect{sec:phase_shifted_heating}, the phase lag of tidal heating (see \eq{phase_forcing_normalised}) does not affect the modulus of surface pressure oscillations. Therefore, to compute this phase lag, we use the imaginary component of the numerical solution of \citet[][Fig.~1c ; see the grey markers in the bottom panel of \fig{fig:comparison_DG2024_ana}]{DG2024}. The phase angle $\phiheating$ is readily deduced from the ratio between the two extrema of the imaginary part using \eqs{ratio_extrema_imag}{phiatm_expr}. Interpolating GCM data with a cubic spline, we find $\Im \left\{ \presstideneq \right\}_{\iplus} = 137.17 $~Pa for the positive peak and $\Im \left\{ \presstideneq \right\}_{\iminus} = -215.62$~Pa for the negative one, which yields $\philag = 0.224 \ {\rm rad}$. As a result, $\phiheating = 0.513 \ {\rm rad}$ in the isentropic model and $\phiheating = 0.387 \ {\rm rad}$ in the isothermal one. 

To estimate $\presstidevc$, we exploit its relationship with the height of the resonant peak (Table~\ref{tab:solution_properties}), retaining only the leading-order term in $\disspar$,
\begin{align}
& \presstidevc \approx \rcp \disspar \, \absdpresplusi{2}{2} && \mbox{(isentropic)}, \\
& \presstidevc \approx \frac{\adiabexp - 1}{2 - \adiabexp} \disspar \absdpresplusi{2}{2} && \mbox{(isothermal)}.
\end{align}
As in \sect{ssec:DG2024}, we express $\presstidevsurfi{2}{2}$ as a function of $\presstideneq$ using \eq{deltaps22_deltapseq}, which yields 
\begin{align}
& \presstidevc \approx \rcp \disspar \sqrt{\frac{16}{15}}  \frac{\cshape  \abs{\presstideneq}_{\rm max}}{\scal{\HoughF{2}{2}{1}}{ \LegFnorm{2}{2}} \psurf}&& \mbox{(isentropic)}, \\
& \presstidevc \approx \frac{\adiabexp - 1}{2 - \adiabexp} \disspar \sqrt{\frac{16}{15}}  \frac{ \cshape \abs{\presstideneq}_{\rm max}}{\scal{\HoughF{2}{2}{1}}{ \LegFnorm{2}{2}} \psurf} && \mbox{(isothermal)}.
\end{align}
The analytical fits, plotted in \fig{fig:comparison_DG2024_ana}, agree very well with the GCM outputs. Although not as good as the numerical fits, they do not differ much from the latter, indicating that analytical scalings can provide accurate estimates of the inferred model parameters even if they are limited to leading-order terms. 

\end{appendix}
\end{document}